\documentclass[sigconf]{acmart}

\usepackage{booktabs} 
\usepackage{subfig}
\usepackage{colortbl}
\usepackage{multirow}
\usepackage{xcolor}
\usepackage{amsfonts}
\usepackage{amsthm}
 
\theoremstyle{definition}

\newcommand\literature[1]{\textcolor{blue}{#1}}

\newcommand\ainote[1]{\textcolor{red}{[AI: #1]}}

\newcommand\shnote[1]{\textcolor{purple}{#1}}

\newcommand{\ignore}[1]{}

\setcopyright{acmcopyright}

\begin{document}
\title{On the Privacy of dK-Random Graphs}


\author{Sameera Horawalavithana}
\orcid{1234-5678-9012}
\affiliation{%
  \institution{University of South Florida}
  \city{Tampa} 
  \state{FL} 
}
\email{sameera1@mail.usf.edu}

\author{Adriana Iamnitchi}
\affiliation{%
	\institution{University of South Florida}
	\city{Tampa} 
	\state{FL} 
}
\email{anda@cse.usf.edu}

%
%
%
%
%


\begin{abstract}

Real social network datasets provide significant benefits for understanding phenomena such as information diffusion or network evolution. 
Yet the privacy risks raised from sharing real graph datasets, even when stripped of user identity information, are significant. 
Previous research shows that many graph anonymization techniques fail against existing graph de-anonymization attacks.
However, the specific reason for the success of such de-anonymization attacks is yet to be understood.
This paper systematically studies the structural properties of real graphs that make them more vulnerable to machine learning-based techniques for de-anonymization. 
More precisely, we study the boundaries of anonymity based on 
the structural properties of real graph datasets in terms of how their $dK$-based anonymized versions resist (or fail) to various types of attacks.
Our experimental results lead to three contributions. 
First, we identify the strength of an attacker based on the graph characteristics of the subset of  nodes from which it starts the de-anonymization attack. Second, we quantify the relative effectiveness of $dK$-series for graph anonymization.
And third, we identify the properties of the original graph that make it more vulnerable to de-anonymization.
\end{abstract}

\maketitle

\section{Introduction}


Social graphs serve as mathematical models of network structures that represent interactions between real-world entities.  
Such graphs are often mined to uncover insights about the structure and function of the interactions represented.  
This substantial scientific value to the research community comes with risks: the release of such data may jeopardize the privacy of individuals. 
The Web-strip~\cite{web-strip-scandal} and AOL~\cite{AOLscandal} scandals are textbook examples on breaching the privacy of individuals by publicly releasing unsanitized data.

Many anonymization methods have been proposed to mitigate the privacy invasion of individuals from the public release of graph data~\cite{ji2016survey}.
Naive anonymization schemes employ methods to scrub identities of nodes without modifying the graph structure.
Structural anonymization methods change the topology of the original graph while attempting to preserve (at least some) resemblance with the original graph characteristics~\cite{liu2008towards,
sala2011sharing,liu2016linkmirage}.  
A main research challenge is to develop a principled understanding of the two conflicting goals of preserving privacy and preserving utility at the release of anonymized graph data.

One approach for preserving utility in anonymized graphs is to generate graphs with particular characteristics (utility).
$dK$-graphs~\cite{mahadevan2006systematic} are one such class of techniques that generate graphs with a given degree distribution.
The node degree distribution has been shown to be a defining property for many real graph characteristics~\cite{orsini2015quantifying}.
While $dK$-graphs model topological constraints systematically, they are known to be less random and more structured, the higher the $d$~\cite{orsini2015quantifying}.
Significant attention has been invested in increasing the utility of $dK$-based graph generation techniques. 
What is not well understood, however, is the interplay between the anonymity guarantees that $dK$-graphs provide vs. the strength of the attack and the particularities of the dataset to be anonymized. 

This work addresses this very interaction by employing a machine learning approach to mount de-anonymization attacks on $dK$ graphs generated from a set of real social graphs. 
We believe that an attacker that uses machine learning to mount its attacks is a realistic attacker in today's context~\cite{sharad2014automated}. 
Under this machine learning framework, inspired by~\cite{Sharad2016benchmark}, we set three objectives. 
First, we plan to understand what makes an attack more successful.  
Second, we want to quantify the power of $dK$-based graph generation techniques for providing anonymity.
And third, we try to understand what makes some graph datasets more resilient to de-anonymization attacks in the specific case of $dK$-based anonymization techniques. 

The remainder of this paper is organized as follows. 
Section~\ref{sec:related_work} outlines the related work. 
The system model to quantify anonymity is presented in Section~\ref{sec:methodology}, while Section~\ref{sec:characteristics} describes the characteristics of the datasets we used in our empirical investigations. 
Section~\ref{sec:experiments} presents our results. 
We discuss and summarize our contributions in Section~\ref{sec:discussions}.


\section{Related Work}
\label{sec:related_work}


The availability of auxiliary data helps reveal the true identities of anonymized individuals, as proven empirically in large privacy violation incidents~\cite{NetflixScandal,griffith2005messin}.
Similarly, in the case of graph deanonymization attacks, information from an auxiliary graph is used to re-identify the nodes in an anonymized graph~\cite{narayanan2009anonymizing}. 
The quality of such an attack is determined by the rate of correct reidentification of the original nodes in the network.
In general, de-anonymization attacks harness structural characteristics of nodes that are uniquely distinguishable~\cite{ji2016survey}.
Many such attacks can be categorized into \emph{seed-based} and \emph{seed-free}, based on the prior seed knowledge available to an attacker~\cite{ji2016survey}. 

In seed-based attacks, the process of de-anonymization is conducted to re-identify nodes and ties with the support of sybil nodes~\cite{backstrom2007wherefore} or some known mappings of nodes in an auxiliary graph~\cite{narayanan2011link, srivatsa2012deanonymizing, ji2014structure, ji2016general, korula2014efficient}. 
The effectiveness of such attacks is influenced by the quality of the seeds~\cite{Sharad2016benchmark}.

In seed-free attacks, the problem of de-anonymization is usually modeled as a graph matching problem. 
Several research efforts have proposed statistical models for the re-identification of nodes without relying on seeds, such as the Bayesian model~\cite{pedarsani2013bayesian} or optimization models~\cite{ji2014structural, ji2016structuralsf}. 
Many heuristics were taken into account for the propagation process of re-identification, exploiting graph characteristics such as degree~\cite{gulyas2016efficient}, k-hop neighborhood~\cite{yartseva2013performance}, linkage-covariance~\cite{aggarwal2011hardness}, eccentricity~\cite{narayanan2009anonymizing}, or community~\cite{nilizadeh2014community}.
However, the success rate of a de-anonymization process as often reported in the literature is dependent on the chosen heuristic of the attack, which is typically designed in knowledge of the anonymization technique.
Comparing the strengths of different anonymization techniques thus becomes challenging, if not impossible. 

Several works proposed theoretical frameworks to examine how vulnerable or de-anonymizable any (anonymized) graph dataset is, given its structure~\cite{pedarsani2011privacy,ji2014structural, ji2015your, ji2016relative}.
However, some techniques are based on Erd{\"o}s-R{\`e}nyi (ER) models~\cite{pedarsani2011privacy}, while others make impractical assumptions about the seed knowledge~\cite{ji2015your}.
Ji et. al.~\cite{ji2016relative} also introduced a configuration model to quantify the de-anonymizablity of graph dataset by considering the topological importance of nodes.

Recently, Sharad~\cite{Sharad2016benchmark} proposed a general threat model to measure the quality of a de-anonymization attack which is independent of the anonymization scheme. 
He proposed a machine learning framework to benchmark perturbation-based graph anonymization schemes.
This framework explores the hidden invariants and similarities to re-identify nodes in the anonymized graphs~\cite{sharad2013anonymizing, sharad2014automated}.
Importantly, this framework can be easily tuned to model various types of attacks. 

Any anonymization scheme is subject to two conflicting forces: privacy and utility.
Shokri et al.~\cite{shokri2015privacy} used game theory to model privacy-utility trade-off on solving the optimization problem for general distortion
and differential privacy metrics.
Further, this model has been generalized to analyze the privacy-utility trade-off in temporal graphs~\cite{theodorakopoulos2014prolonging}.
It has been shown experimentally that utility is degraded much faster than privacy is achieved~\cite{shokri2015privacy,aggarwal2011hardness}.


Differential privacy (DP), initially proposed in the context of databases~\cite{dwork2014algorithmic}, was also adapted to graph anonymization in conjunction with $dK$-graph generators~\cite{dimitropoulos2009graph, sala2011sharing,proserpio2012workflow,proserpio2014calibrating,wang2013preserving}.
Differential privacy techniques enforce levels of statistical noise to the $dK$-distribution before generating respective $dK$-private graphs.
Sala et. al.~\cite{sala2011sharing} observe that $dK$-series are highly sensitive, thus requiring high level of noise to reach a required level of privacy.
Wang et. al.~\cite{wang2013preserving} study this inherent tension as a variable of privacy enforced by the $dK$ level.
They observe that $1K$ graphs require smaller magnitude of noise to have strong privacy guarantee, while $2K$ graphs have weak privacy guarantee at the same level of noise.

In our work, 
we study the inherent conditions in $dK$-graphs that resist (or fail) to general de-anonymization attacks based on machine learning techniques. 
To the best of our knowledge, ours is the first study to analyze the anonymization power of $dK$-graph generation techniques and in conjuction with the type of seed knowledge the attacker has access to.


\section{Methodology}
\label{sec:methodology}
The methodology for measuring the anonymity of $dK$-random graphs using machine learning techniques is presented in this section. 
$dK$-random graphs~\cite{mahadevan2006systematic} are synthetic graphs generated such that they preserve a given degree distribution of size $d$, where $d$ is typically a small number. 
For example, $1K$-random graphs preserve the degree distribution of the original graph, but create a different graph structure with possibly other graph properties.
$2K$-random graphs preserve the co-joint degree distribution, and are shown to better preserve the characteristics of the original graph than $1K$-random graphs~\cite{orsini2015quantifying}.
However, $2K$-random graphs expose more structural information than $1K$-random graphs (encapsulated in a larger $d$), and hence our research questions in this paper: \emph{Are $dK$-based graph generators appropriate for graph anonymization and if so, for what values of $d$, for what types of graphs, and under what attack conditions?}

To answer these questions, we developed a methodology based on that presented in~\cite{Sharad2016benchmark}. 
We use the same threat model (Section~\ref{sec:threat-model}) that aims at finding a bijective mapping between nodes in two different, already anonymized graphs. 
Unlike in~\cite{Sharad2016benchmark}, as anonymization technique we used $dK$-graphs, for $d$ equal to 1, 2, and 2.5 (Section~\ref{sec:dk_random_graphs}). 
We mount a machine-learning based attack (Section~\ref{sec:random-forest-classification}), in which the algorithm learns the correct mapping between some pairs of nodes from the two graphs, and estimates the mapping of the rest of the dataset. 
In order to represent the nodes in the leaning algorithms, we used the same feature vector as in~\cite{Sharad2016benchmark} (described in Section~\ref{sec:node-signatures}). 

\subsection{The Threat Model}
\label{sec:threat-model}
In this scenario, the attacker aims to match (unlabeled) nodes from two structurally anonymized, undirected graphs. 
For example, an attacker has two unlabeled networks of individuals in an organization that represent the communication patterns (e.g., email) and friendship information available from an online social network. 
Both networks have been anonymized by removing the original node labels and by perturbing edges.
These graphs are structurally overlapping, in that some individuals are present in both graphs, even if their identities have been removed, their links are different and have possibly been perturbed.  
The attacker's task is to find a bijective mapping between the two subsets of nodes in the two graphs that correspond to the individuals present in both networks. 
It has been demonstrated empirically that when more nodes are common to the two graphs (i.e., the size $\alpha$ of the overlap increases), more nodes can be identified since the attacker has access to more side information~\cite{Sharad2016benchmark,ji2015secgraph}.

\subsection{Anonymization via $dK$-Random Graph Generation}
\label{sec:dk_random_graphs}


The $dK$-series represent a set of descriptive statistic metrics that capture the original graph structure at multiple levels of detail~\cite{sala2011sharing}.
Specifically, the $dK$-series summarizes the structure of a graph from the degree distribution of a subgraph pattern of size $d$. 
Thus, $0K$-graphs are random graphs with a given average node degree, $1K$-graphs are random graphs with a given degree distribution, $2K$-graphs are random graphs with a given joint degree distribution, $3K$-graphs are random graphs with a given interconnectivity of triplets of nodes, and so on. 
Intermediate steps in the series can be defined, such as the $2.5K$ graph, which is a relaxed version of $3K$-graphs that reproduces both joint degree distribution and degree-dependent clustering coefficient \cite{gjoka201325k}.  



We use three graph generation techniques (for $1K,\ 2K,$ and $2.5K$, respectively) proposed in previous work. 
$1K$-graphs are graphs randomly chosen from the space of graphs with a given degree distribution. 
To generate them, we used Orbis~\cite{mahadevan2006systematic} with $dK$ randomizing rewiring.  
The algorithm works by rewiring random pairs of edges from the original graph in order to preserve the $1K$-distribution.
As suggested in~\cite{mahadevan2006systematic}, we allow a sufficient number of initial isomorphic rewirings, and then follow a subsequent set of rewirings to verify that the $1K$-series remain intact.
This method results in an unbiased sampling of $1K$-graphs with a significant variation of properties, such as degree-correlations and clustering~\cite{orsini2015quantifying}. 

The $2K$-graphs preserve a given joint degree distribution, including the degree correlations. 
We used $dK$-targeting $d'K$-preserving rewiring based on Metropolis dynamics as implemented by Gjoka et al.~\cite{gjoka201325k}. 
$2K$-graphs are generated by beginning with any $1K$-graph and rewiring edges only if such modifications move the graph closer to $2K$-series.

For generating $2.5K$ graphs we used an improved version of the Metropolis method (MCMC) proposed in~\cite{gjoka201325k}. 
$2.5K$ graphs are $2K$ graphs that preserve a given clustering coefficient in addition to the $2K$ distribution. 
This method starts by adding triangles to a $2K$ graph to reach a clustering coefficient threshold and then destroys them to recreate the desired joint degree distribution. 
For generating these graphs, we chose the parameters to ensure uniform sampling over the $dK$-space. (It has been shown experimentally~\cite{orsini2015quantifying} that the number of $dK$-swaps required for the MCMC method is $O(|E|)$).

We limit our study for the $dK$-spaces to $1K,\ 2K$ and $2.5K$ graphs 
because algorithms exist that almost surely sample uniformly at random elements of the corresponding set of graphs~\cite{orsini2015quantifying}.  
No such algorithms are known for steps higher in the series~\cite{orsini2015quantifying}.

\subsection{Experimental Framework}
\label{sec:experimental-framework}
We assume that the adversary has a sanitized graph $G_{san}$ that could be associated with an auxiliary graph $G_{aux}$ for the re-identification attack. 
As discussed before, $G_{san}$ could be the anonymized version of the communication (e.g., email) network, while $G_{aux}$ is the friendship network. 

In order to model this scenario, we split a real dataset graph $G=(V,E)$ into two subgraphs $G_1=(V_1,E_1)$ and $G_2=(V_2,E_2)$, such that $V_1 \subset V$, $V_2 \subset V$ and $V_1 \cap V_2 = V_\alpha$, where $V_\alpha \ne \phi$. 
The fraction of the overlap $\alpha$ is measured by the Jaccard coefficient of two subsets: $\alpha=\frac{|V_1 \cap V_2|}{|V_1 \cup V_2|}$. 
In the shared subgraph induced by the nodes in $V_\alpha$, nodes will preserve their edges with nodes from that subgraphs but might have different edges to nodes that are part of $V_1 - V_{alpha}$ or part of $V_2 - V_{alpha}$.

\begin{figure}[t]
	\centering
	\includegraphics[width=1\linewidth]{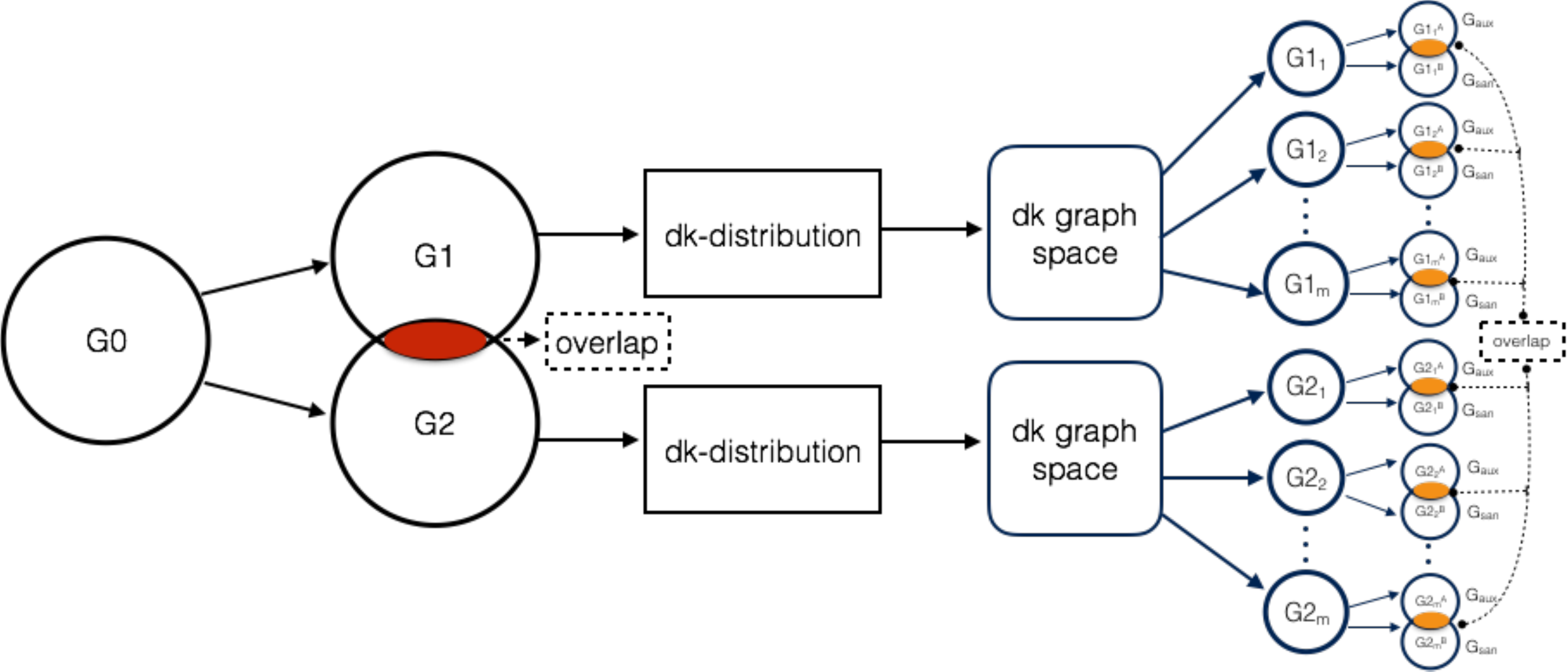}
	\caption{Graph mining system flow to generate node pairs with the ground truth of identical and non-identical pairs. 
	}
	\label{fig:dkminingflow}
\end{figure}

In an optimistic scenario, an attacker has access to a part of the original graph (e.g., $G_1$) as auxiliary data and to an unperturbed subgraph (e.g., $G_2$) as the sanitized data whose nodes the attacker wants to re-identify.
However, this is unrealistic, since researchers and practitioners are well aware now that without perturbing the structure of the graph, even with identifiable node and edge attributes removed, there is no anonymity~\cite{narayanan2009anonymizing}.
Hence, in order to reproduce a realistic scenario, we anonymize both $G_1$ and $G_2$ graphs (as depicted in Figure~\ref{fig:dkminingflow}).
Note that this is a more challenging scenario for the attacker.
The (not anonymized) graphs $G_1$ and $G_2$ will be used as baseline to measure the anonymization power of the $dK$-based techniques. 

We produced $m$ versions of the anonymized graphs, in order to increase the dataset used for experimentation. 
The anonymized graphs are then split again into two overlapping graphs, maintaining the same values of overlap parameters as above.
The resulting graphs are now the equivalent of the email/friendship networks we used as example above. 
The overlap is the knowledge repository that the attacker uses for de-anonymization~\cite{henderson2011s}. 
Part of this knowledge will be made available to the machine learning algorithms. 

Previous work shows that the larger $\alpha$, the more successful the attack. 
However, the relative success of attacks under different anonymization schemes is observed to be independent from $\alpha$~\cite{Sharad2016benchmark}.
In order to experiment with various strengths of the attack for a set value of $\alpha=0.2$, we build $V_\alpha$ in four different ways: i)~as a random collection of nodes from the original graph $G$ (R); 
ii)~by selecting the highest degree nodes from $G$ (HD); iii)~ by building a breadth-first-search tree starting from a randomly selected node in $G$ (BFS-R); and iv)~by building a breadth-first-search tree starting from the highest degree node in $G$ (BFS-HD). 
While other alternatives are certainly possible, these four approaches cover an important set of realistic assumptions on the attacker.  



\subsubsection{\textbf{Node Signatures}}
\label{sec:node-signatures}
Since we are employing machine learning techniques to "label" nodes in a graph, we need to represent the nodes as feature vectors. 
We define the node $u$'s features using its neighborhood degree distribution (NDD). 
NDD is a vector of positive integers where $NDD^q_u[k]$ represents the number of $u$'s neighbors at distance $q$ with degree $k$.
The decision of using NDD is supported by the empirical observation that a large proportion of nodes in a graph are uniquely distinguishable by their NDD~\cite{Sharad2016benchmark}.


We employ a binning strategy to capture the aggregate structure of ego networks which is expected to be robust against edge perturbation due to anonymization~\cite{sharad2016learning}.
Thus, we represent the signature of a node $u$ as the concatenation of the binned version of $NDD^1_u$ with the binned version of $NDD^2_u$. 
We bin $NDD^q_u$ by using a bin size of $50$, which was shown empirically~\cite{Sharad2016benchmark} to capture the high degree variations of large social graphs.
For each $q$, we use $21$ bins, which would correspond to a larger node degree of $1050$. All larger values are binned in the last bin.

\subsubsection{\textbf{Random Forest Classification}}
\label{sec:random-forest-classification}

Note that the nodes in $G_{san} \cap G_{aux}$, common to both graphs, can be recognized as being the same node in the two graphs based on their (anonymized) node identifier. 
This is the ground truth against which we measure the accuracy of the learning algorithms. 

We generate examples for the training phase of the deanonymization attack by randomly picking node pairs from  the sanitized $(G_{san})$ and the auxiliary $(G_{aux})$ graphs, respectively. 
In most cases, we have an unbalanced dataset, with the degree of imbalance depending on the overlap parameter $\alpha$, where the majority is non-identical node pairs.
We use reservoir sampling technique~\cite{haas2016data} to take $\ell$ balance sub-samples from the population $S$, and SMOTE algorithm as an over-sampling technique for each sub-sample.
Each sample is trained by a forest of 100 random decision tress that allows to learn features.
Accuracy is measured on unseen examples from the same sample space.
Gini-index is used as an impurity measure for the random forest classification.
Given the size $\alpha$ and choice of the overlap, we measure the quality of the classifier on the task of differentiating two nodes as identical or not. 
We thus use $2 \times m \times \ell \times 100$ decision trees per $dK$-space for the learning process in each graph space.
Table~\ref{tbl:sample_space} summarizes the statistics of the sample population.

We measure the accuracy of the classifier in determining whether a randomly chosen pair of nodes (with one node in $G_{san}$ and another in $G_{aux}$) are identical or not. 
Accuracy is being measured by the distribution of F1-score, which is the harmonic mean between the precision and recall of a local classifier.

\begin{table*}[bthp]
	
	\caption{
Graph properties of the real network datasets. All graphs are undirected. density $(\bar{d})$ is the fraction of all possible edges, transitivity $(C)$ is the fraction of triangles of all possible triangle in the network while the clustering $(\bar{C})$ is the average local clustering coefficients of the nodes. degree-assortativity $(r)$ measures the similarity of relations depending on the associated node degree. avg-path-length $(\kappa)$ depicts the average shortest path length between any pairs of nodes
and degree-1 represents the percentage of nodes in the network with degree exactly 1. 
}
	\label{tbl:dataset}
	\begin{tabular}{|r|r|r|r|r|r|r|r|r|}
		\hline 
		Network & $|N|$ & $|E|$ & $\bar{d}$ & $C$ & $\bar{C}$ & $r$ &  $\kappa$ & degree-1 (\%)\\ 
		\hline \hline
		fb107 & 1034 &  26749 & 0.1000 &  0.2534 & 0.5264  & 0.4316 &  2.9517 & 1.45 \\ 
		\hline 
		caGrQc & 5242 & 14496 & 0.0011 & 0.3621 & 0.5295 & 0.6592 & 3.8047 & 22.83 \\ 
		\hline 
		soc-anybeat & 12645 & 49132 & 0.0006 & 0.0073 & 0.2037 & -0.1234 & 3.1715 & 49.51 \\ 
		\hline 
		soc-gplus & 23628 &  39194 & 0.0001 & 0.0012 & 0.1741 & -0.3885 & 4.0277 & 69.16 \\ 
		\hline 
		wikinews & 25042 & 68679 & 0.0002 & 0.0001 & 0.1133 & -0.2544 & 2.9666 & 33.52 \\ 
		\hline 
	\end{tabular} 
\end{table*}

\begin{table}[bthp]
	\centering
	\caption{Basic statistics of generated node pairs over $dK$-spaces under the overlap choices of Random (R), High Degree (HD) and BFS-trees (rooted in the highest degree node BFS-HD and, respectively, a random node, BFS-R). We generated a total of $\approx$ 1 billion identical and non-identical node pairs over three $dK$-spaces of the five real social network datasets, and took  $364K$ balanced samples.}
	\label{tbl:sample_space}
	\begin{tabular}{|c|c|r|r|r|r|c|}
		\hline
		\multicolumn{2}{|c|}{\multirow{2}{*}{Network}} & \multicolumn{4}{c|}{No. of node pairs (\textit{millions})} & \multirow{2}{*}{ $\ell$}\\
		\cline{3-6}
		\multicolumn{2}{|c|}{} & R & HD & BFS--HD & BFS--R & \\
		\hline
		\multirow{3}{*}{fb107} & 1k & 0.85 & 0.84 & 0.88 & 0.88 & 12800 
		\\
		\cline{2-7}
		& 2k & 0.85 & 0.84 & 0.88 & 0.89 & 12800 
		\\
		\cline{2-7}
		& 2.5k & 0.86 & 0.85 & 0.88 & 0.89 & 12800
		\\
		\hline
		\hline
		\multirow{3}{*}{caGrQc} & 1k & 2.36 & 2.30 & 3.92 & 3.98 & 12800 
		\\
		\cline{2-7}
		& 2k & 3.10 & 3.07 & 4.96 & 3.75 & 12800 
		\\
		\cline{2-7}
		& 2.5k & 3.13 & 3.07 & 4.96 & 5.00 & 12800 
		\\
		\hline
		\hline
		\multirow{3}{*}{soc-anybeat} & 1k & 10.95 & 10.85 & 24.35 & 26.06 & 32000\\
		\cline{2-7}
		& 2k & 10.65 & 12.60 & 24.50 & 26.20 & 32000\\
		\cline{2-7}
		& 2.5k & 11.18 & 13.56 & 24.55 & 26.20 & 32000 \\
		\hline
		\hline
		\multirow{3}{*}{soc-gplus} & 1k & 10.55 & 11.53 & 45.04 & 43.47 & 32000\\
		\cline{2-7}
		& 2k & 12.33 & 11.40 & 45.06 & 49.80 & 32000 \\
		\cline{2-7}
		& 2.5k & 12.56 & 10.40 & 45.06 & 49.80 & 32000 \\
		\hline
		\hline
		\multirow{3}{*}{wikinews} & 1k & 23.58 & 23.16 & 46.80 & 46.67 & 32000\\
		\cline{2-7}
		& 2k & 22.63 & 25.16 & 46.87 & 46.72 & 32000 \\
		\cline{2-7}
		& 2.5k & 21.55 & 25.43 & 46.87 & 46.72 & 32000 \\
		\hline
	\end{tabular}
\end{table}

\section{Data Characteristics}
\label{sec:characteristics}

Our objective is to quantify the graph-anonymization benefits of graph generation techniques that preserve the $dK$ distribution. 
To this end, we select a number of real network datasets (presented in Section~\ref{sec:datasets}) and extract their various $dK$-distributions. 
We then generate random graphs from these $dK$ distributions (as presented in Section~\ref{sec:dk_random_graphs}) to play the roles of auxiliary and sanitized graphs.
Note that a pair of auxiliary and sanitized graph (that is, generated in a particular $dK$-space and for a specific degree distribution) are connected by a fraction $\alpha$ of shared nodes, as presented above. 

\subsection{Network Datasets}
\label{sec:datasets}

We chose five publicly available datasets that represent real social networks of various types. 
\texttt{fb107}~\cite{leskovec2012learning} represents social circles of an ego in Facebook. 
\texttt{caGrQc}~\cite{leskovec2007graph} is a co-authorship network between the authors of papers in general relativity and quantum cosmology.
\texttt{soc-anybeat}~\cite{Fire2012} is an interaction network available in the Anybeat online community, which is a public gathering place across the world.
\texttt{soc-gplus}~\cite{networkrepo} is a follower network from Google+.
Finally, \texttt{wikinews}~\cite{networkrepo} is a discussion network of Wikipedia pages. 
Table~\ref{tbl:dataset} summarizes properties of these datasets.

\subsection{Characteristics of $dK$-generated Graphs}
We confirmed that the $m=4$ instances of $dK$-generated graphs for every set of parameters are distinct. 
We also measured the structural distance between the original graph and the generated graphs. 
For this, we used D-measure~\cite{schieber2017quantification}, a dissimilarity metric between 0 and 1 that quantifies the topological differences based on network node dispersion (NND). 
A low D-measure ($\approx 0$) suggests that graphs have the same distance distribution, the same NND and the same $\beta$-centrality vector~\cite{schieber2017quantification}. 
Non-zero D-measure indicates graphs are non-isomorphic. 

\begin{figure}[bthp]
	\centering
	\includegraphics[width=\linewidth]{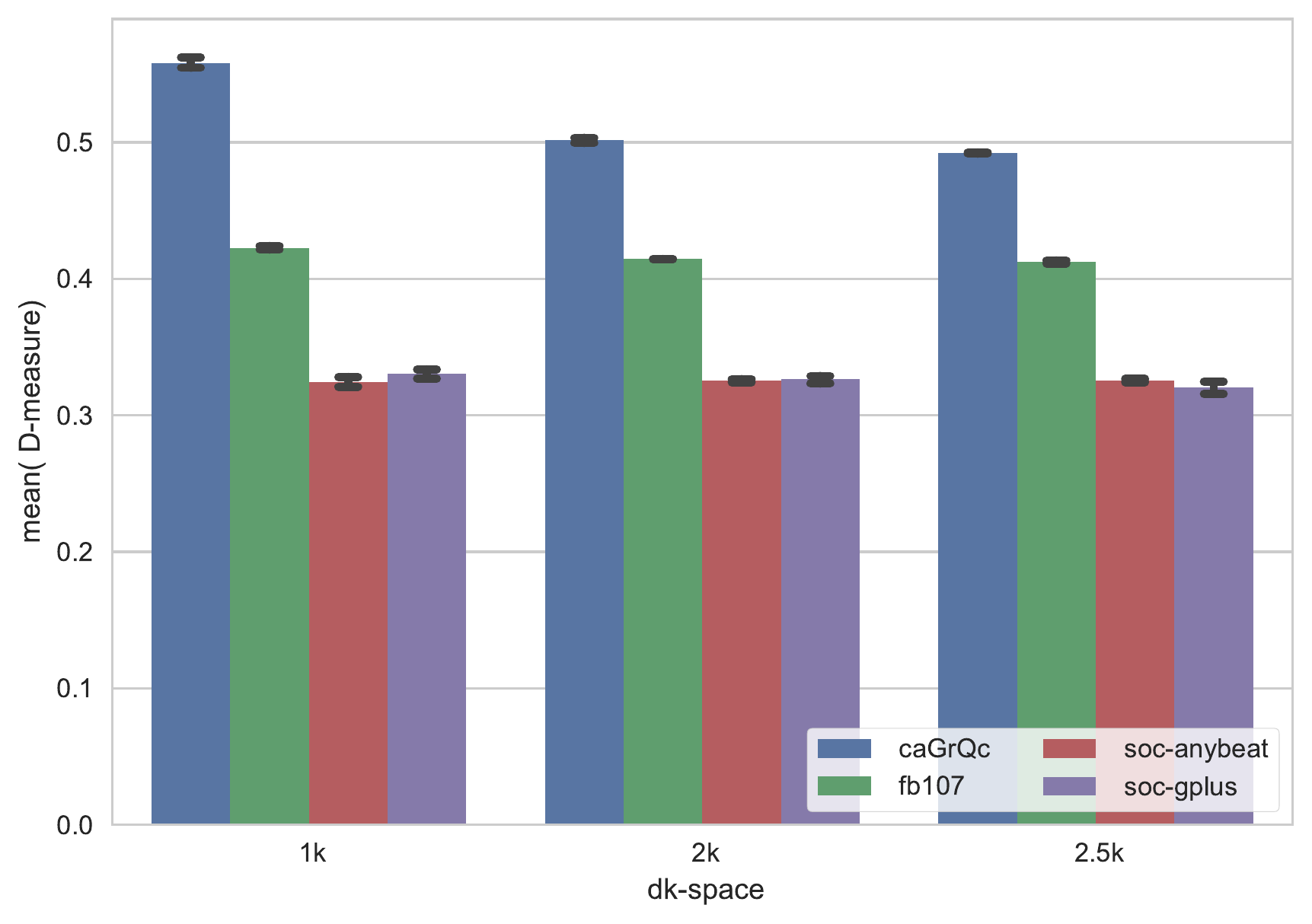}
	\caption{
	Average structural distance between the original network and a $dK$-generated network 
	using D-measure~\cite{schieber2017quantification}. The structural distance D-measure is calculated between the original network and 4 different instances of its corresponding $dK$-random graphs and the average value is reported, along with the confidence intervals}
	\label{fig:d_distance_all}
\end{figure}

Figure~\ref{fig:d_distance_all} shows the average distance between each of the real networks 
and its corresponding $dK$-generated graph in each of the $dK$ spaces we consider. 
The plot suggests the generated networks are relatively close to the original dataset for sparser graphs (e.g., for the \texttt{soc-anybeat} dataset the average D-measure is $0.325$, while for \texttt{fb107}, the average D-measure is $0.415$). 
This observation is supported by intuition, as sparse graphs tend to have lower node degrees, thus relatively few opportunities for rewiring to lead to significantly different topology.

\ignore{
\begin{figure}[bthp]
	\centering
	\subfloat[High Degree]{
		\includegraphics[scale=0.2]{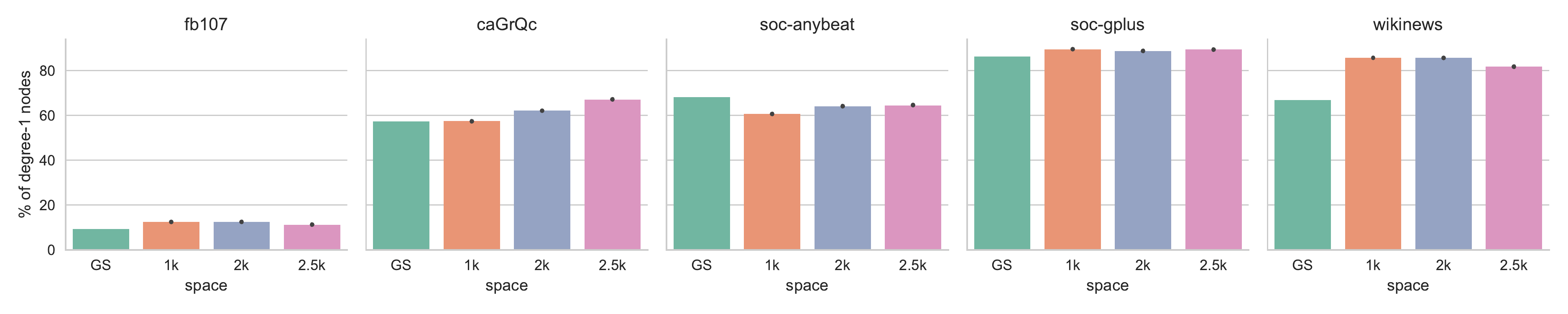}
			\label{fig:low_degree_deg}
	}
	\hspace{0mm}
	\subfloat[BFS-HD]{
		\includegraphics[scale=0.2]{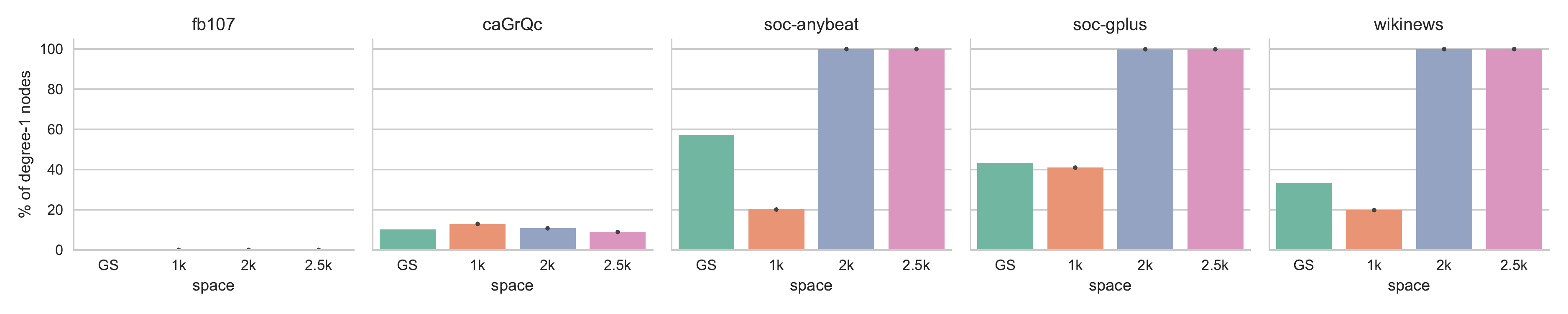}
			\label{fig:low_degree_bfs1}
	}
	\caption{Average percentage of nodes with degree 1 in the overlap subgraph across dk-spaces (including GS). The overlaps are averaged over 8 sample instances in each dk space for two overlap strategies: highest-degree nodes (a) and BFS rooted in the highest degree node (b). Figure also shows the confidence intervals of average measurements.}
	\label{fig:low_degree}
\end{figure}
}
\section{EMPIRICAL RESULTS}
\label{sec:experiments}
We focus our experimental evaluation on measuring the strength of an attacker, understanding the $dK$-space vulnerability for various $d$ values, and identifying the inherent properties of graphs that resist (or fail) a de-anonymization attack.

\subsection{The Strength of the Attacker}
We ask what type of information would help most an attacker under the threat model we consider. 
Intuitively, it is not only the size of the subgraph that the attacker has access to that matters (i.e., the seed knowledge base), but also the "quality" of the subgraph: for example, a disjoint set of low degree nodes (which would be the case of a randomly chosen set of nodes from a power-law graph) carries much less structural information than a connected subgraph of the same number of nodes. 

\begin{figure}[!tb]
	\centering
	\subfloat[fb107: $1K$]{
		\includegraphics[width=0.3\linewidth,height=0.1\textheight, keepaspectratio]{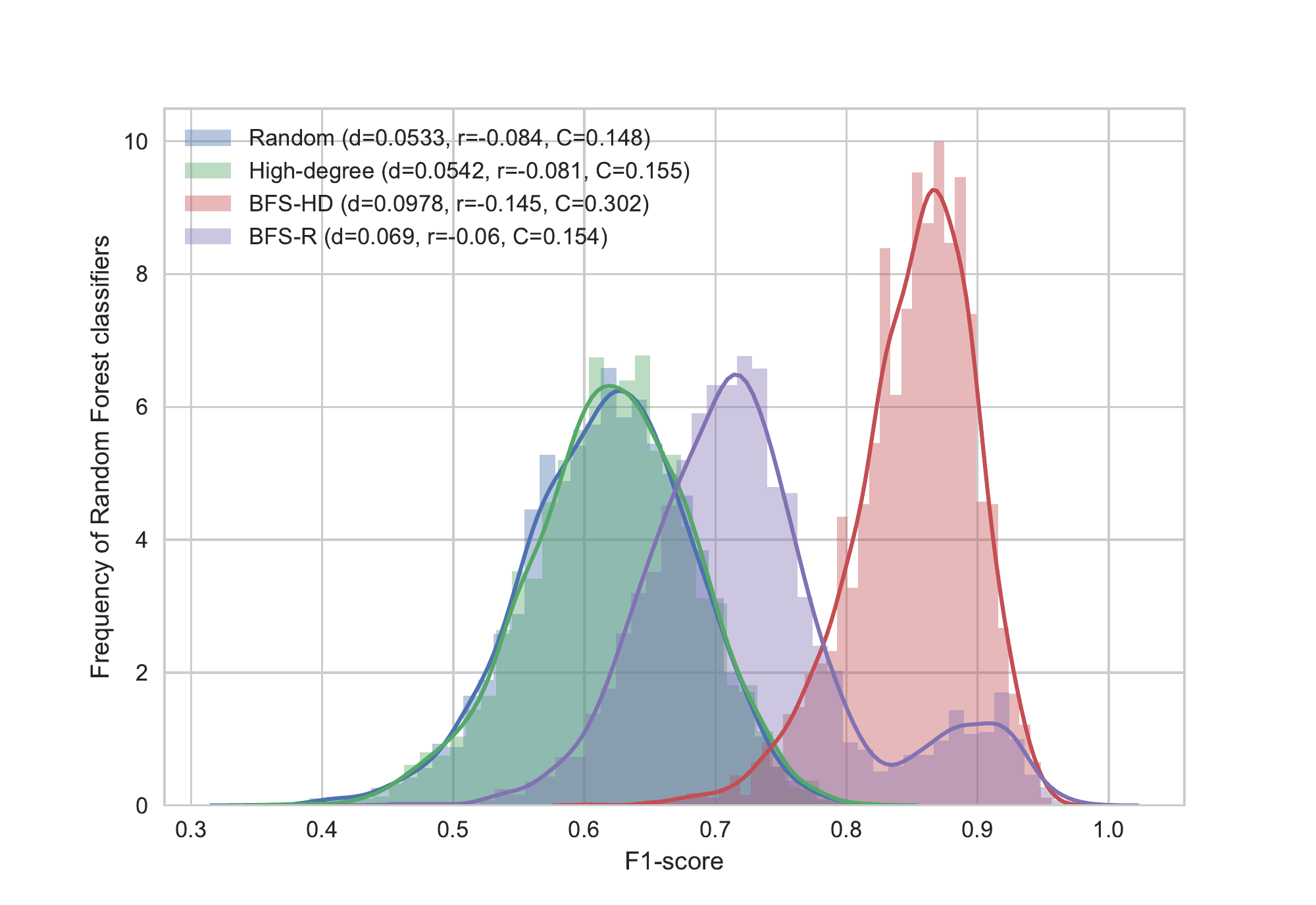}
			\label{fig:attack_strength_fb107_1k}
	}
	\subfloat[fb107: $2K$]{
		\includegraphics[width=0.3\linewidth,height=0.1\textheight, keepaspectratio]{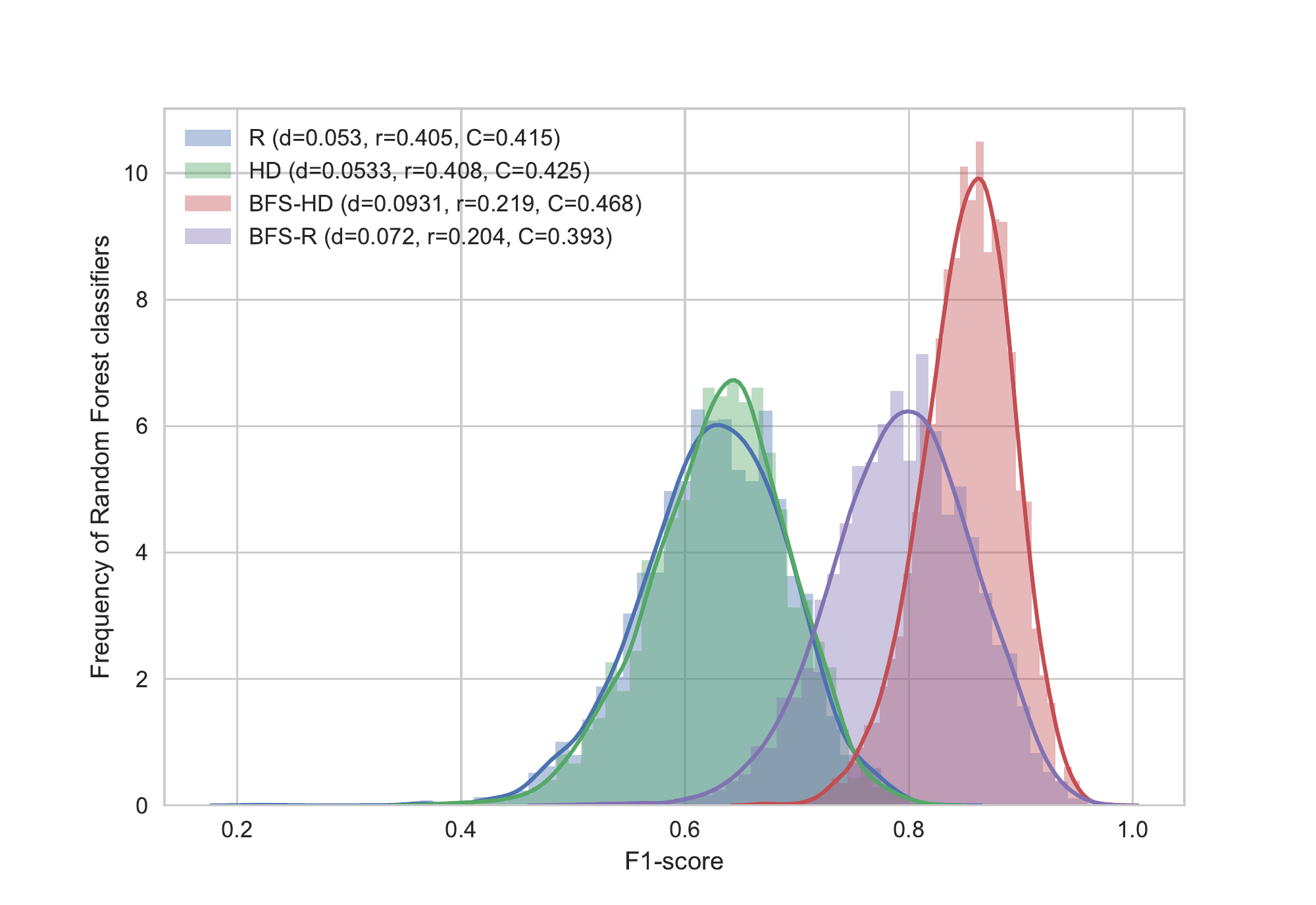}
			\label{fig:attack_strength_fb107_2k}
	}
	\subfloat[fb107: $2.5K$]{
		\includegraphics[width=0.3\linewidth,height=0.1\textheight, keepaspectratio]{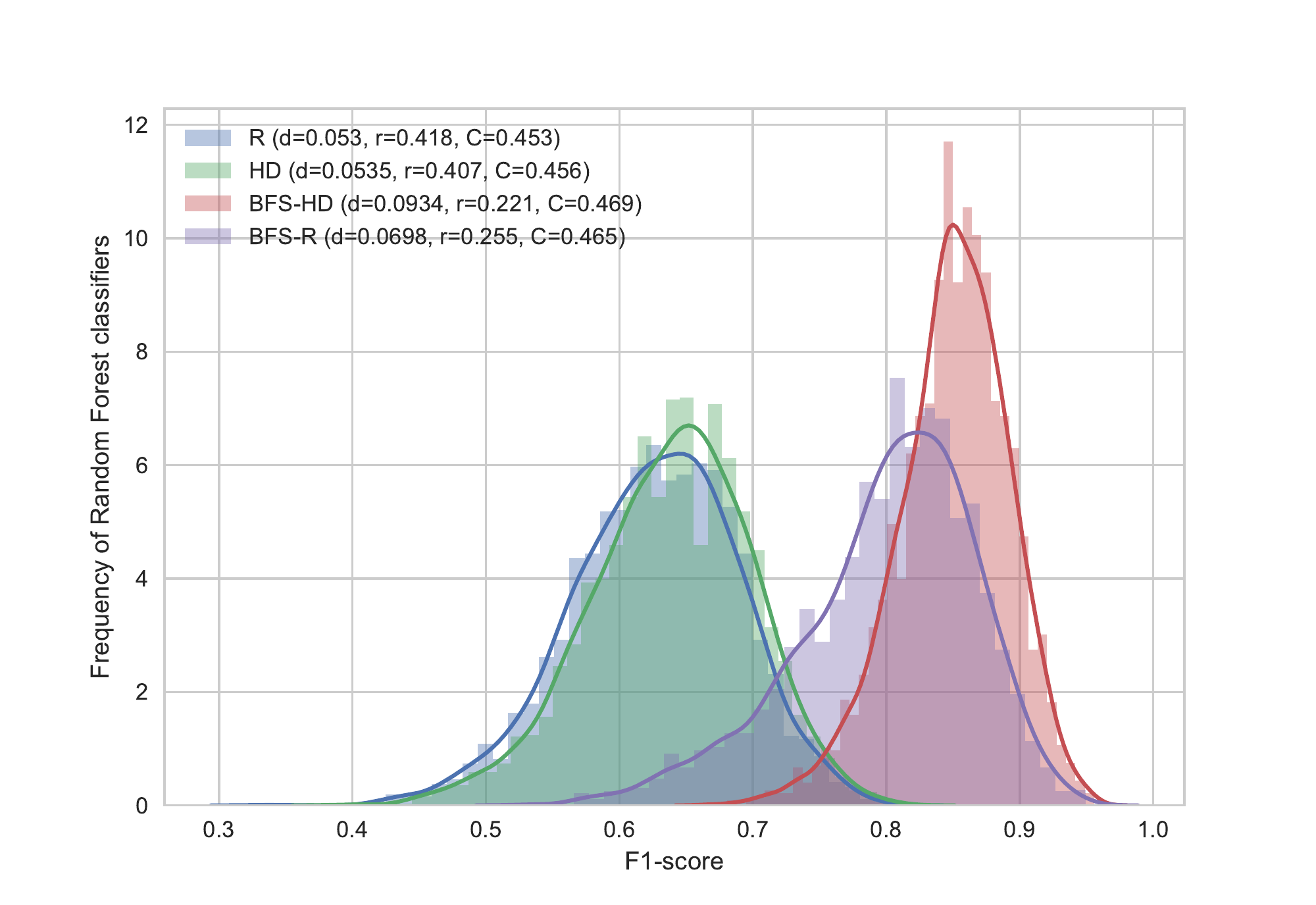}
		\label{fig:attack_strength_fb107_25k}
	}
	\hspace{0mm}
	\subfloat[caGrQc: $1K$]{
		\includegraphics[width=0.3\linewidth,height=0.1\textheight, keepaspectratio]{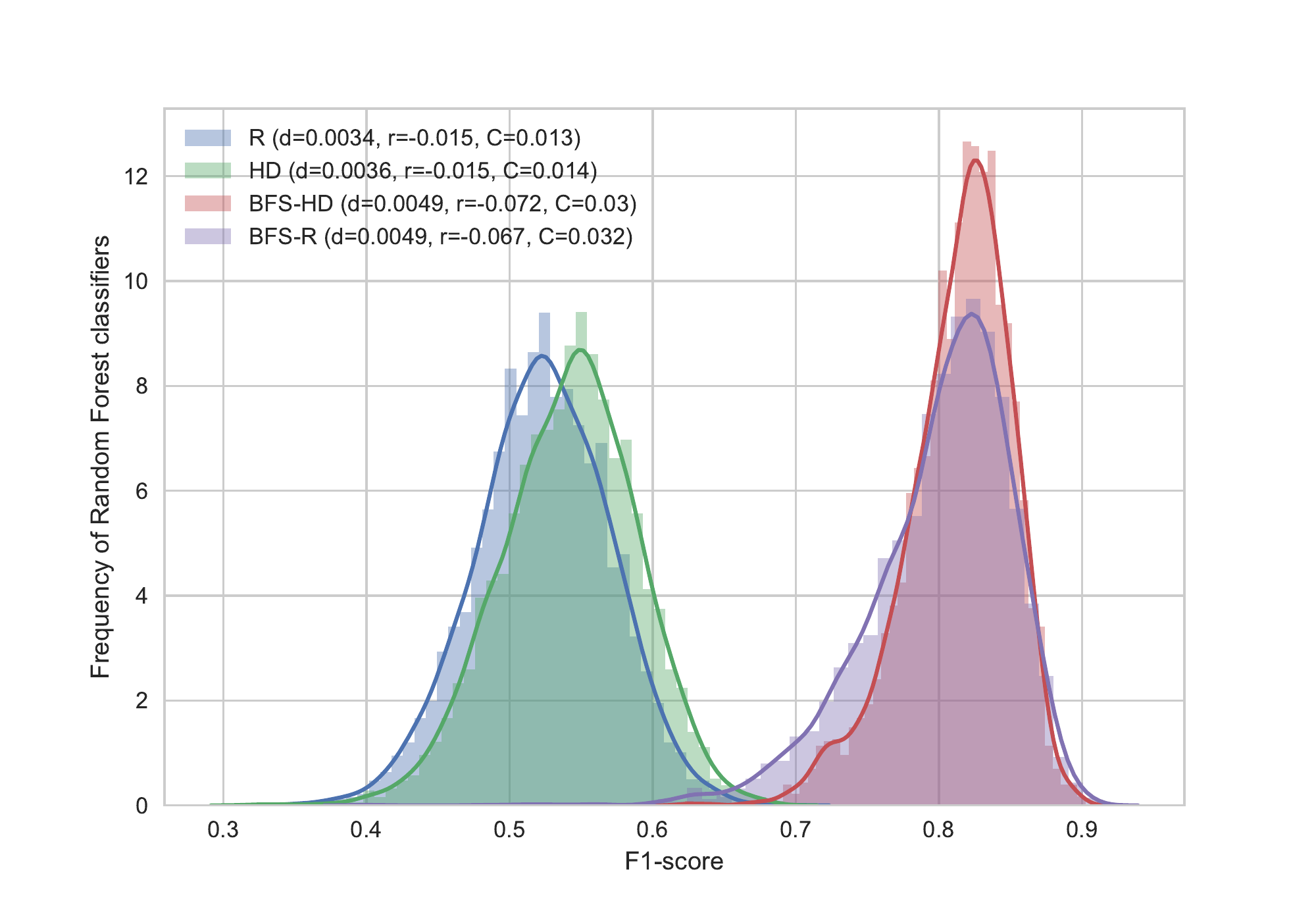}
		\label{fig:attack_strength_caGrQc_1k}
	}
	\subfloat[caGrQc: $2K$]{
		\includegraphics[width=0.3\linewidth,height=0.1\textheight, keepaspectratio]{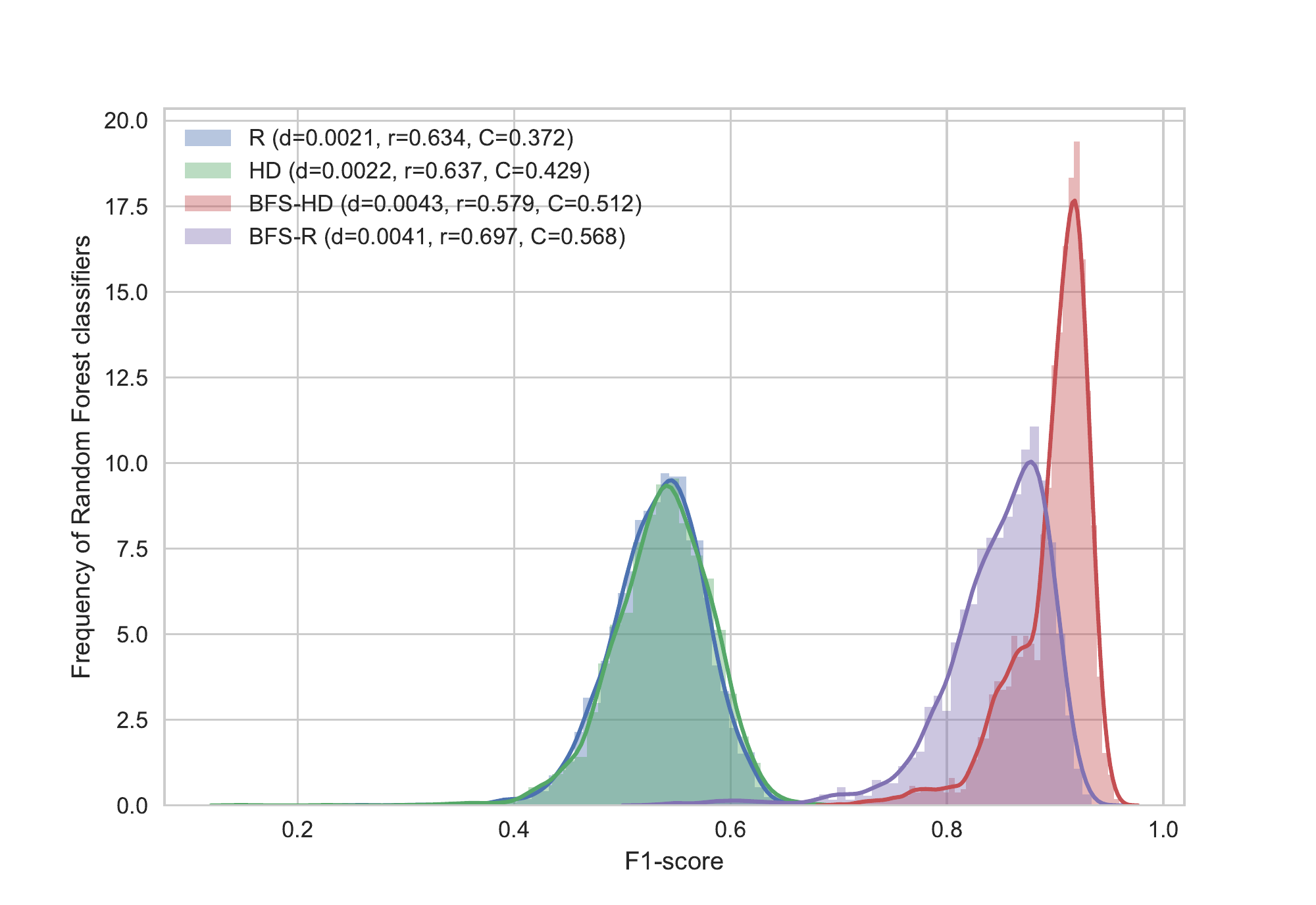}
		\label{fig:attack_strength_caGrQc_2k}
	}
	\subfloat[caGrQc: $2.5K$]{
		\includegraphics[width=0.3\linewidth,height=0.1\textheight, keepaspectratio]{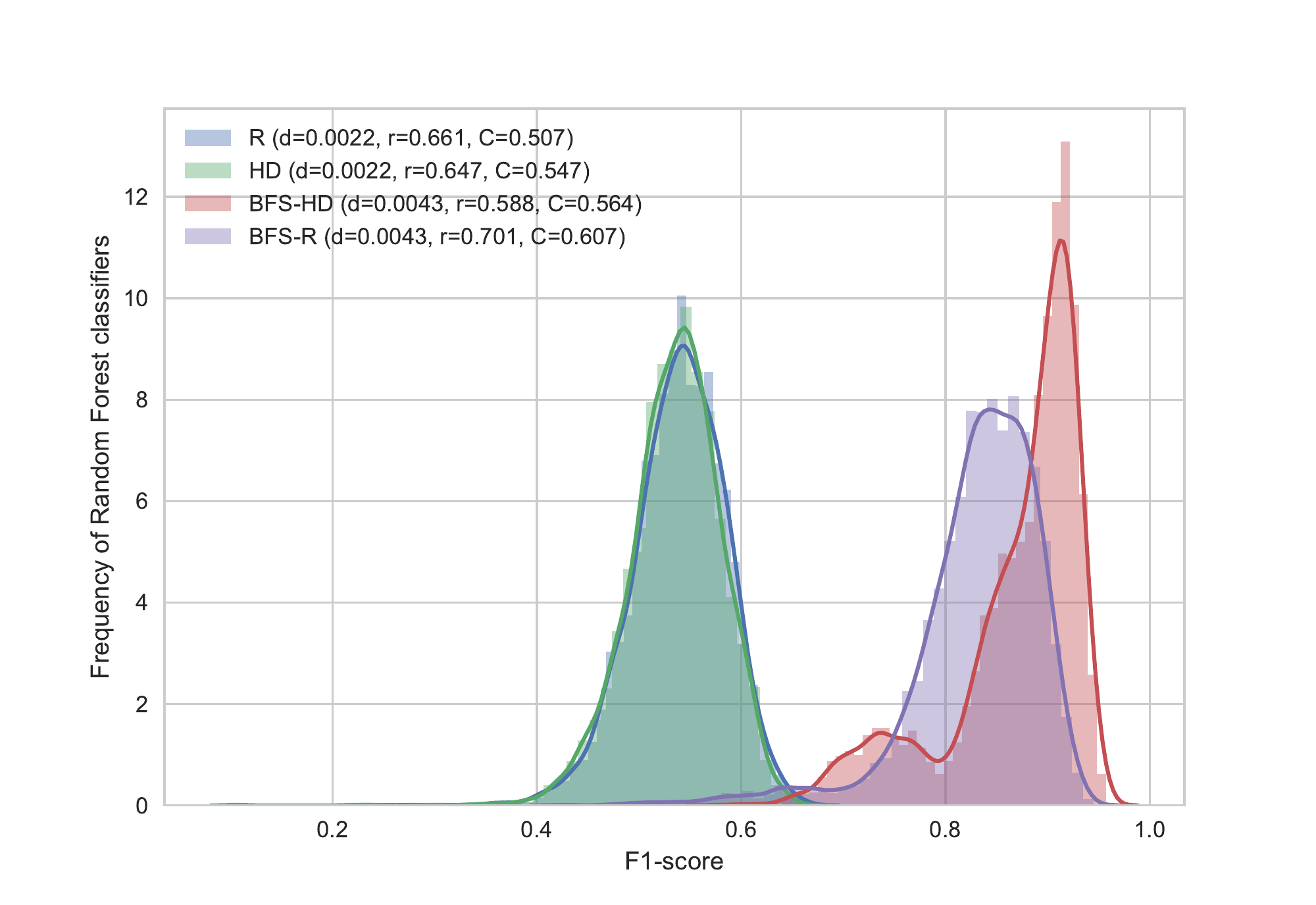}
			\label{fig:attack_strength_caGrQc_25k}
	}
	\hspace{0mm}
	\subfloat[soc-anybeat: $1K$]{
		\includegraphics[width=0.3\linewidth,height=0.1\textheight, keepaspectratio]{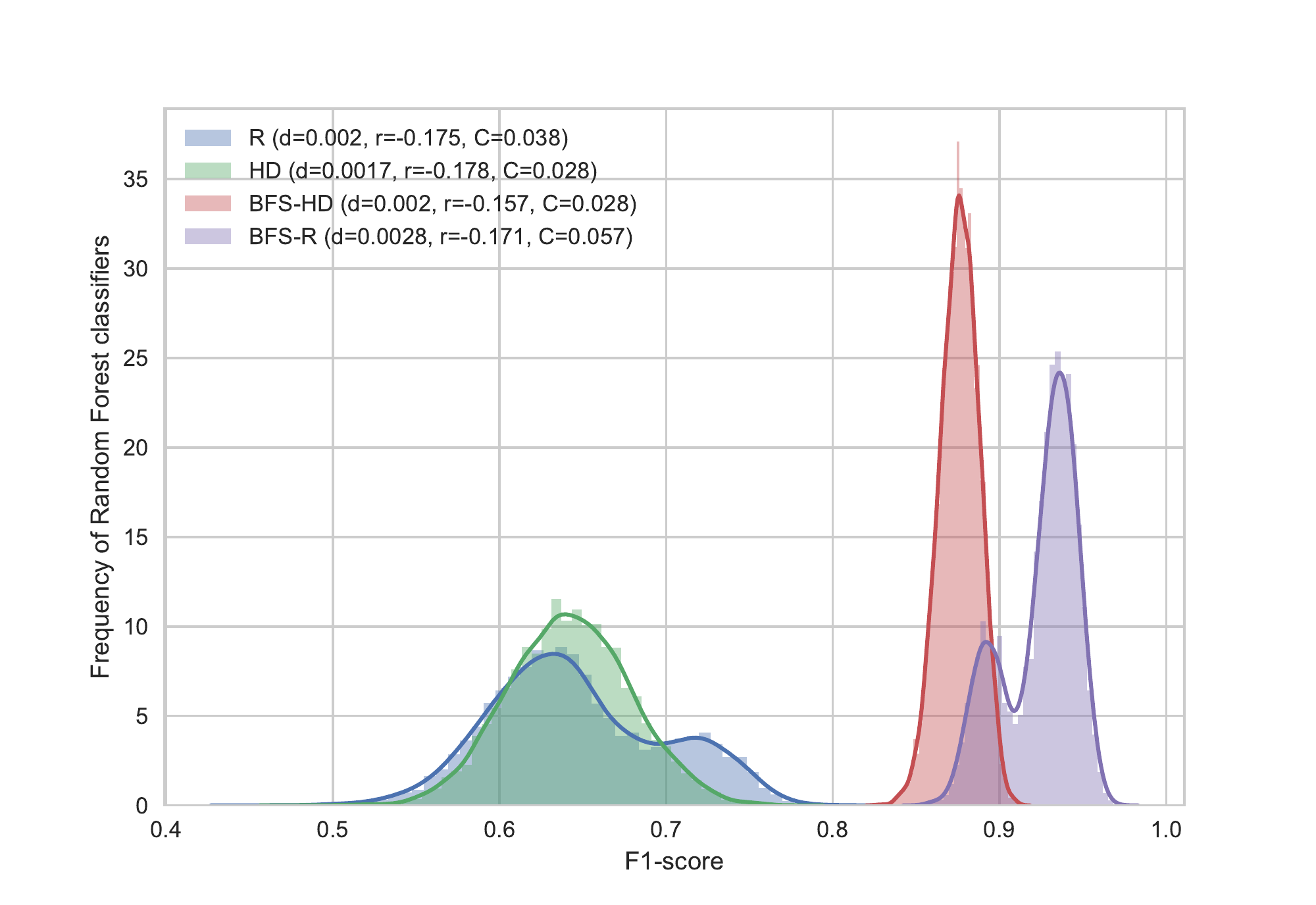}
			\label{fig:attack_strength_anybeat_1k}
	}
	\subfloat[soc-anybeat: $2K$]{
		\includegraphics[width=0.3\linewidth,height=0.1\textheight, keepaspectratio]{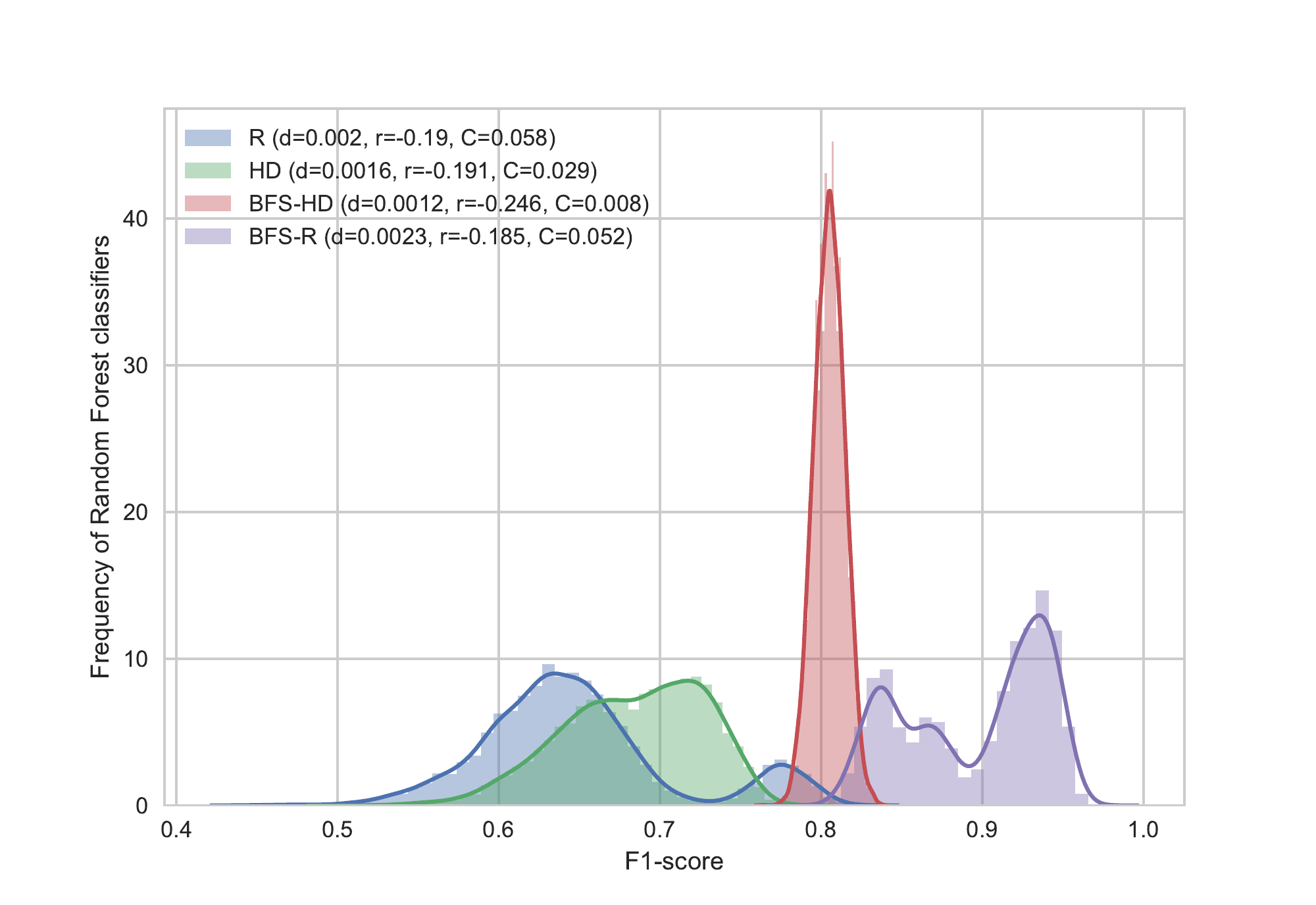}
			\label{fig:attack_strength_anybeat_2k}
	}
	\subfloat[soc-anybeat: $2.5K$]{
		\includegraphics[width=0.3\linewidth,height=0.1\textheight, keepaspectratio]{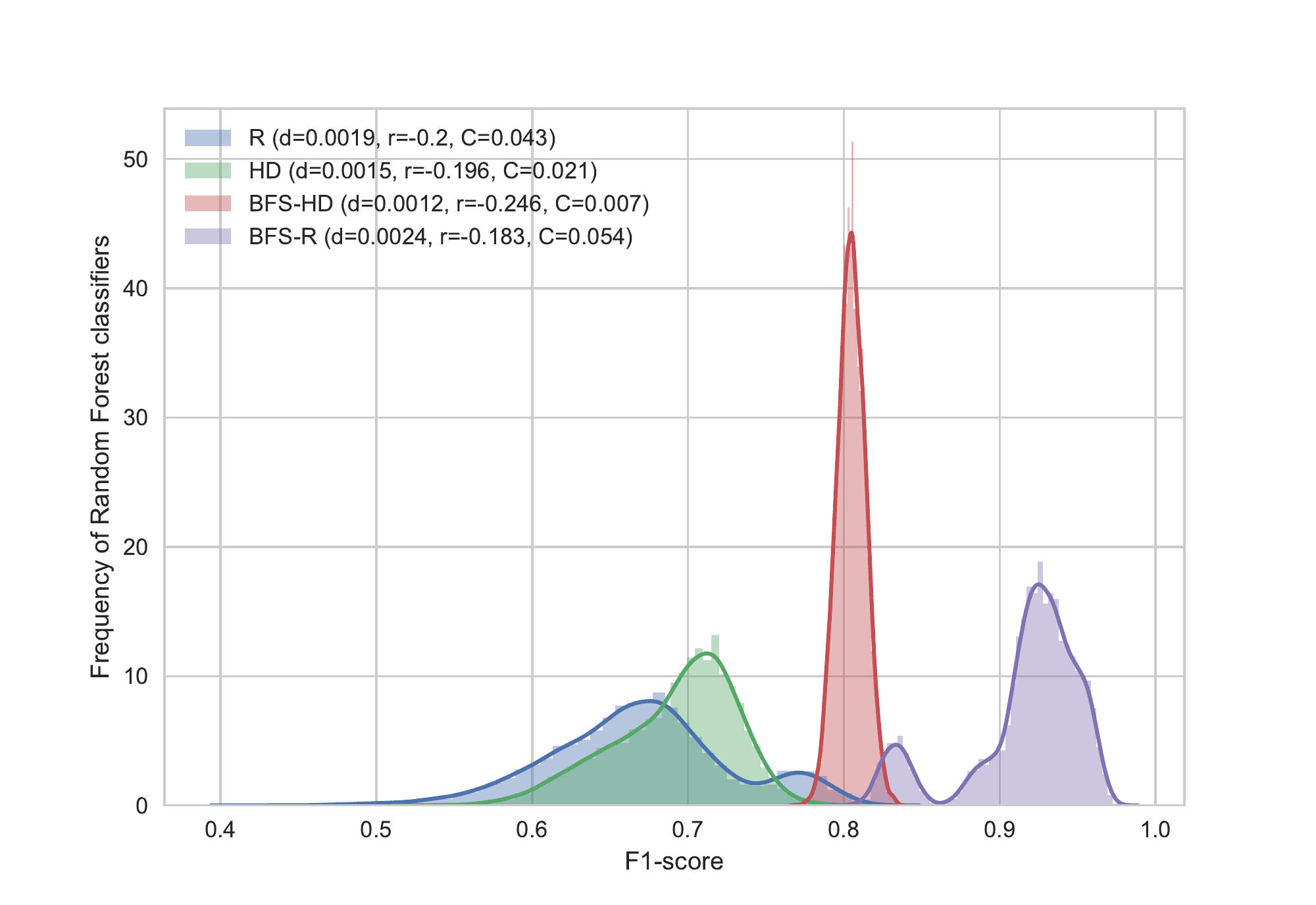}
		\label{fig:attack_strength_anybeat_25k}
	}
	\hspace{0mm}
	\subfloat[soc-gplus: $1K$]{
		\includegraphics[width=0.3\linewidth,height=0.1\textheight, keepaspectratio]{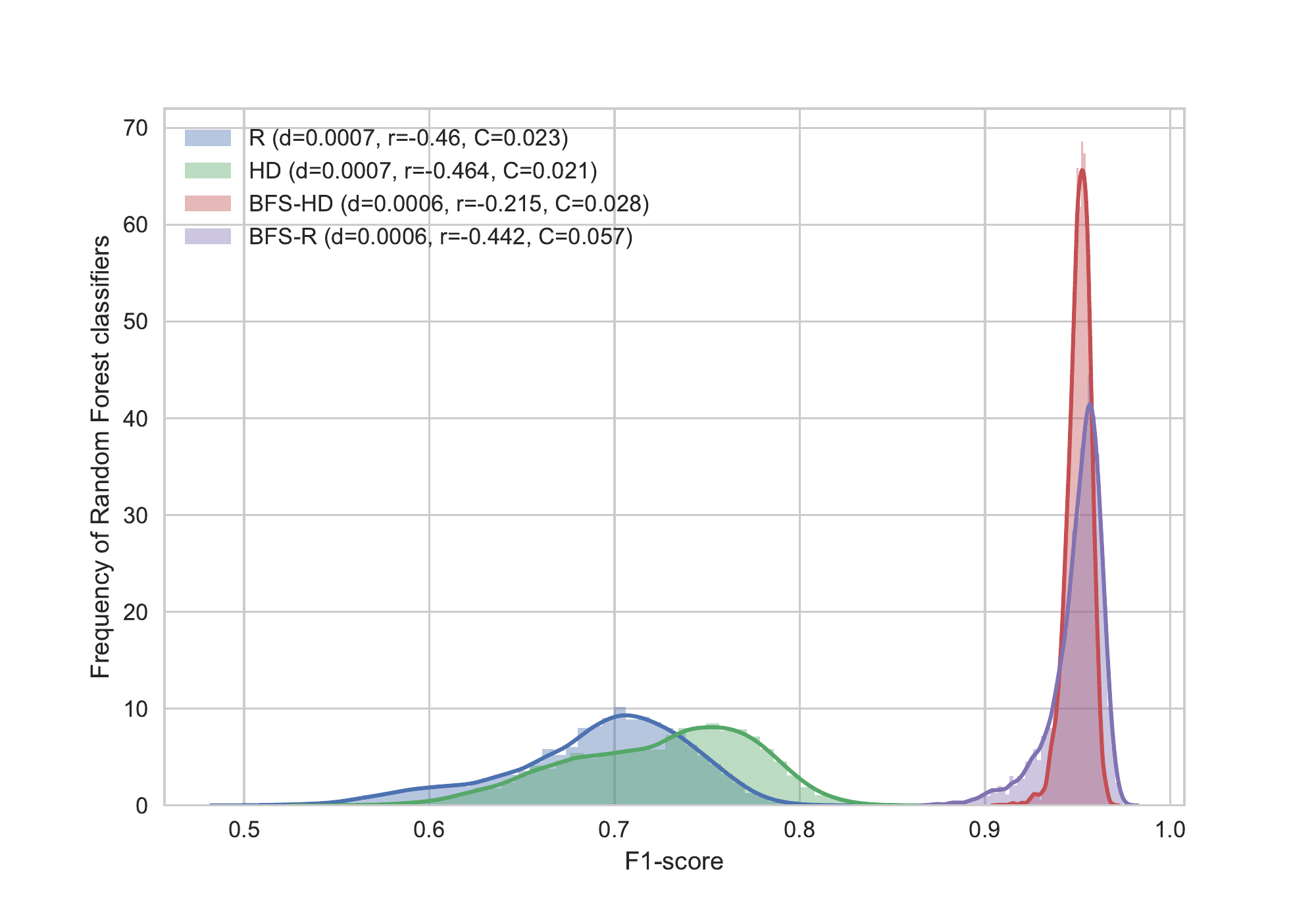}
		\label{fig:attack_strength_gplus_1k}
	}
	\subfloat[soc-gplus: $2K$]{
		\includegraphics[width=0.3\linewidth,height=0.1\textheight, keepaspectratio]{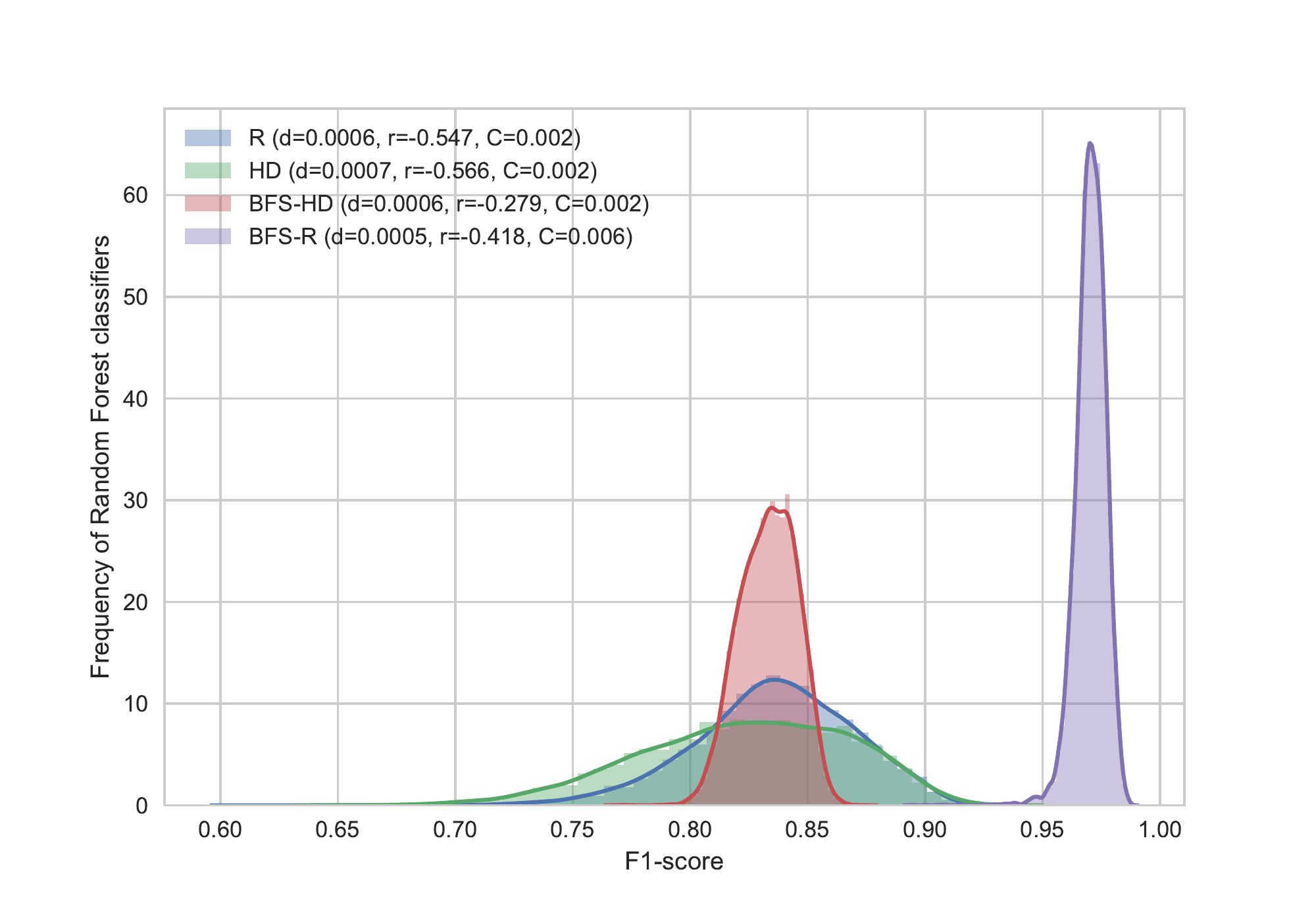}
		\label{fig:attack_strength_gplus_2k}
	}
	\subfloat[soc-gplus: $2.5K$]{
		\includegraphics[width=0.3\linewidth,height=0.1\textheight, keepaspectratio]{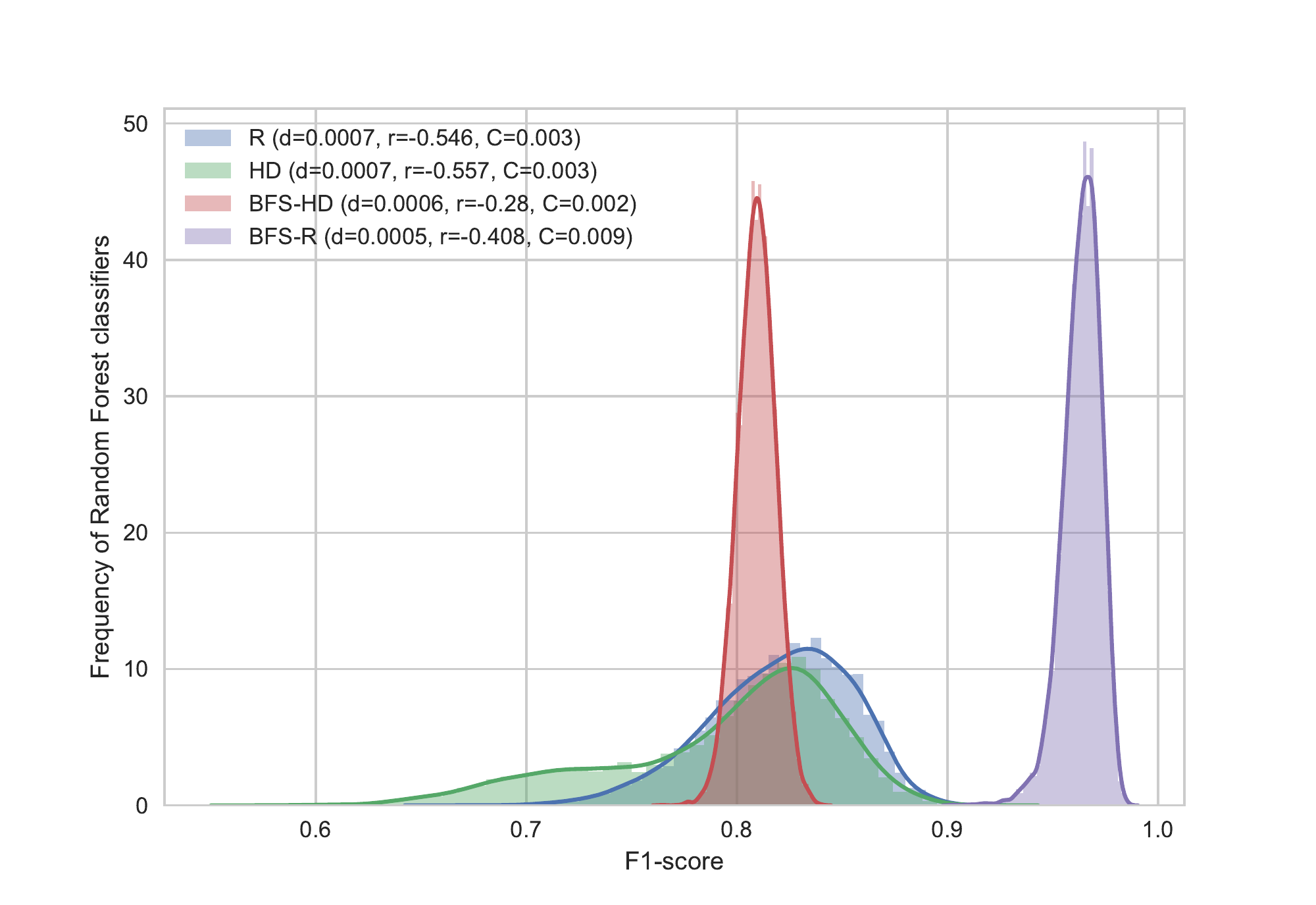}
		\label{fig:attack_strength_gplus_25k}
	}
	\hspace{0mm}
	\subfloat[wikinews: $1K$]{
		\includegraphics[width=0.3\linewidth,height=0.1\textheight, keepaspectratio]{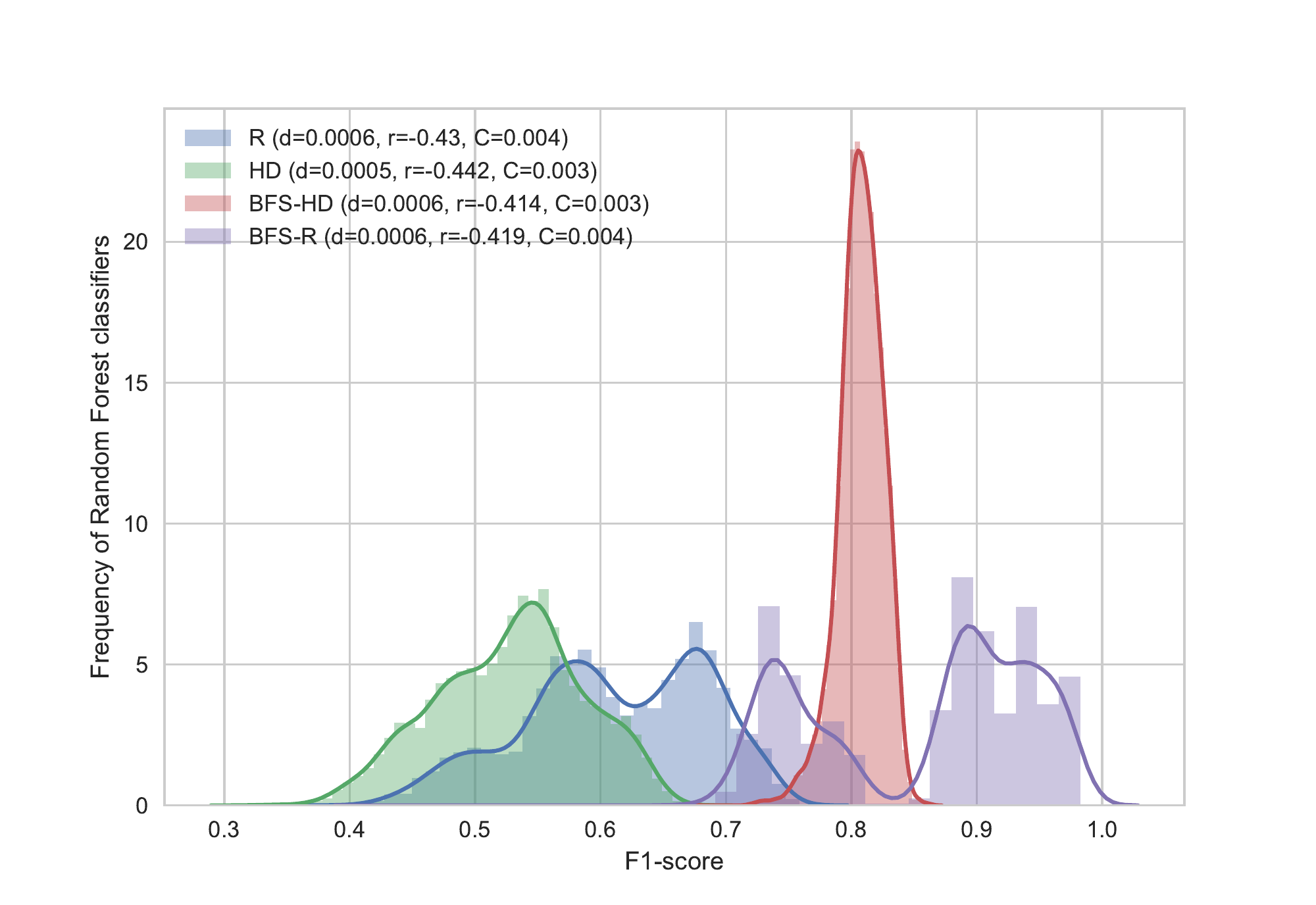}
		\label{fig:attack_strength_web_1k}
	}
	\subfloat[wikinews: $2K$]{
		\includegraphics[width=0.3\linewidth,height=0.1\textheight, keepaspectratio]{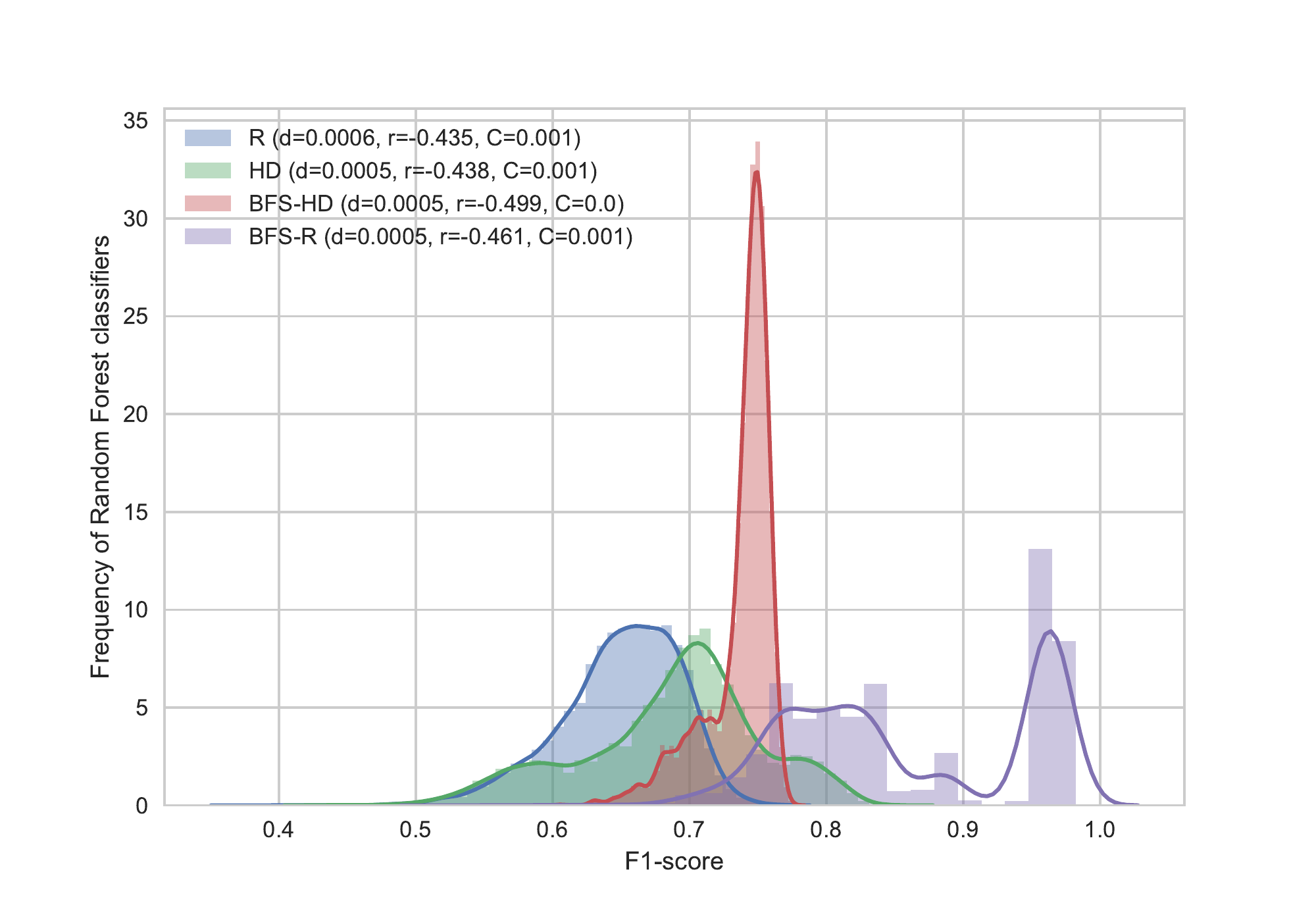}
		\label{fig:attack_strength_web_2k}
	}
	\subfloat[wikinews: $2.5K$]{
		\includegraphics[width=0.3\linewidth,height=0.1\textheight, keepaspectratio]{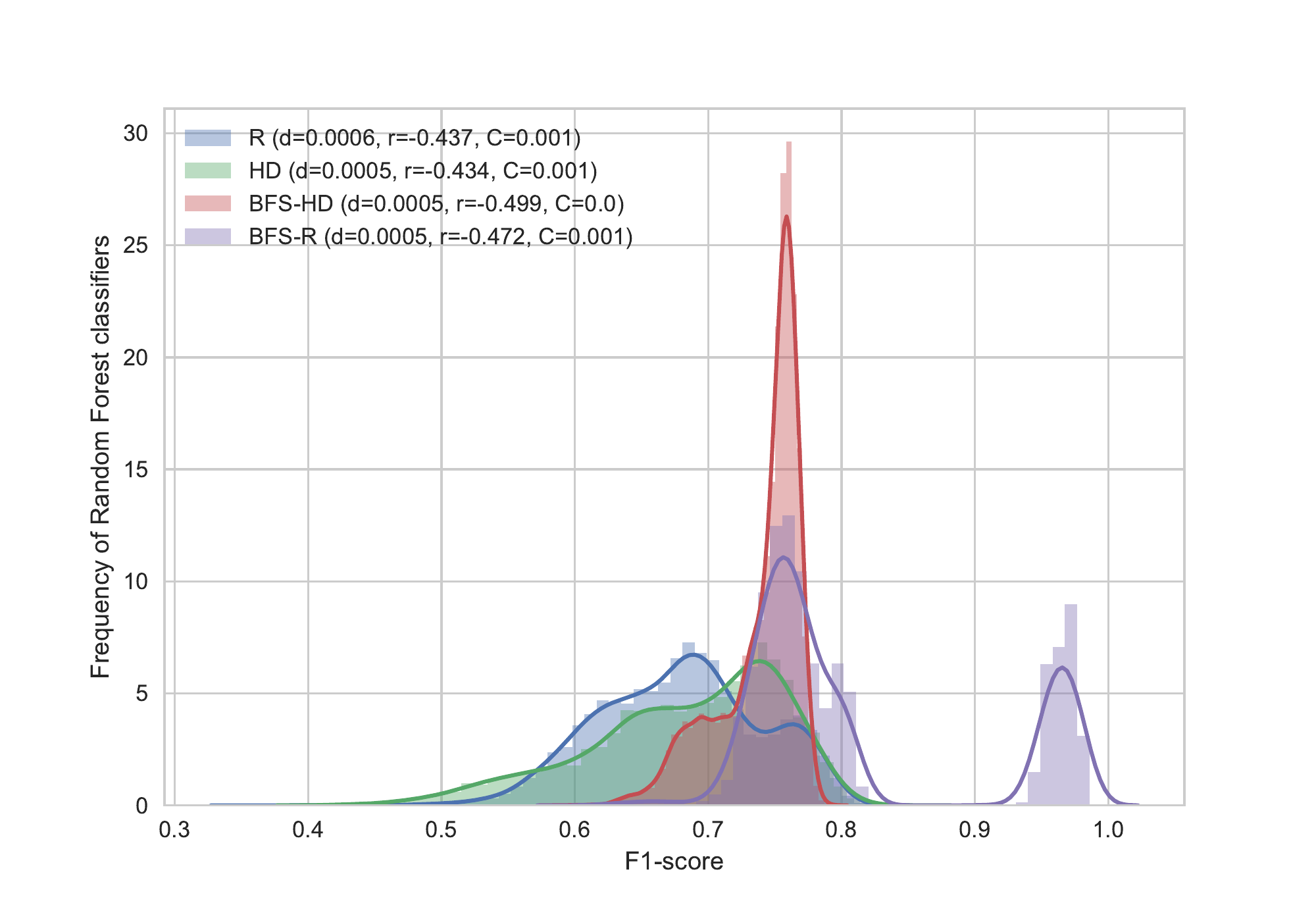}
		\label{fig:attack_strength_web_25k}
	}
	
	\caption{Accuracy of prediction of identical pairs over different $dK$ spaces. Graph properties of density $(\bar{d})$, assortativity $(r)$, and transitivity $(C)$ are averaged over 8 subgraphs per $dK$-space that are associated with the given overlap. 
	}
	\label{fig:attack_strength}
\end{figure}

Figure~\ref{fig:attack_strength} shows the performance of the classifier on re-identifying identical nodes over different $dK$-spaces and different overlap subgraph topologies.
When the overlap is a strongly connected neighborhood, more nodes are re-identified accurately. 
As an example, when the nodes are part of an BFS-Tree (red and purple colors in the plot), 
they are more vulnerable to re-identification. 

We analyzed the structural properties of the $G_{san}$/$G_{aux}$ graphs to explain this phenomenon.
Before discussing our observations, we note that the $G_{san}$ and $G_{aux}$ graphs generated as part of our methodology are structurally dependent on the choice of overlap, because they are subgraphs of the $dK$-graph biased by the choice of the overlap.
Consequently, a higher density overlap choice (e.g., BFS-based) will likely lead to a higher density $G_{san}$/$G_{aux}$ graphs. 
We compute the density and average transitivity over the 80 sample graphs generated for each $dK$-space by grouping the less vulnerable graphs (according to Figure~\ref{fig:attack_strength}) and the more vulnerable graphs together. 
We found that density and transitivity are positively correlated with the levels of vulnerability.
That is, the mean density of $dK$-graphs with an overlap with weakly connected nodes (i.e., R or HD) is always lower than the $dK$-graphs with an overlap of strongly connected neighborhood (i.e., generated by a BFS tree) and also less vulnerable to re-identification.

At the same time, we notice that across all $dK$-spaces, the average density remains constant (e.g., 0.012 for R/HD-based $dK$-graphs and 0.018 for BFS-based $dK$-graphs). 
We believe this is due to the fact that density is highly constrained by degree distribution which is preserved in all $dK$-spaces.
Not the same happens with transitivity across $dK$-spaces. 
Higher $dK$-spaces preserve the connectivity of $d$-subgraphs ($d\geq2$) that, in turn, preserve the local clustering.
Hence, it appears that attackers benefit from a higher clustering coefficient.
Also, for sparser graphs, transitivity is a telling metric for quantifying the strength of an attack: a connected neighborhood which has a relatively high transitivity can put more nodes in danger in a sparse graph (see the BFS-R plots in Figures~\ref{fig:attack_strength_anybeat_1k}--\ref{fig:attack_strength_web_25k}). 

\ignore{
\begin{figure}[t]
	\subfloat[Density]{
		\includegraphics[scale=0.425]{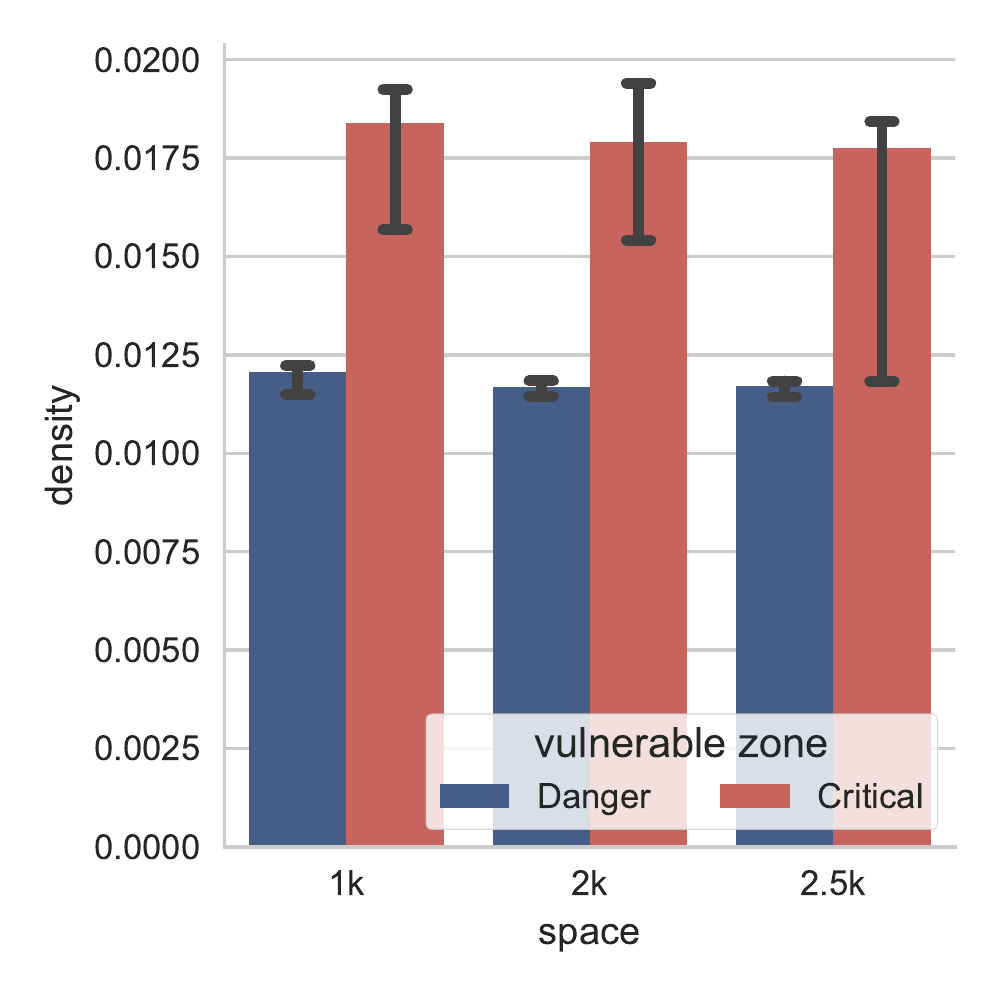}
		\label{fig:vulnerable_zones_density}
	}
	\subfloat[Transitivity]{
		\includegraphics[scale=0.425]{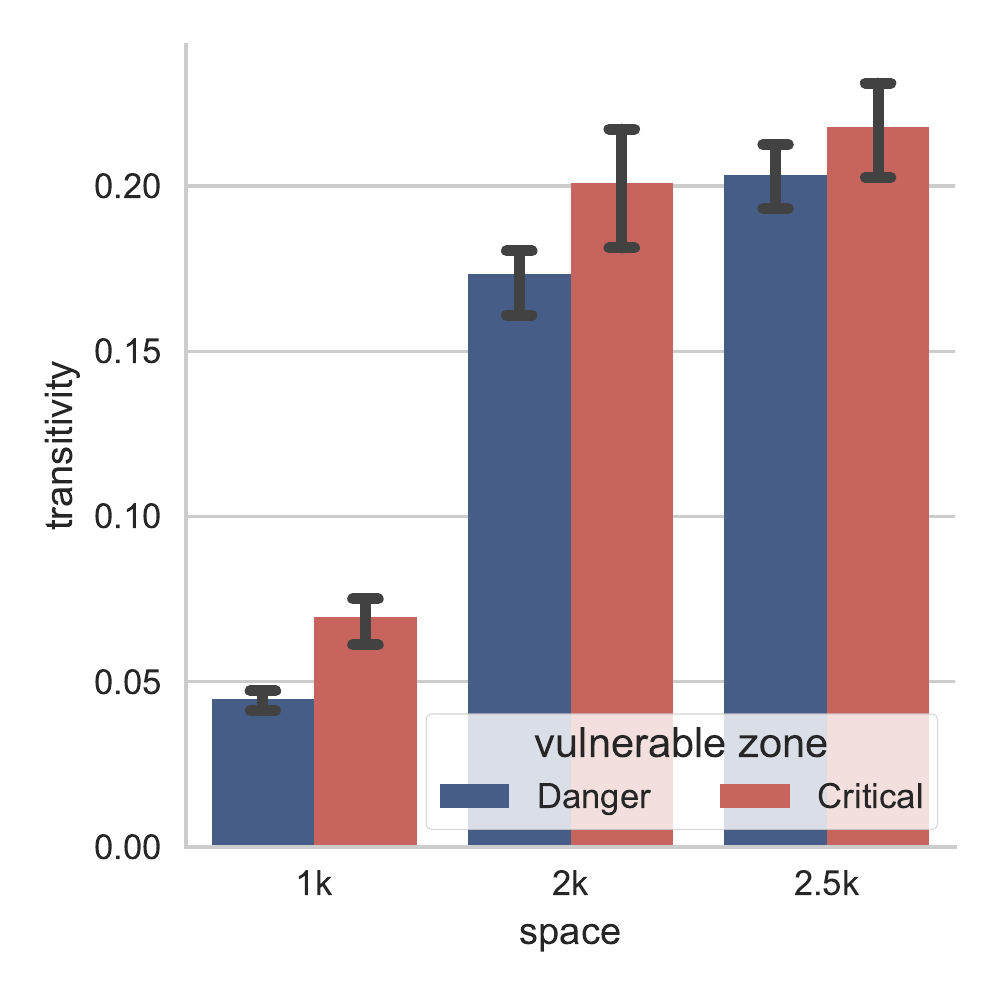}
		\label{fig:vulnerable_zones_transitivity}
	}
	\caption{\ainote{I don't get this picture for the life of me. Let's talk.} \shnote{check now, I change the style of the figure to a barplot} A classification of $dK$-subgraph metrics over the zones of vulnerability, where danger zone represents $dK$-graphs under the overlap of a partially connected neighborhood (e.g., High Degree and Random) and critical zone represents $dK$-graphs under the overlap of a strongly connected neighborhood (e.g., BFS-Tree). Critical zone depicts more vulnerable graph spaces. a) shows the mean density b) shows the mean transitivity over $80$ graphs for each $dK$-space, while both figures indicate the confidence intervals using error bars.}
	\label{fig:vulnerable_zones}
\end{figure}
}

\subsection{$dK$-space Vulnerability}
\begin{figure*}
	\begin{tabular}{cccc}
		\subfloat[fb107: R]{\includegraphics[width=0.25\textwidth,height=0.2\textheight,keepaspectratio]{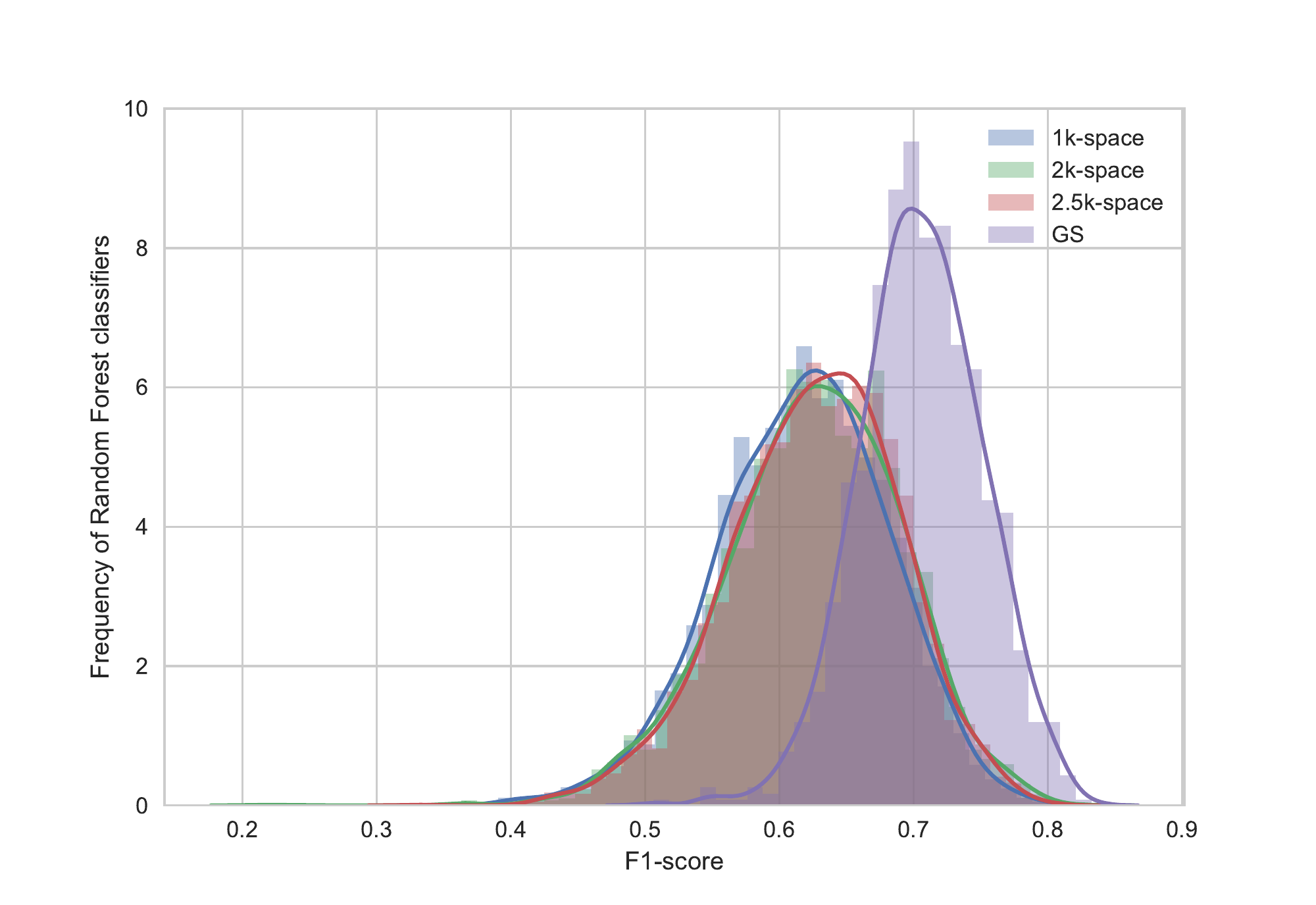}\label{fig:dk_space_vulnerability_fb107_ran1}} &
		\subfloat[fb107: HD]{\includegraphics[width=0.25\textwidth,height=0.2\textheight,keepaspectratio]{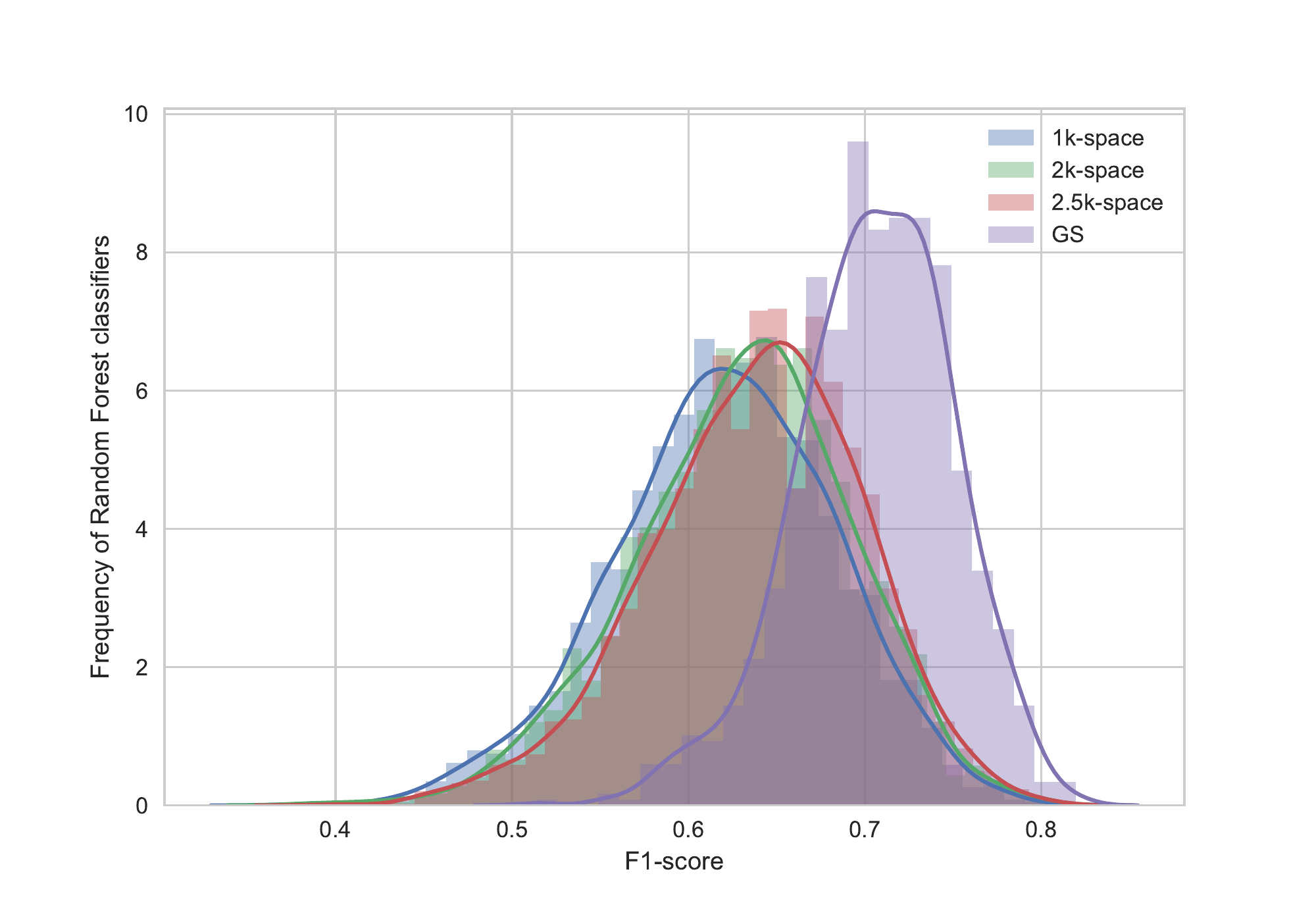}\label{fig:dk_space_vulnerability_fb107_deg}} &
		\subfloat[fb107: BFS-HD]{\includegraphics[width=0.25\textwidth,height=0.2\textheight,keepaspectratio]{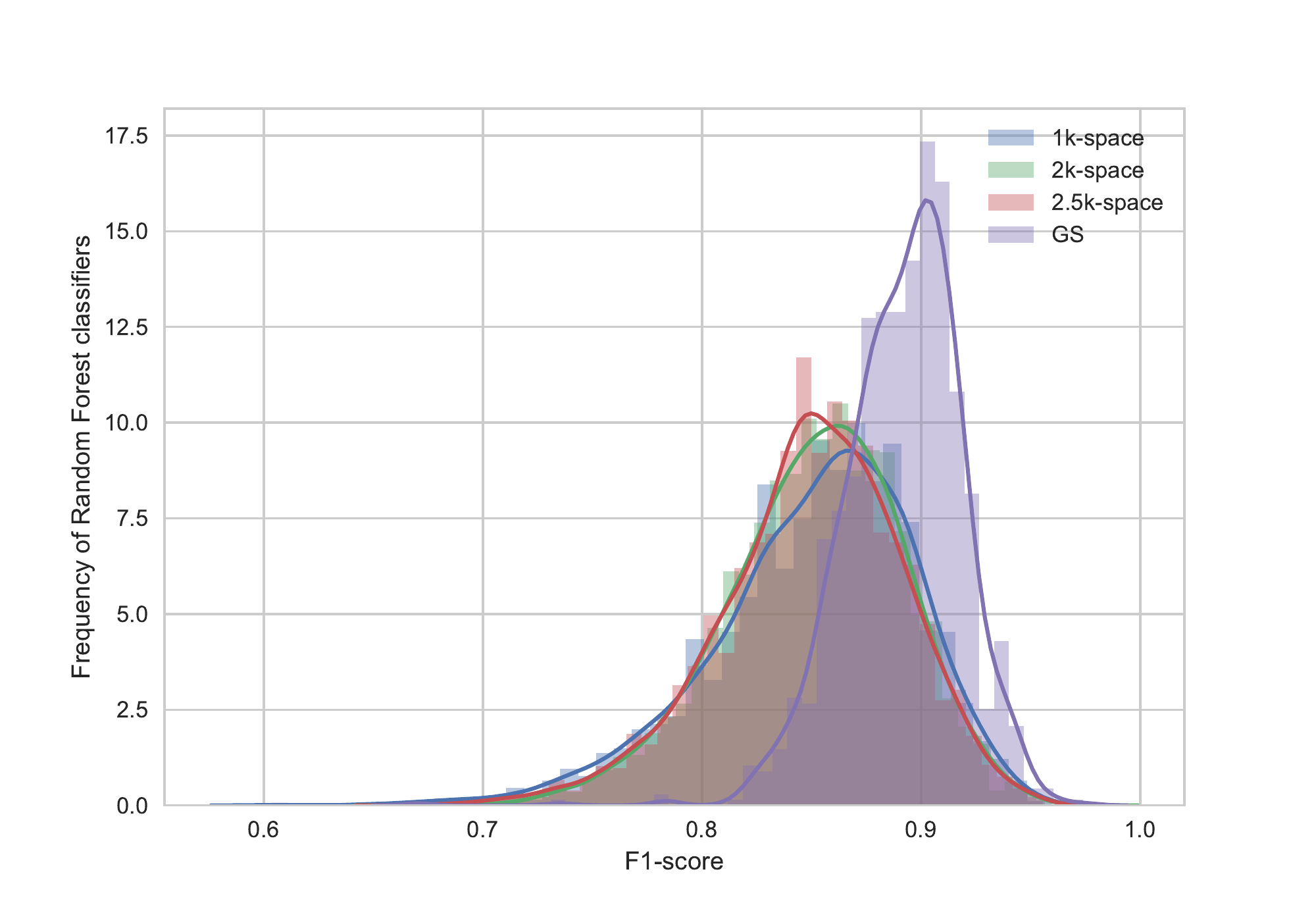}\label{fig:dk_space_vulnerability_fb107_bfs1}} &
		\subfloat[fb107: BFS-R]{\includegraphics[width=0.25\textwidth,height=0.2\textheight,keepaspectratio]{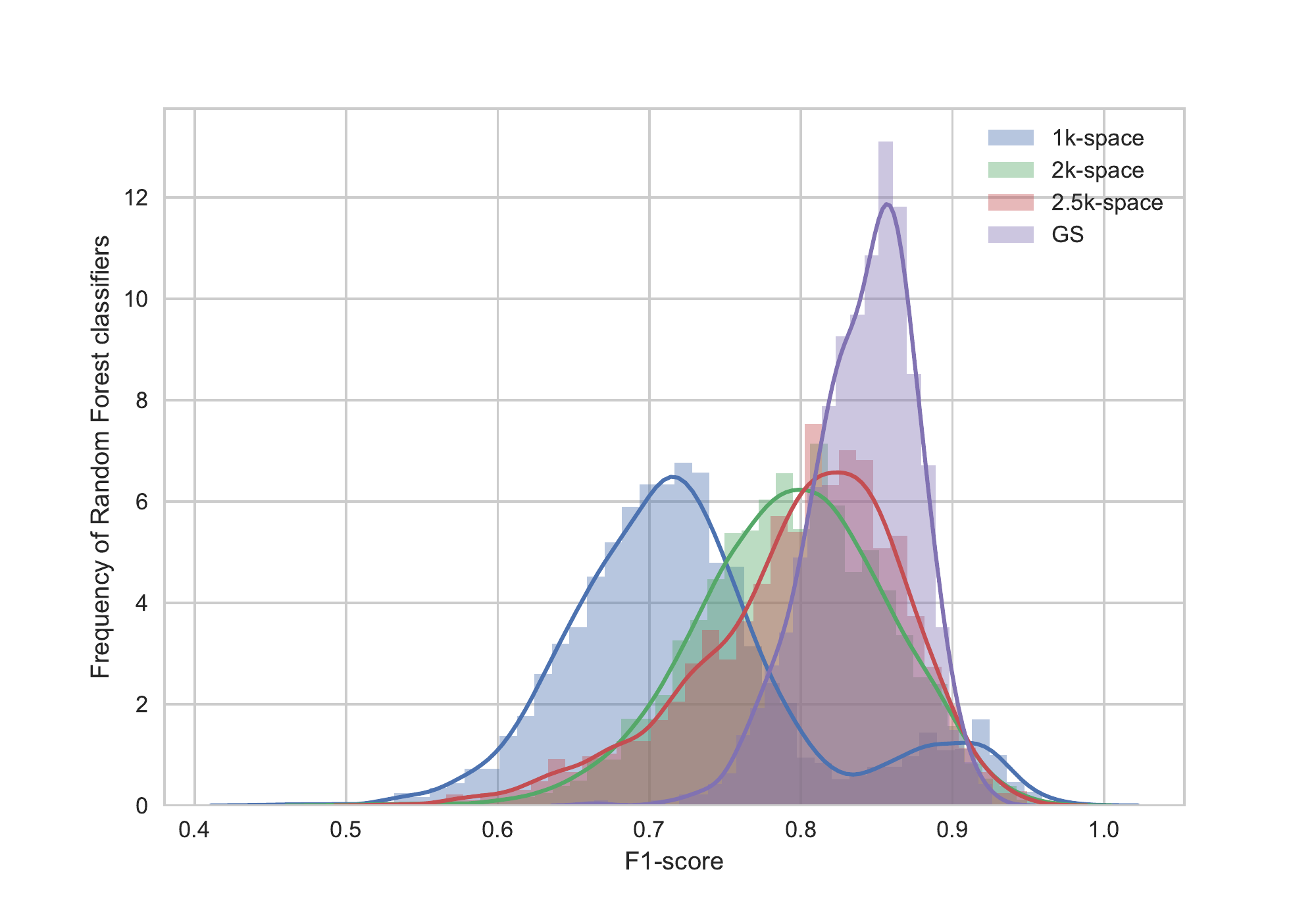}\label{fig:dk_space_vulnerability_fb107_bfs2}}\\
\subfloat[caGrQc: R]{\includegraphics[width=0.25\textwidth,height=0.15\textheight,keepaspectratio]{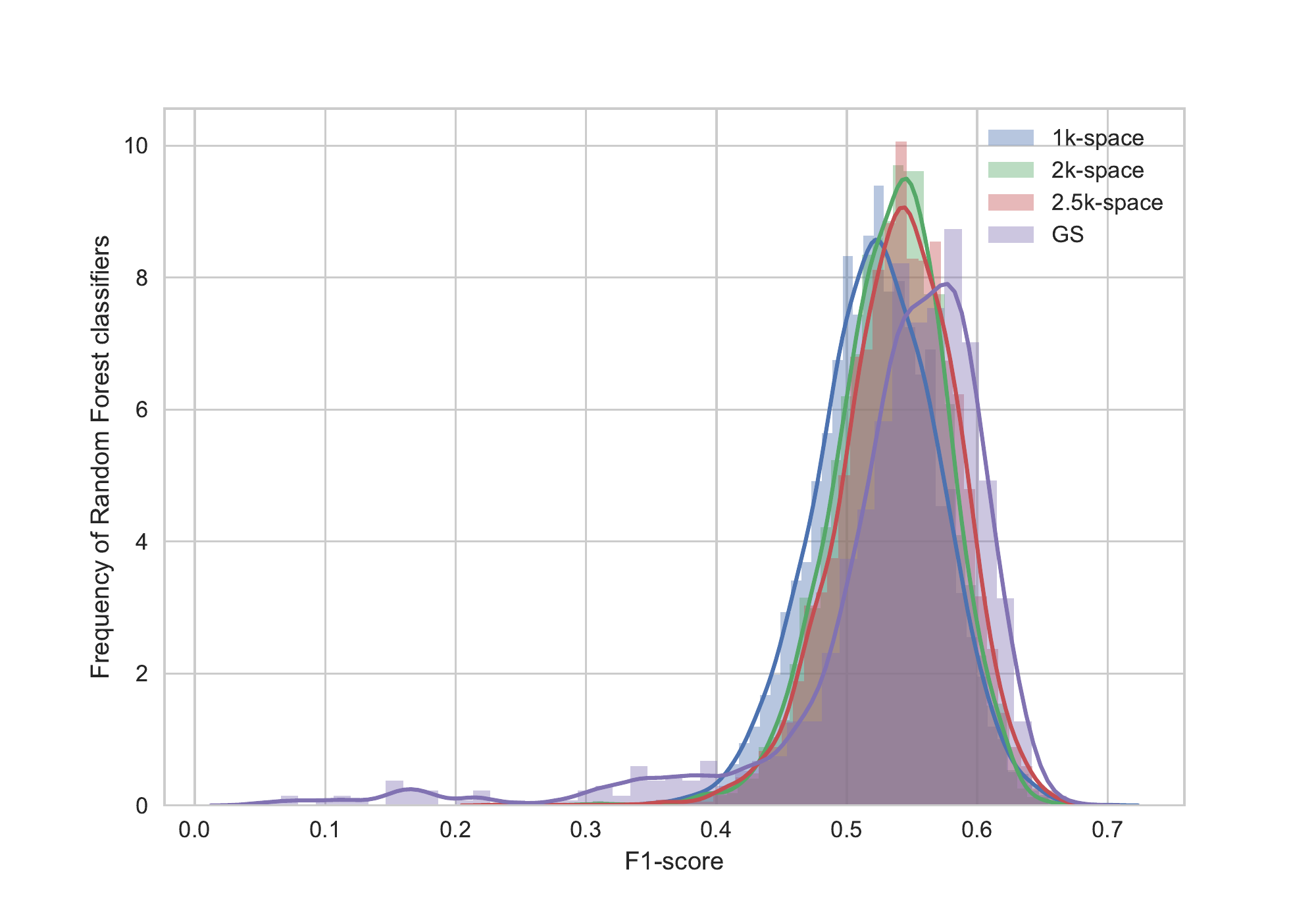}\label{fig:dk_space_vulnerability_caGrQc_ran1}} &
\subfloat[caGrQc: HD]{\includegraphics[width=0.25\textwidth,height=0.15\textheight,keepaspectratio]{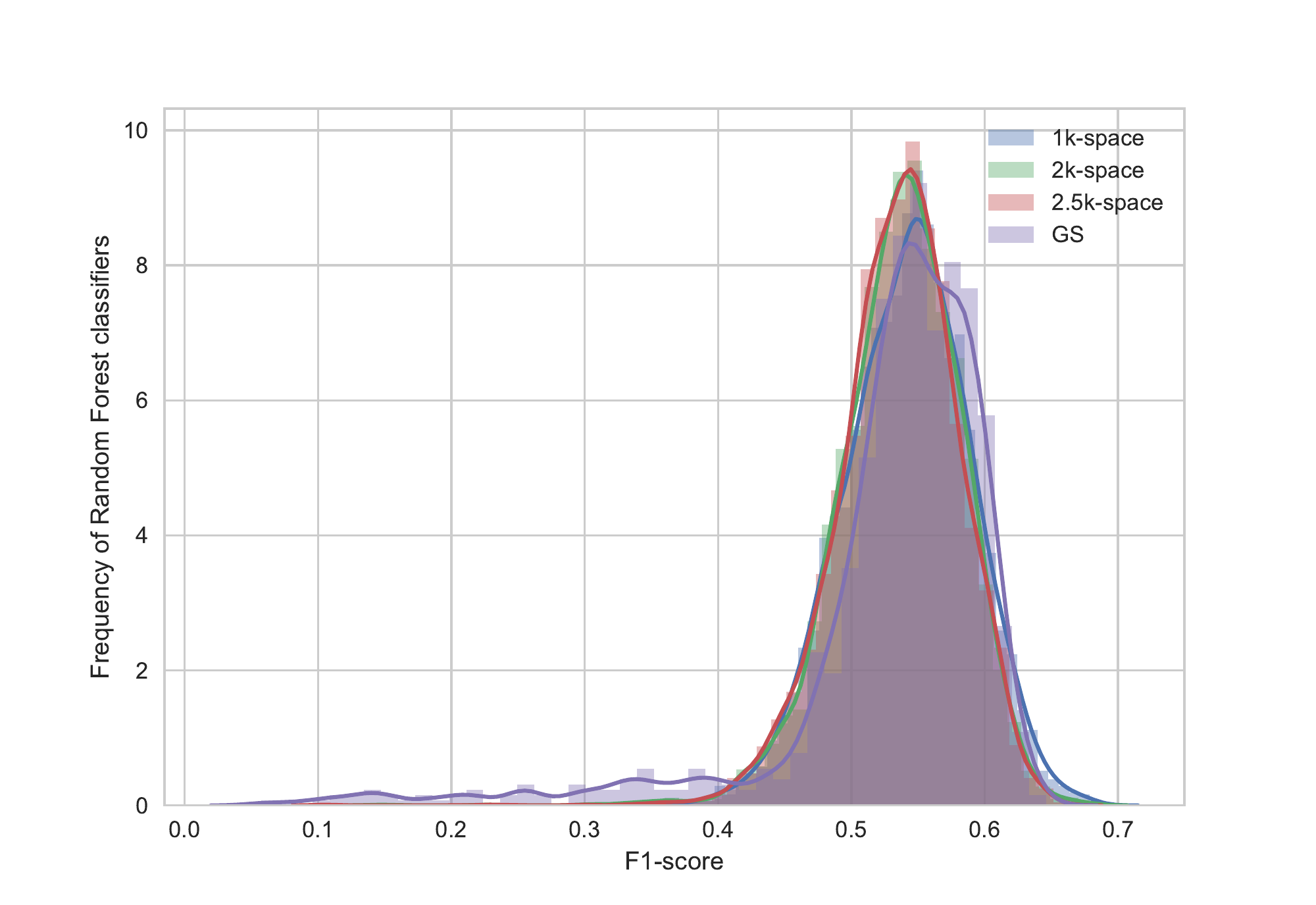}\label{fig:dk_space_vulnerability_caGrQc_deg}} &
\subfloat[caGrQc: BFS-HD]{\includegraphics[width=0.25\textwidth,height=0.15\textheight,keepaspectratio]{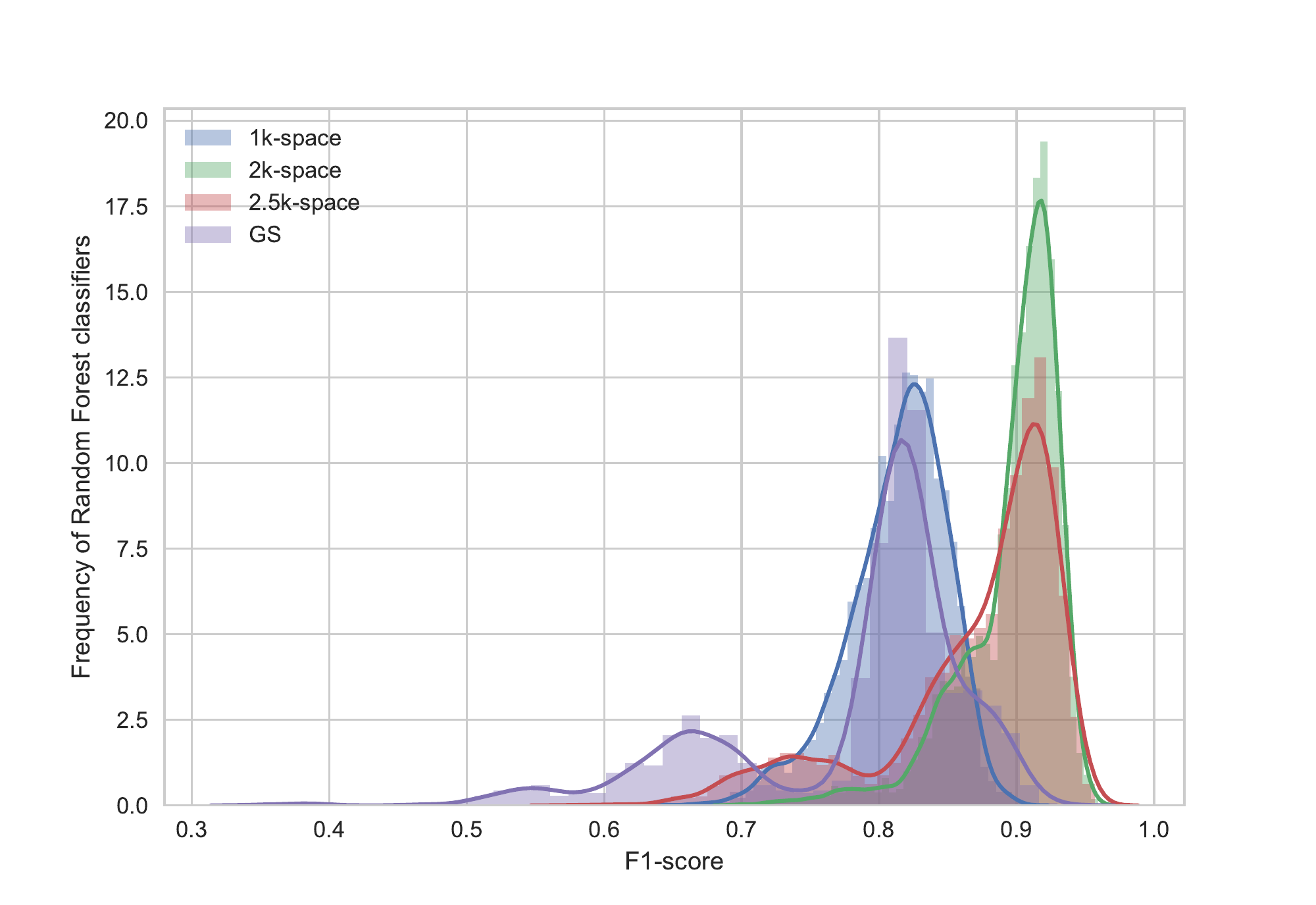}\label{fig:dk_space_vulnerability_caGrQc_bfs1}} &
\subfloat[caGrQc: BFS-R]{\includegraphics[width=0.25\textwidth,height=0.15\textheight,keepaspectratio]{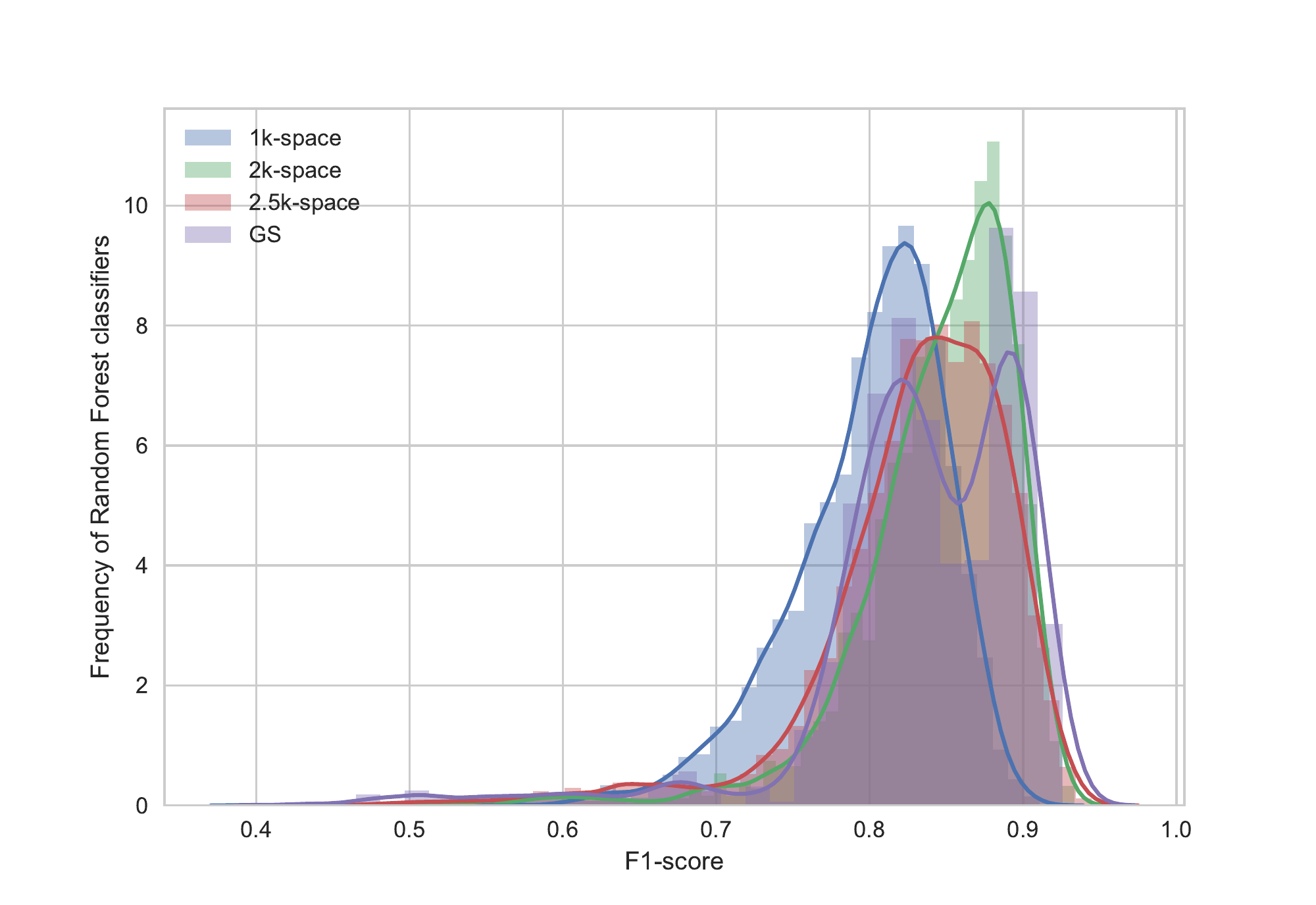}\label{fig:dk_space_vulnerability_caGrQc_bfs2}}\\
	\subfloat[soc-anybeat: R]{\includegraphics[width=0.25\textwidth,height=0.15\textheight,keepaspectratio]{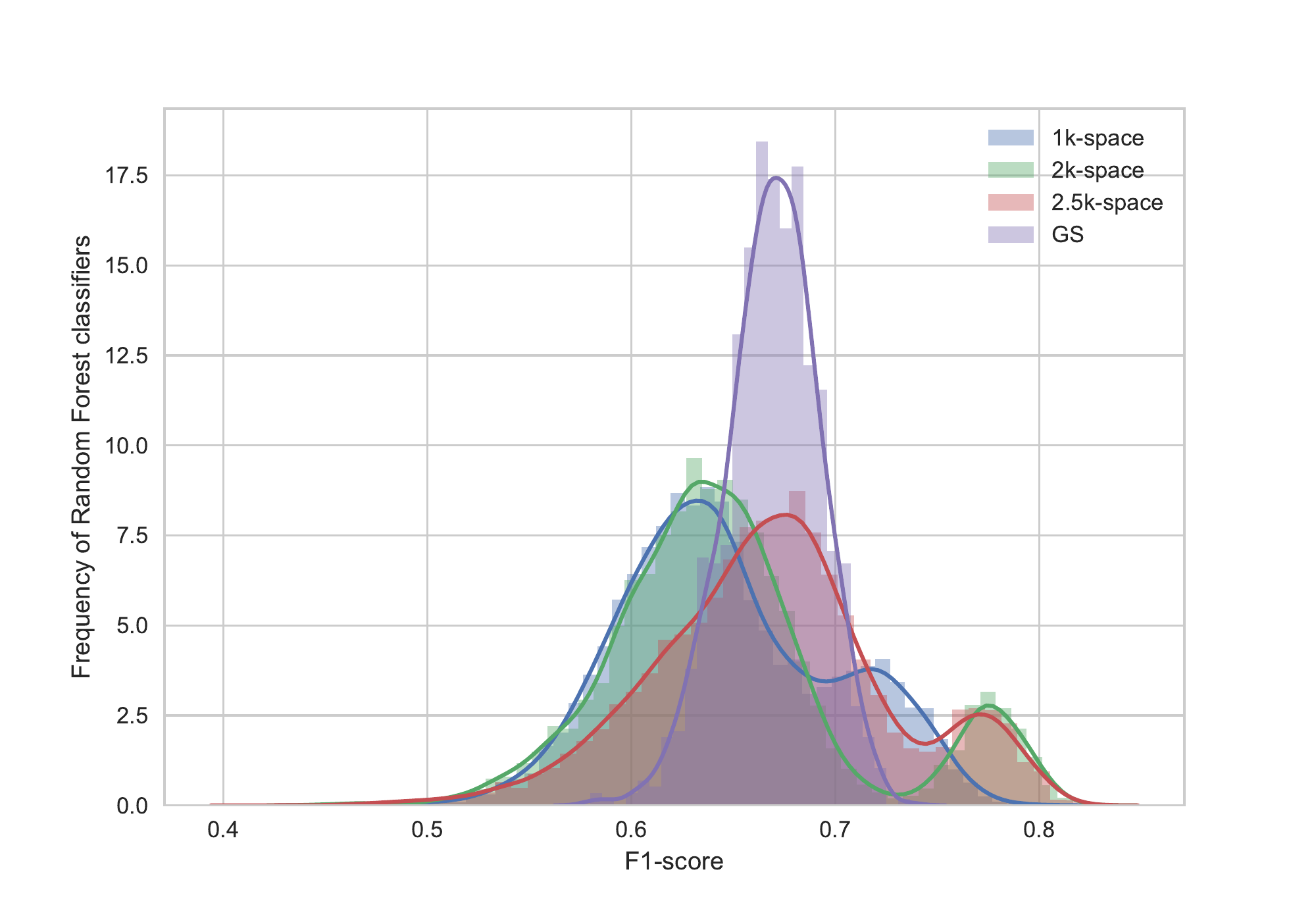}\label{fig:dk_space_vulnerability_anybeat_ran1}} &
	\subfloat[soc-anybeat: HD]{\includegraphics[width=0.25\textwidth,height=0.15\textheight,keepaspectratio]{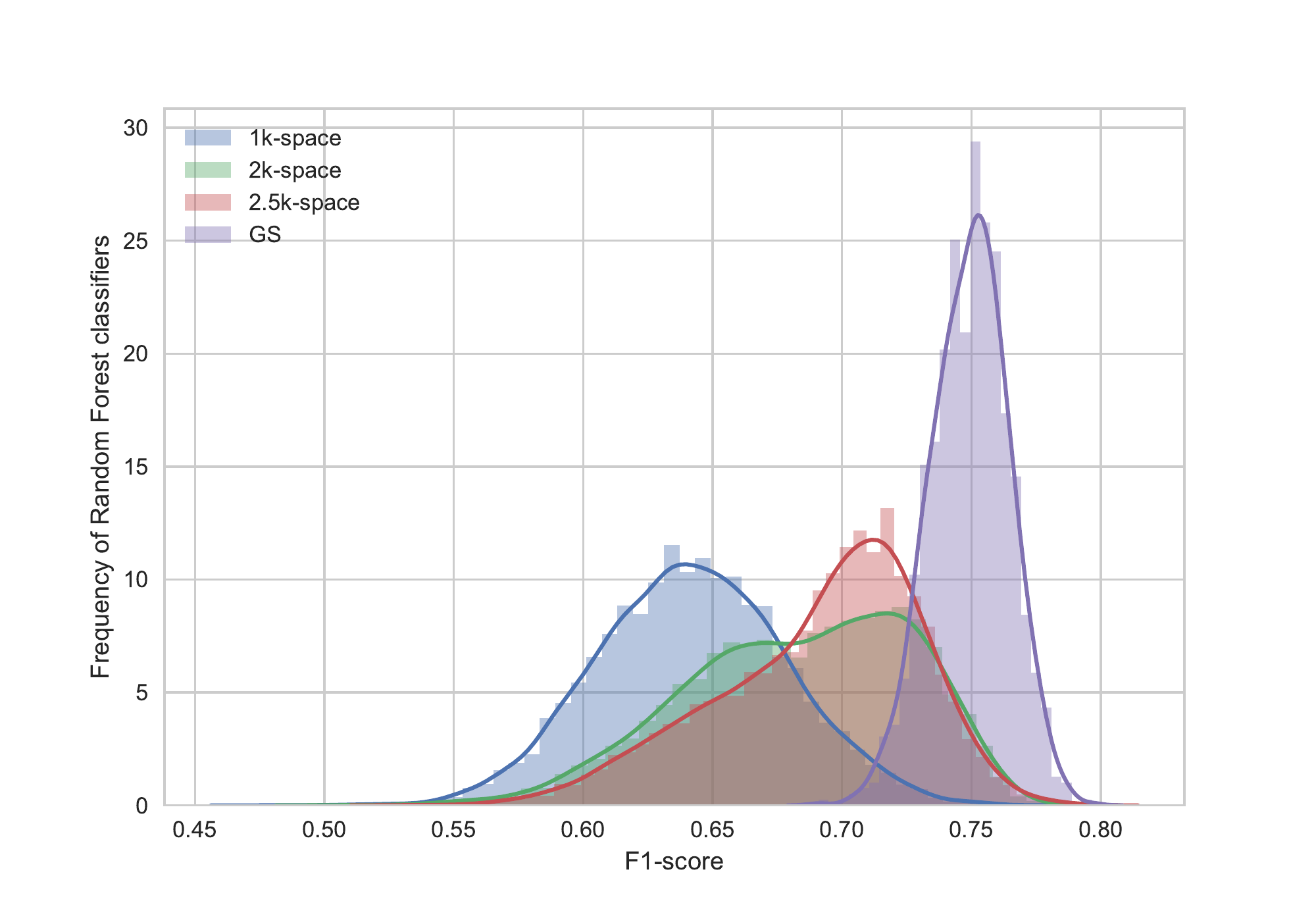}\label{fig:dk_space_vulnerability_anybeat_deg}} &
	\subfloat[soc-anybeat: BFS-HD]{\includegraphics[width=0.25\textwidth,height=0.15\textheight,keepaspectratio]{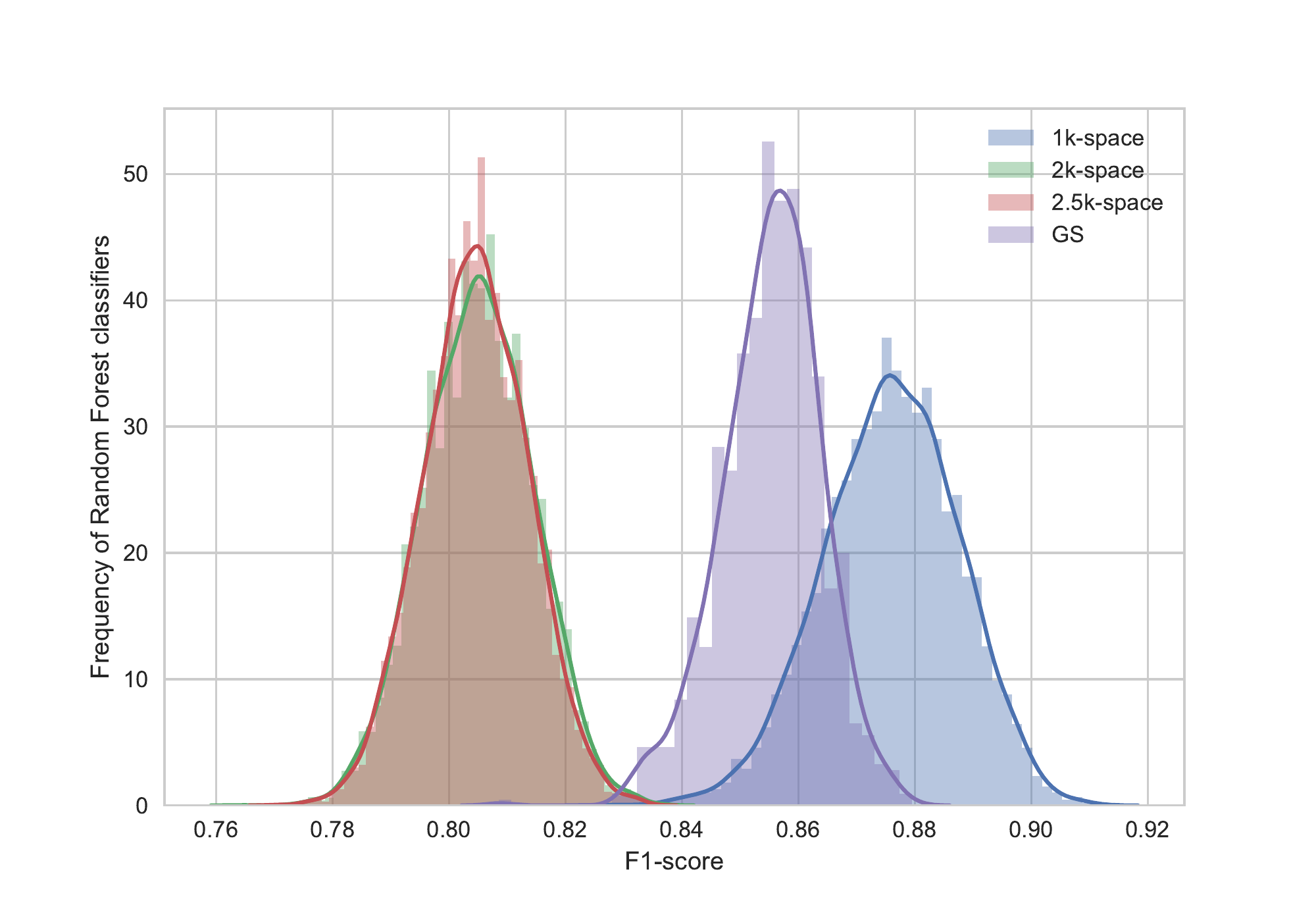}\label{fig:dk_space_vulnerability_anybeat_bfs1}} &
	\subfloat[soc-anybeat: BFS-R]{\includegraphics[width=0.25\textwidth,height=0.15\textheight,keepaspectratio]{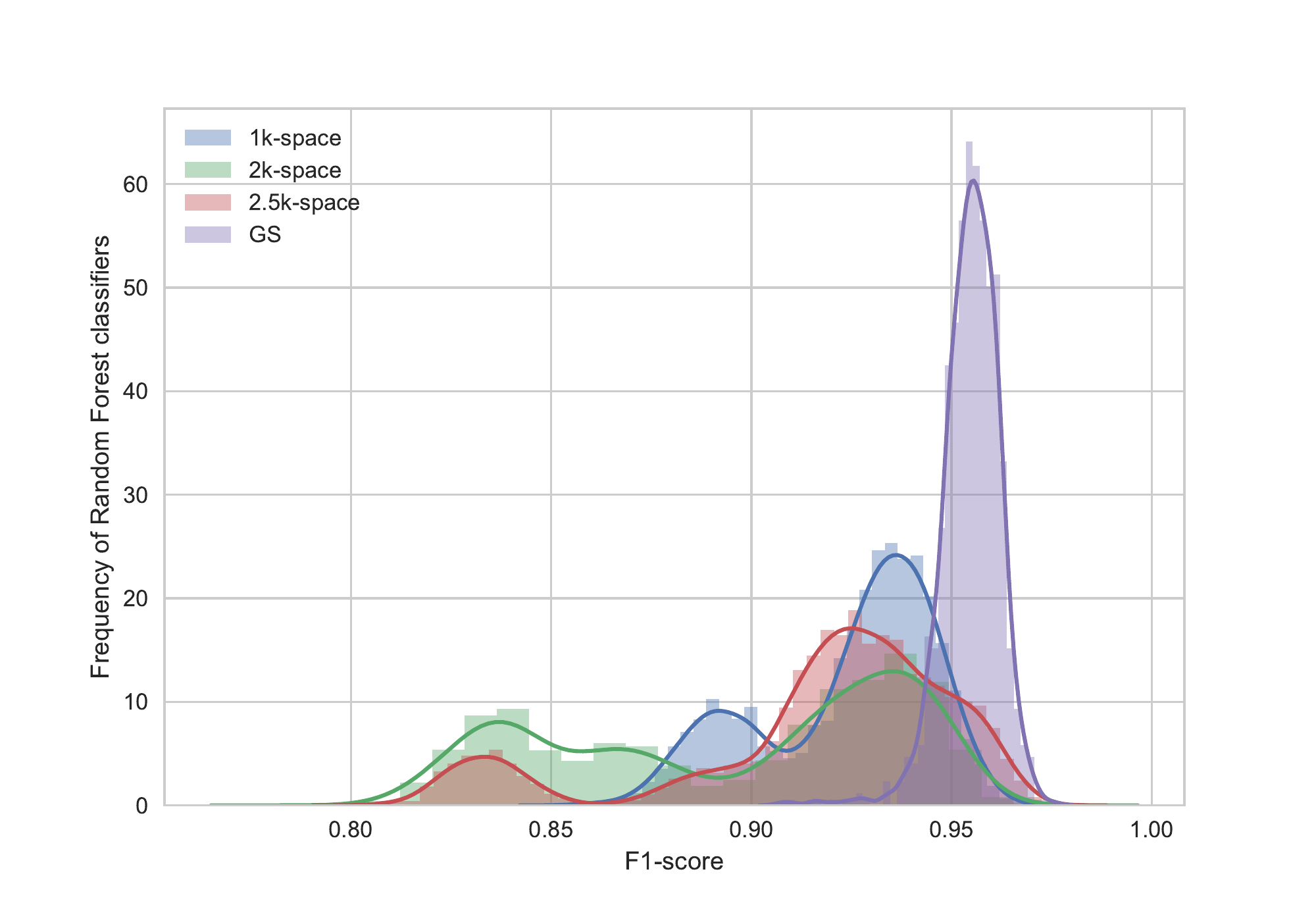}	\label{fig:dk_space_vulnerability_anybeat_bfs2}}\\
	\subfloat[soc-gplus: R]{\includegraphics[width=0.25\textwidth,height=0.15\textheight,keepaspectratio]{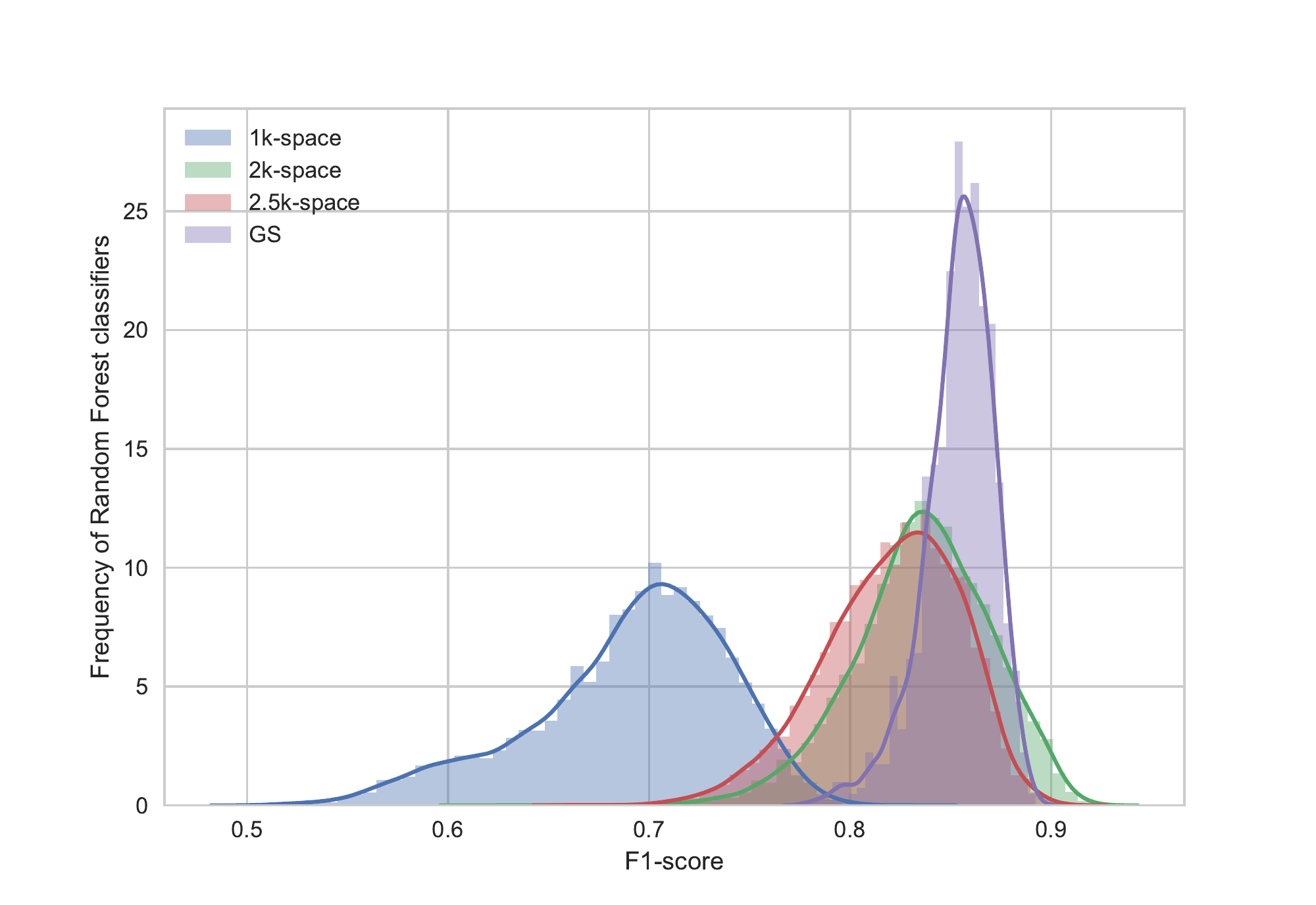}	\label{fig:dk_space_vulnerability_gplus_ran1}} &
	\subfloat[soc-gplus: HD]{\includegraphics[width=0.25\textwidth,height=0.15\textheight,keepaspectratio]{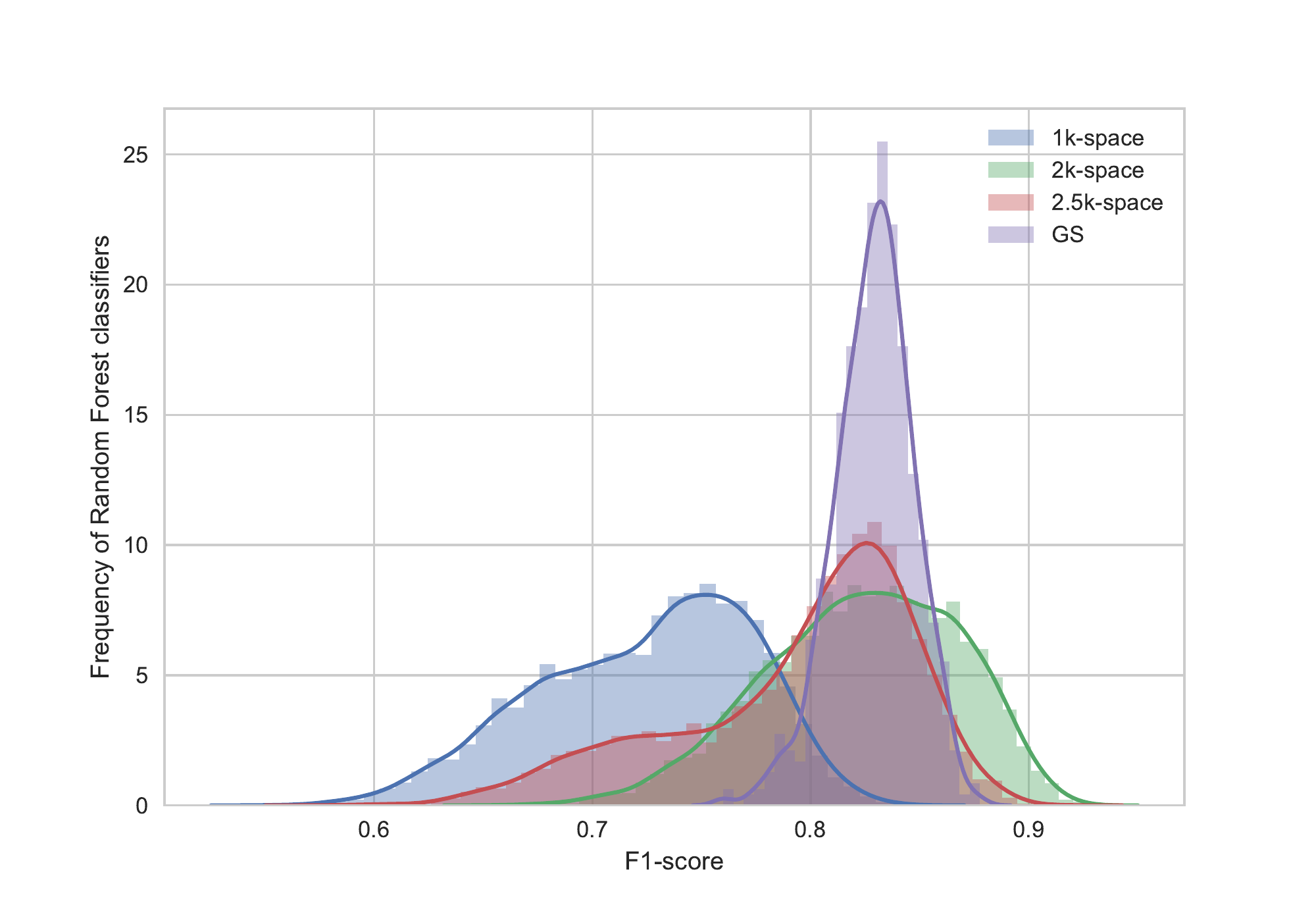}	\label{fig:dk_space_vulnerability_gplus_deg}} &
	\subfloat[soc-gplus: BFS-HD]{\includegraphics[width=0.25\textwidth,height=0.15\textheight,keepaspectratio]{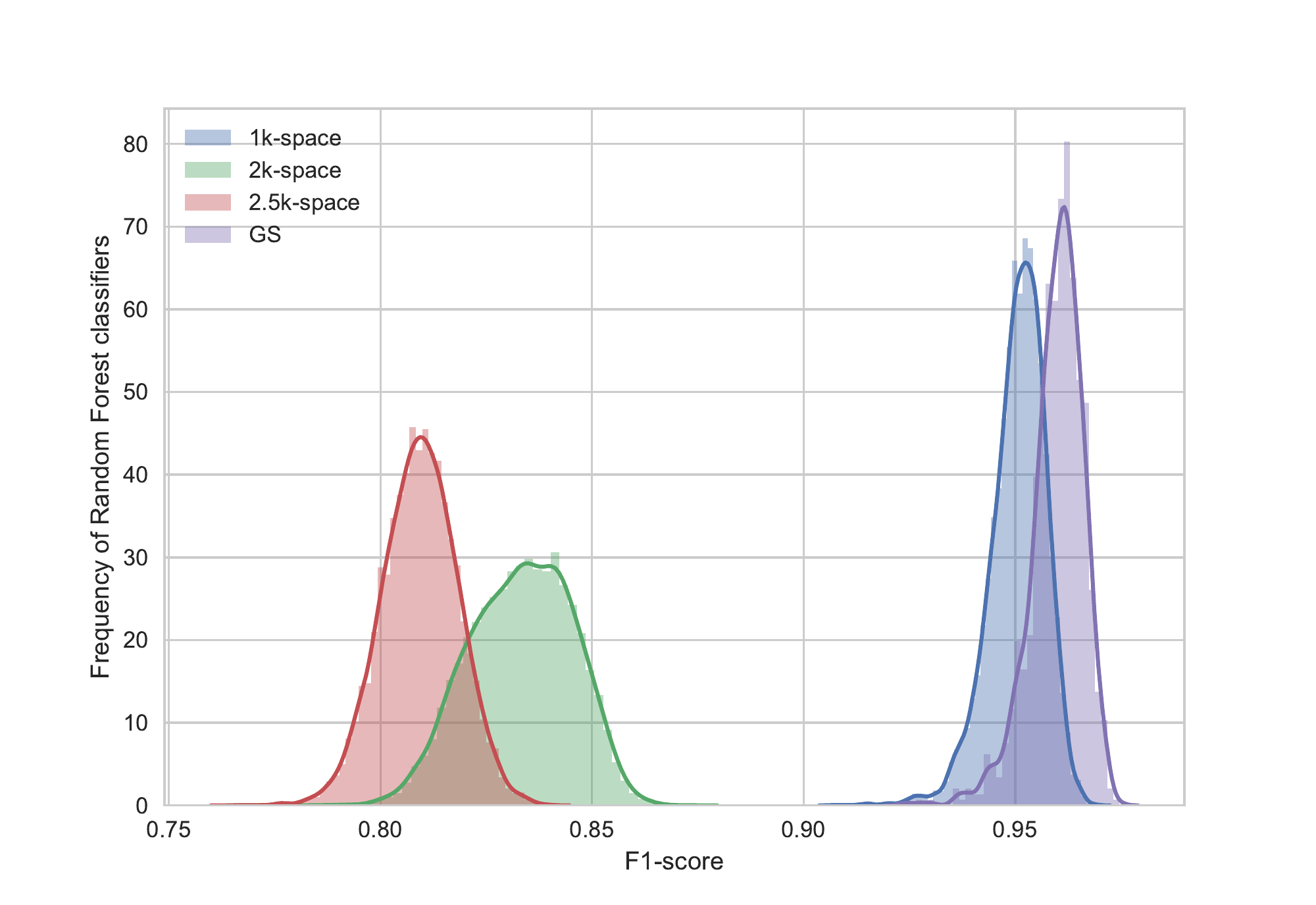}	\label{fig:dk_space_vulnerability_gplus_bfs1}} &
	\subfloat[soc-gplus: BFS-R]{\includegraphics[width=0.25\textwidth,height=0.15\textheight,keepaspectratio]{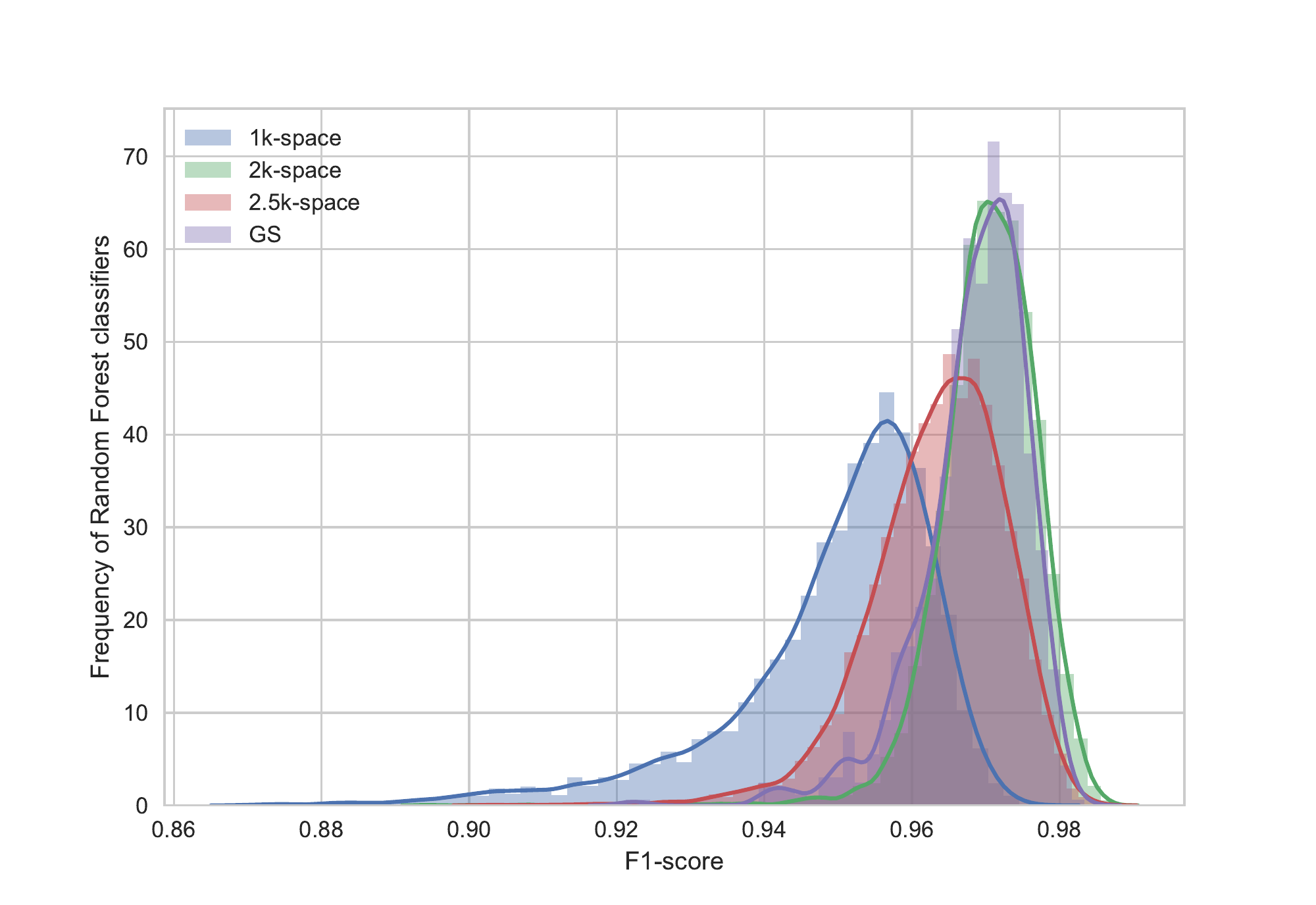}
	\label{fig:dk_space_vulnerability_gplus_bfs2}}\\
	\subfloat[wikinews: R]{\includegraphics[width=0.25\textwidth,height=0.15\textheight,keepaspectratio]{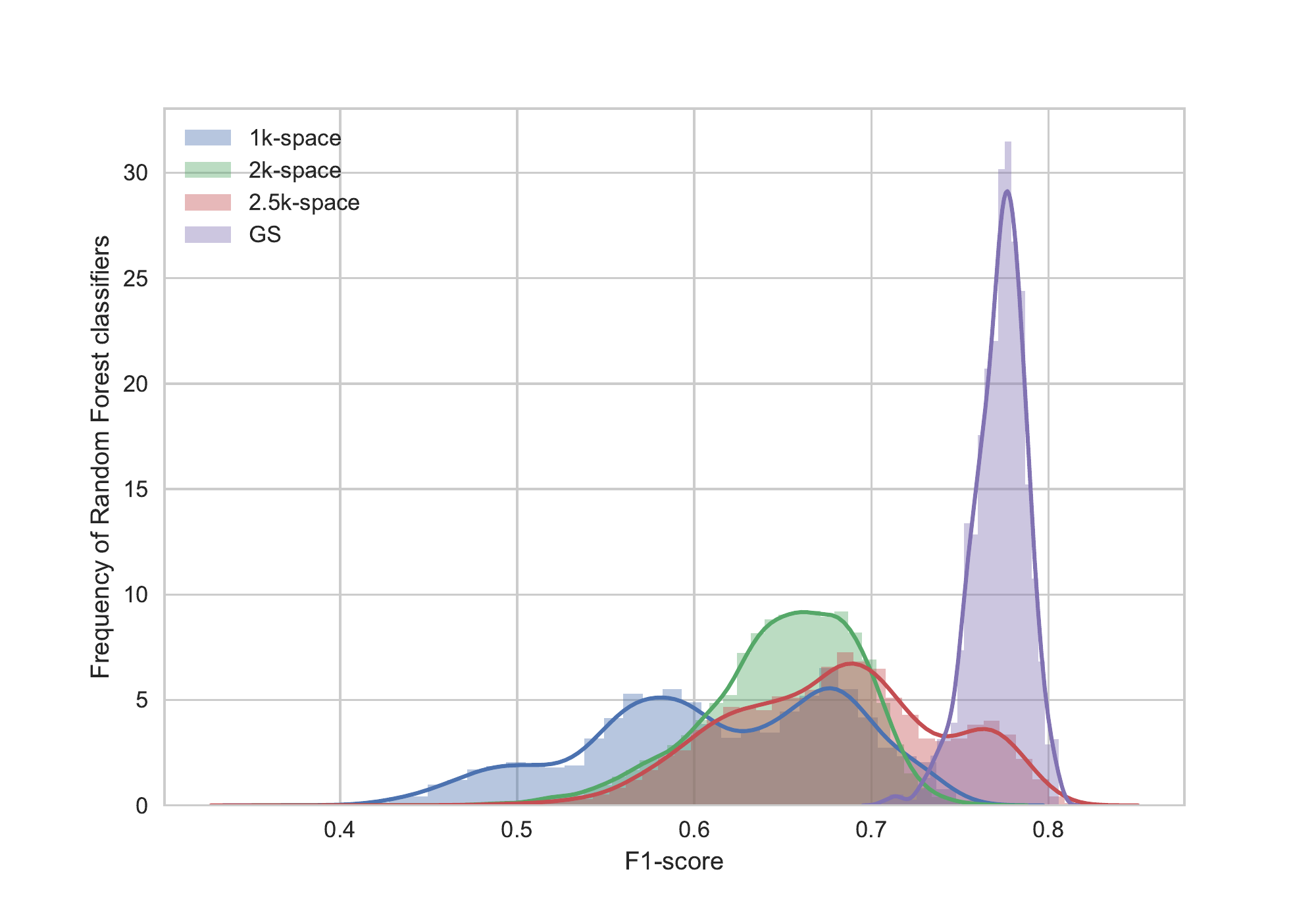}\label{fig:dk_space_vulnerability_web_ran1}} &
	\subfloat[wikinews: HD]{\includegraphics[width=0.25\textwidth,height=0.15\textheight,keepaspectratio]{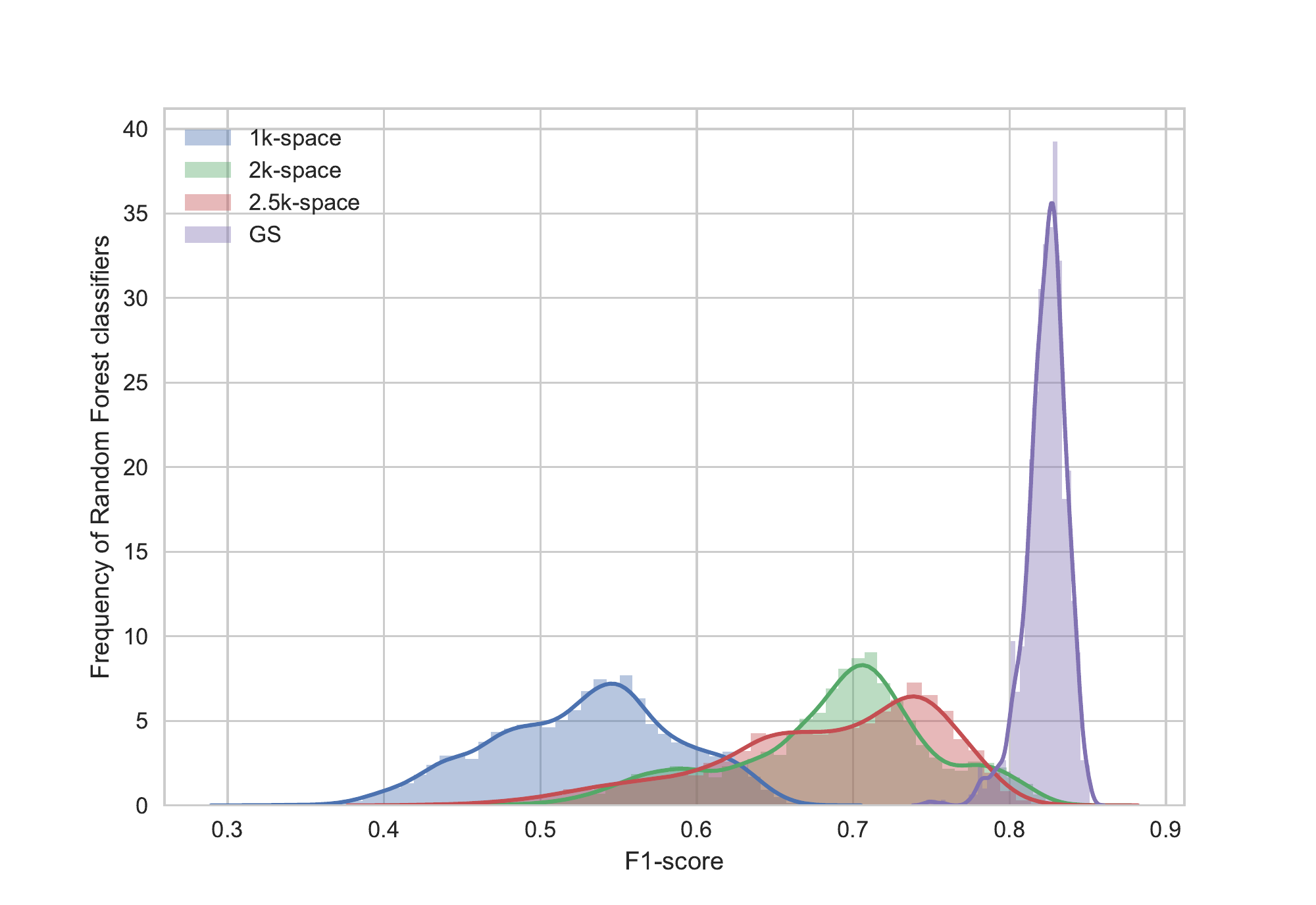}\label{fig:dk_space_vulnerability_web_deg}} &
	\subfloat[wikinews: BFS-HD]{\includegraphics[width=0.25\textwidth,height=0.15\textheight,keepaspectratio]{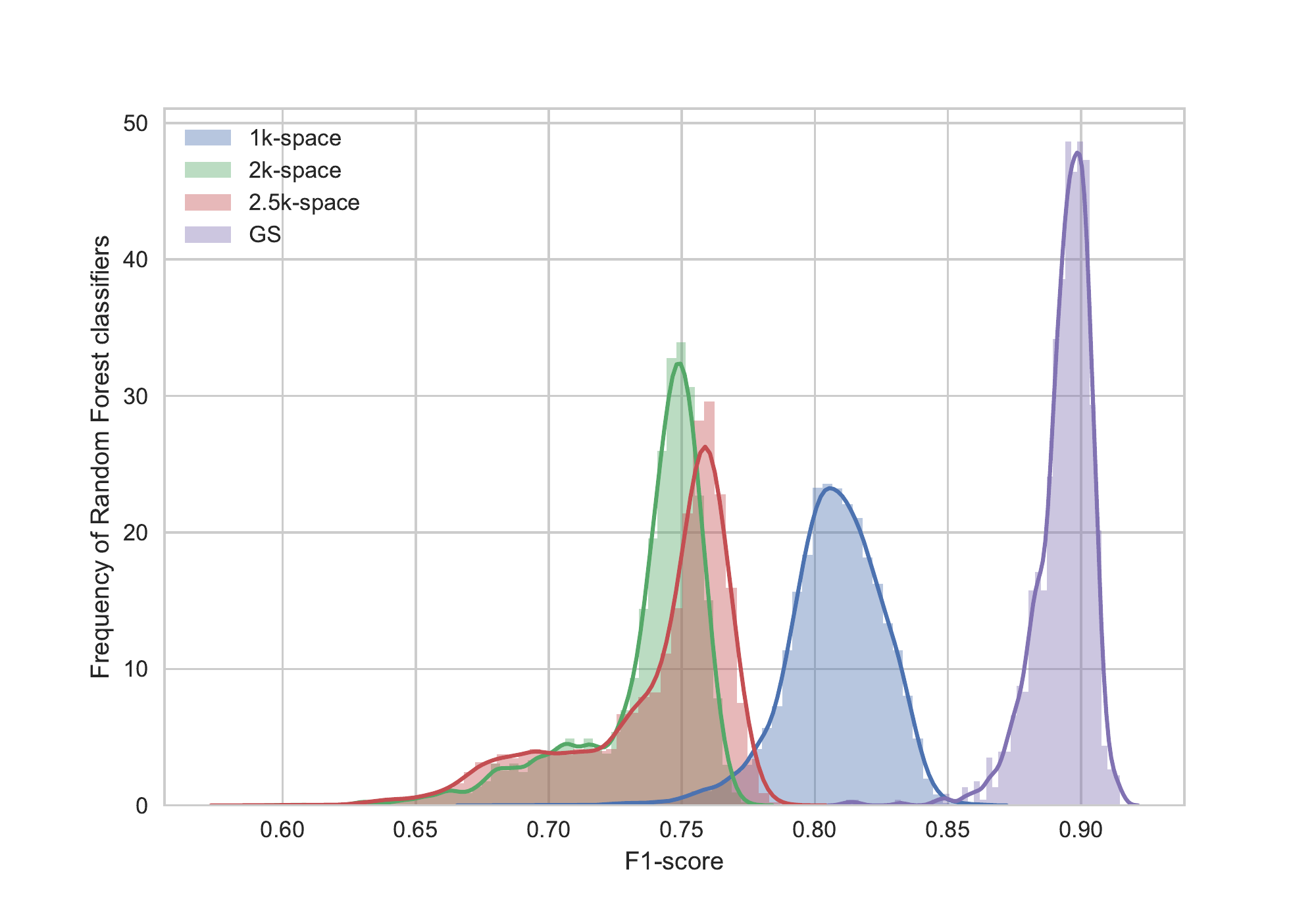}\label{fig:dk_space_vulnerability_web_bfs1}} &
	\subfloat[wikinews: BFS-R]{\includegraphics[width=0.25\textwidth,height=0.15\textheight,keepaspectratio]{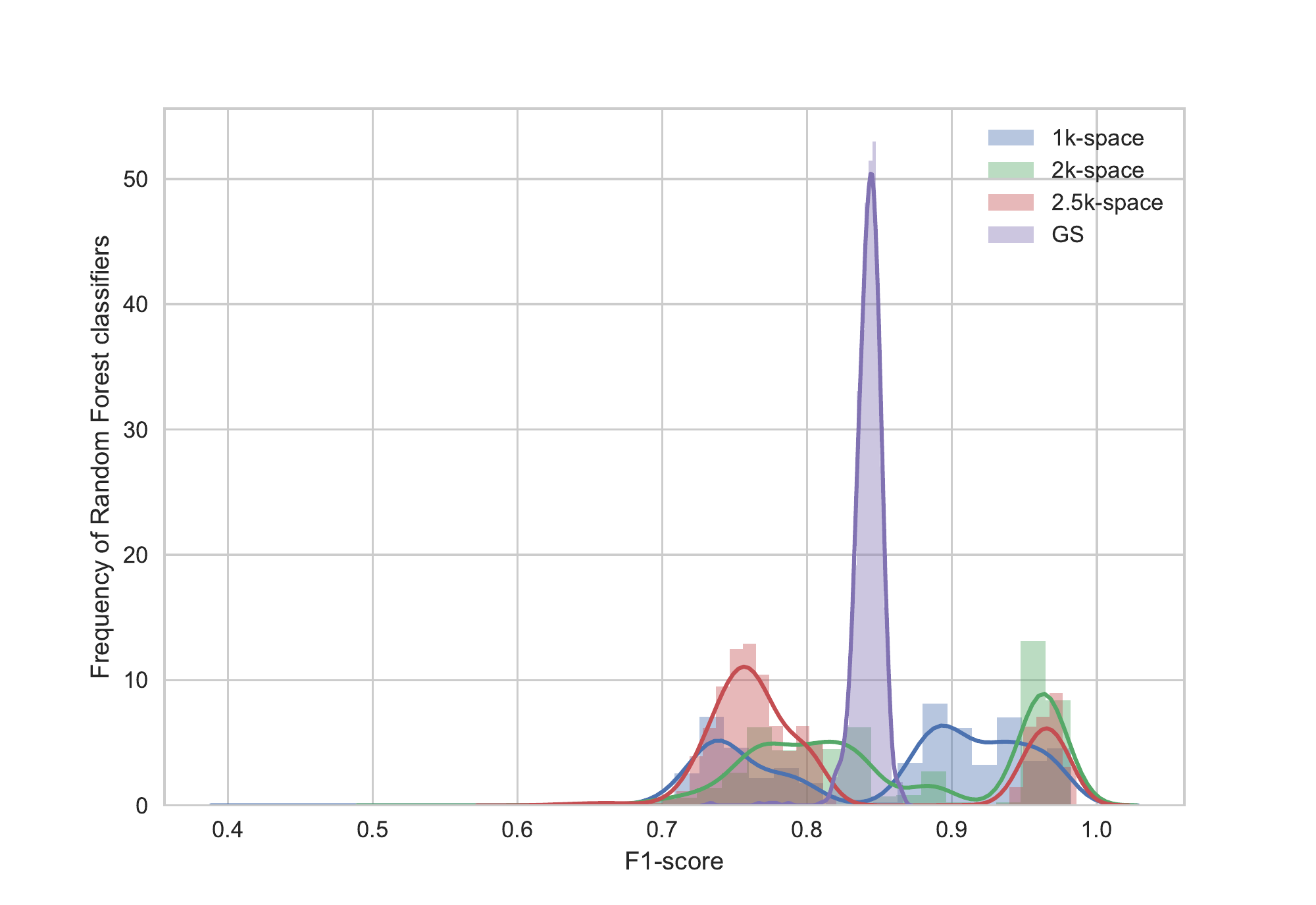}
	\label{fig:dk_space_vulnerability_web_bfs2}}\\
	\end{tabular}
	\caption{Accuracy of prediction of identical pairs across $dK$ spaces. GS denotes the instances of original subgraphs at an attacker's disposal, which represents a strong attack model.
	Note four overlap strategies; random selection of nodes (R),  highest degree nodes (HD), while BFS-HD and BFS-R denote the breadth-first-search tree starting from a random selection and the highest degree based selection respectively. 
	}
	\label{fig:dk_space_vulnerability}
\end{figure*}

Our second objective is to evaluate the relative power of $dK$-based graph generation techniques for anonymization. 
Figures~\ref{fig:dk_space_vulnerability} presents the accuracy of re-identification across $dK$-spaces. 
As baseline, we also include the results of classifiers on the non-anonymized original graph, denoted by "GS".
As expected, the re-identification of nodes in GS is always better than for the "anonymized" $dK$-space graphs. 
This clearly shows there is some anonymization power in the $dK$ graph-generation techniques. 
Also expected, there is a clear increase in the classifier accuracy of classifiers towards higher $dK$-spaces (Figures~\ref{fig:dk_space_vulnerability_anybeat_ran1}-\ref{fig:dk_space_vulnerability_web_bfs2}).
In general, the $1K$-space is the stronger anonymization approach, as it leads to the poorest re-identification. 
This confirms the tradeoff between utility and anonymity, since $1K$-graphs maintain the lowest utility.

However, this transition of vulnerability between $dK$ spaces is not significant for the networks of \texttt{fb107} (Figures~\ref{fig:dk_space_vulnerability_fb107_ran1}-\ref{fig:dk_space_vulnerability_fb107_bfs2}) and \texttt{caGrQc} (Figures~\ref{fig:dk_space_vulnerability_caGrQc_ran1}-\ref{fig:dk_space_vulnerability_caGrQc_bfs2}). 
These graphs are denser than the others.
We reason that even a weak attack is helped by graph density, compensating thus for a stronger anonymization approach. 
This is independently supported by Sharad~\cite{Sharad2016benchmark} who observed that the properties of dense graphs are more resilient to edge perturbation, and thus more structural knowledge is preserved in the anonymized graph topology.

We also notice some unexpected cases: $1K$-space tends to be more vulnerable than their counterparts $2K$ and $2.5K$ in sparse graphs with the BFS-HD overlap (Figures~\ref{fig:dk_space_vulnerability_anybeat_bfs1},\ref{fig:dk_space_vulnerability_gplus_bfs1} and~\ref{fig:dk_space_vulnerability_web_bfs1}). 
In order to understand this behavior, we analyze the properties of the connected neighborhood, especially the proportion of nodes with degree one (Figure~\ref{fig:low_degree}). 
This is because nodes with low degree have less structural information, thus they are more difficult to re-identify~\cite{ji2014structural}.

\begin{figure}[bthp]
	\centering
	\subfloat[HD]{
		\includegraphics[scale=0.2]{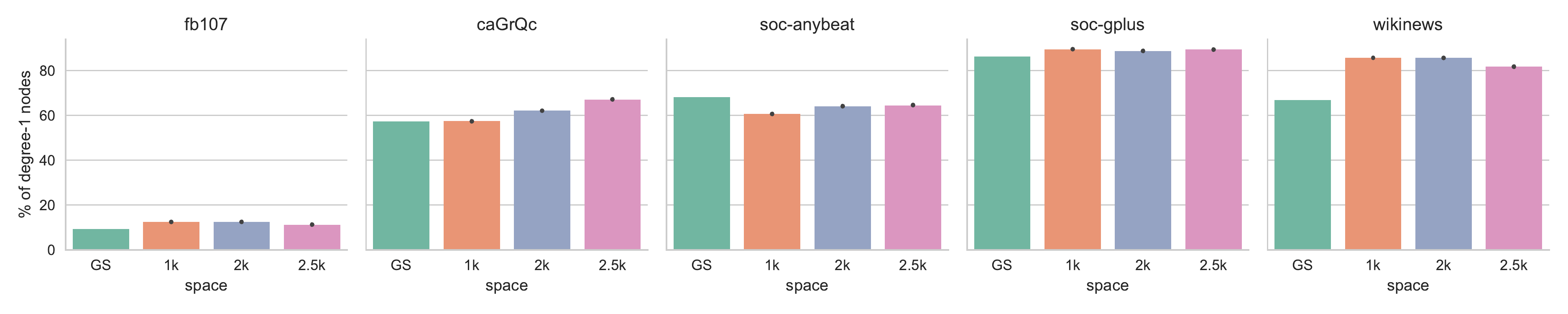}
			\label{fig:low_degree_deg}
	}
	\hspace{0mm}
	\subfloat[BFS-HD]{
		\includegraphics[scale=0.2]{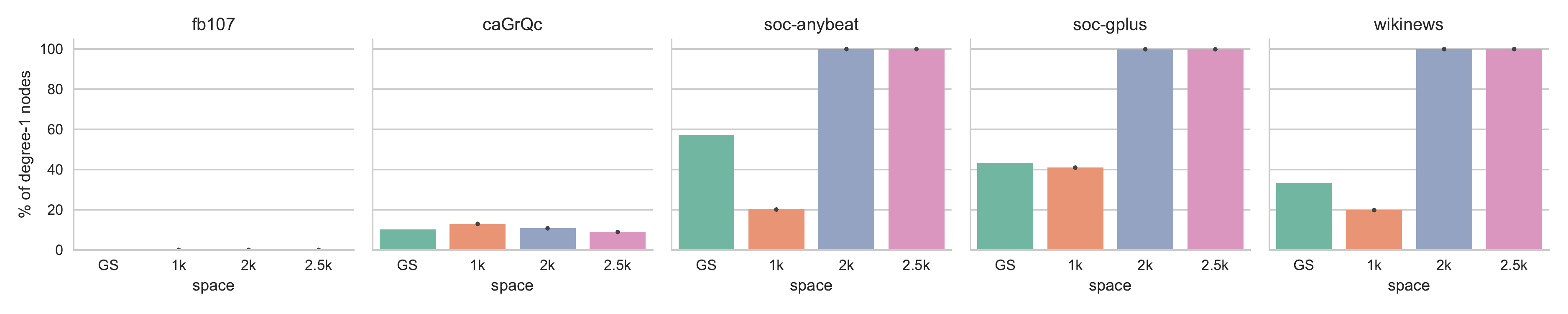}
			\label{fig:low_degree_bfs1}
	}
	\caption{Average percentage of nodes with degree 1 in the overlap subgraph across $dK$-spaces (including GS). The overlaps are averaged over 8 sample instances in each $dK$ space for two overlap strategies: HD (highest-degree nodes) (a) and BFS rooted in the highest degree node (b). Figure also shows the confidence intervals of average measurements.}
	\label{fig:low_degree}
\end{figure}

We observe that the proportion of degree-1 nodes in the overlap has a strong negative correlation with the accuracy of the classifier in $dK$ levels.
As an example, \texttt{soc-anybeat} has an average $20 \%$ of degree-1 nodes in the overlap in the $1K$-space and $99 \%$ in $2K$ and $2.5K$ spaces (Figure~\ref{fig:low_degree_bfs1}).
$1K$-space turns out to be more vulnerable to the re-identification attack than $2K$ and $2.5K$ spaces (Figure~\ref{fig:dk_space_vulnerability_anybeat_bfs1}).




\subsection{Properties of Vulnerable Graphs}

\begin{figure*}[htb!]
	\centering
	\subfloat[BFS-HD: GS]{
		\includegraphics[width=0.25\linewidth]{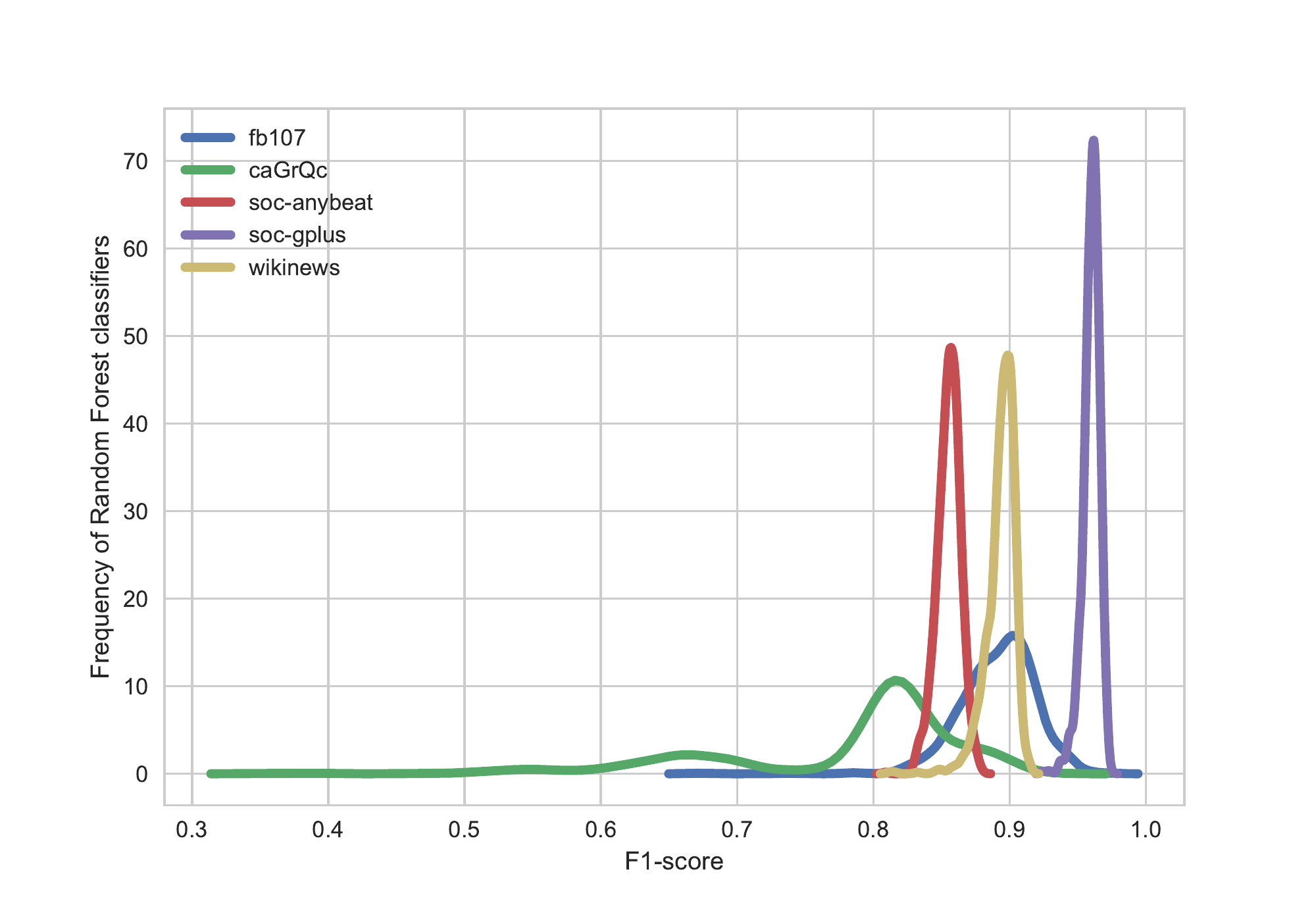}
		\label{fig:vulnerable_graphs_0-2_bfs1_gs}
	}
	\subfloat[BFS-HD: $1K$ space]{
		\includegraphics[width=0.25\linewidth]{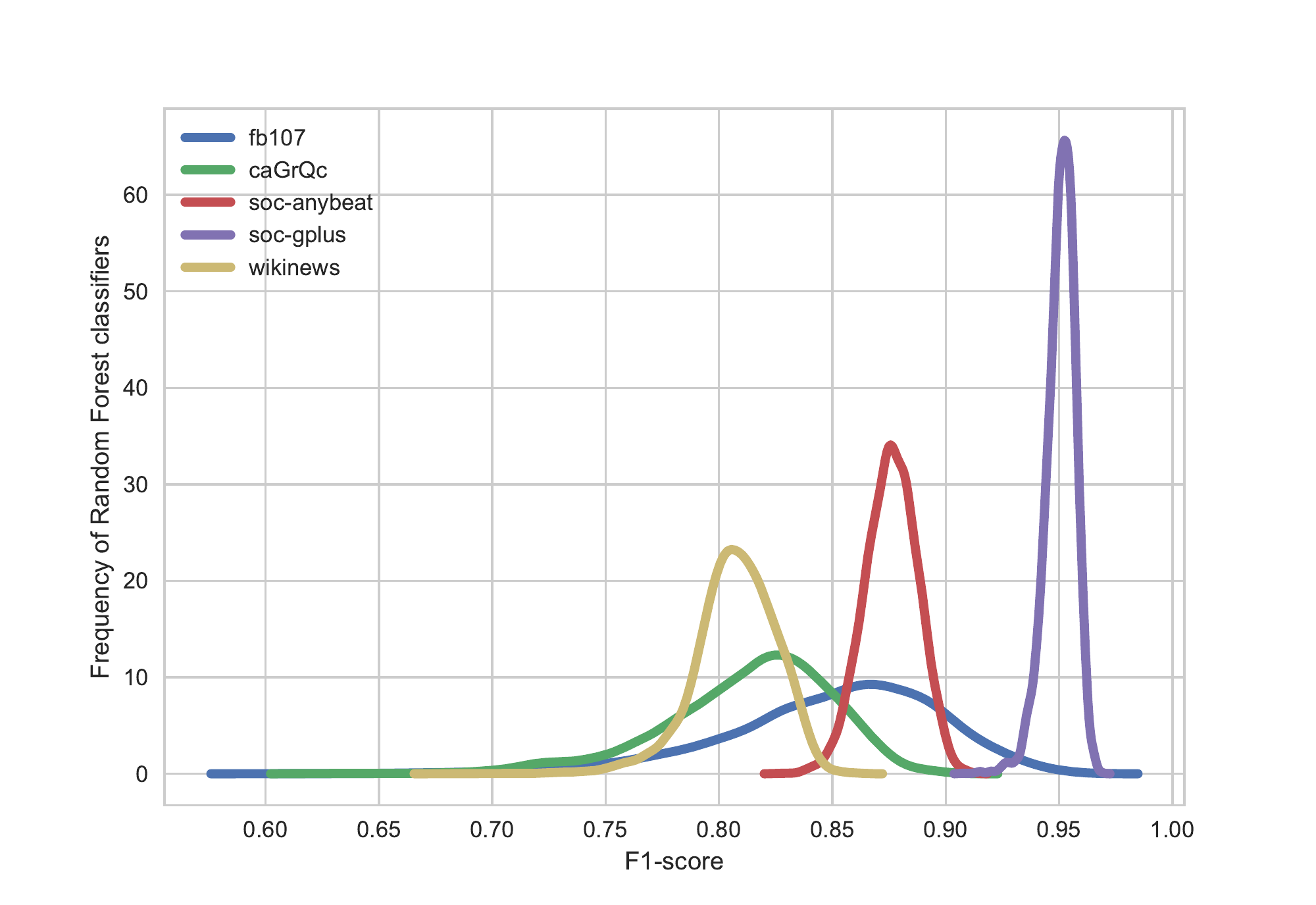}
		\label{fig:vulnerable_graphs_0-2_bfs1_1k}
	}
	\subfloat[BFS-HD: $2K$ space]{
		\includegraphics[width=0.25\linewidth]{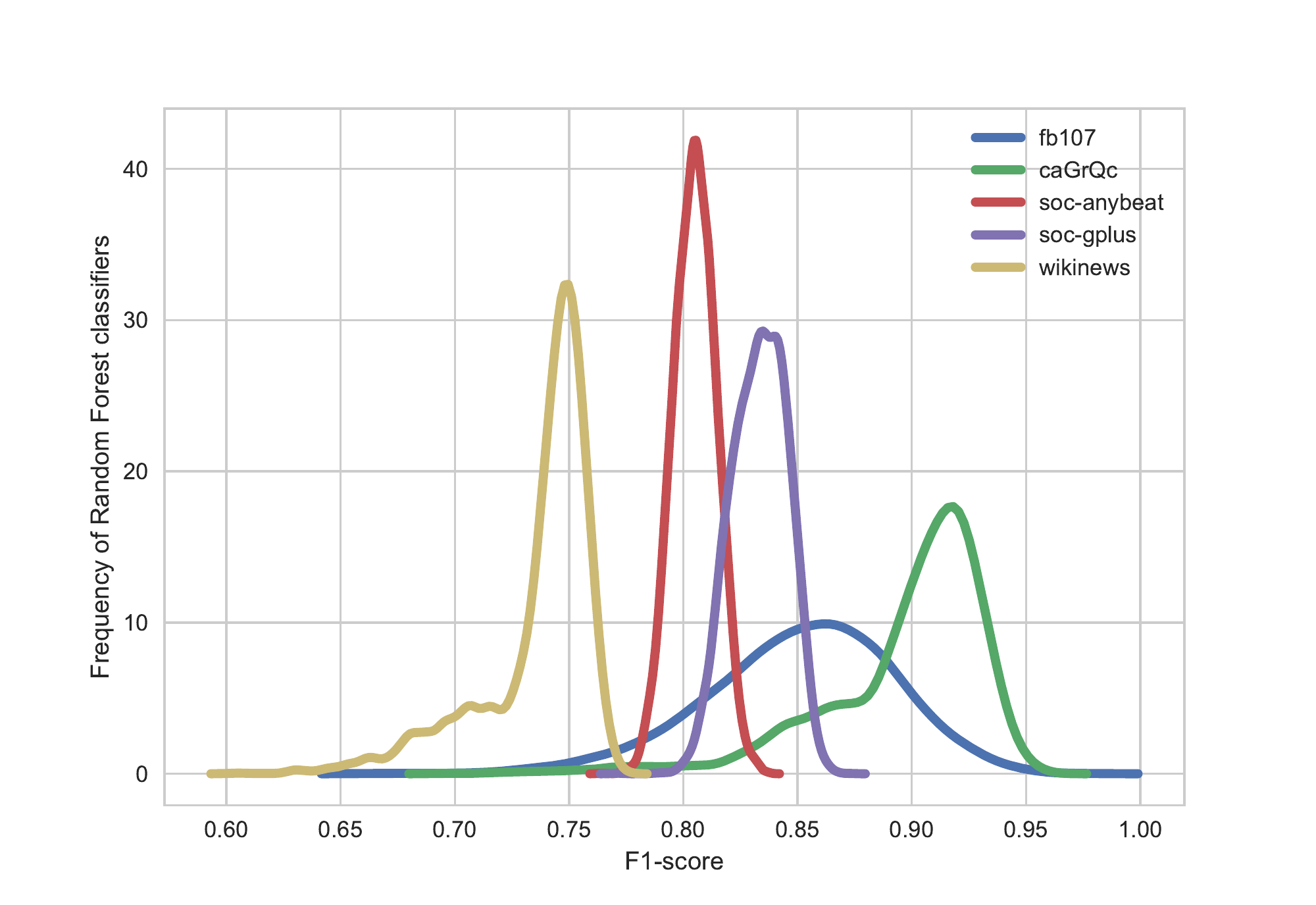}
		\label{fig:vulnerable_graphs_0-2_bfs1_2k}
	}
	\subfloat[BFS-HD: $2.5K$ space]{
		\includegraphics[width=0.25\linewidth]{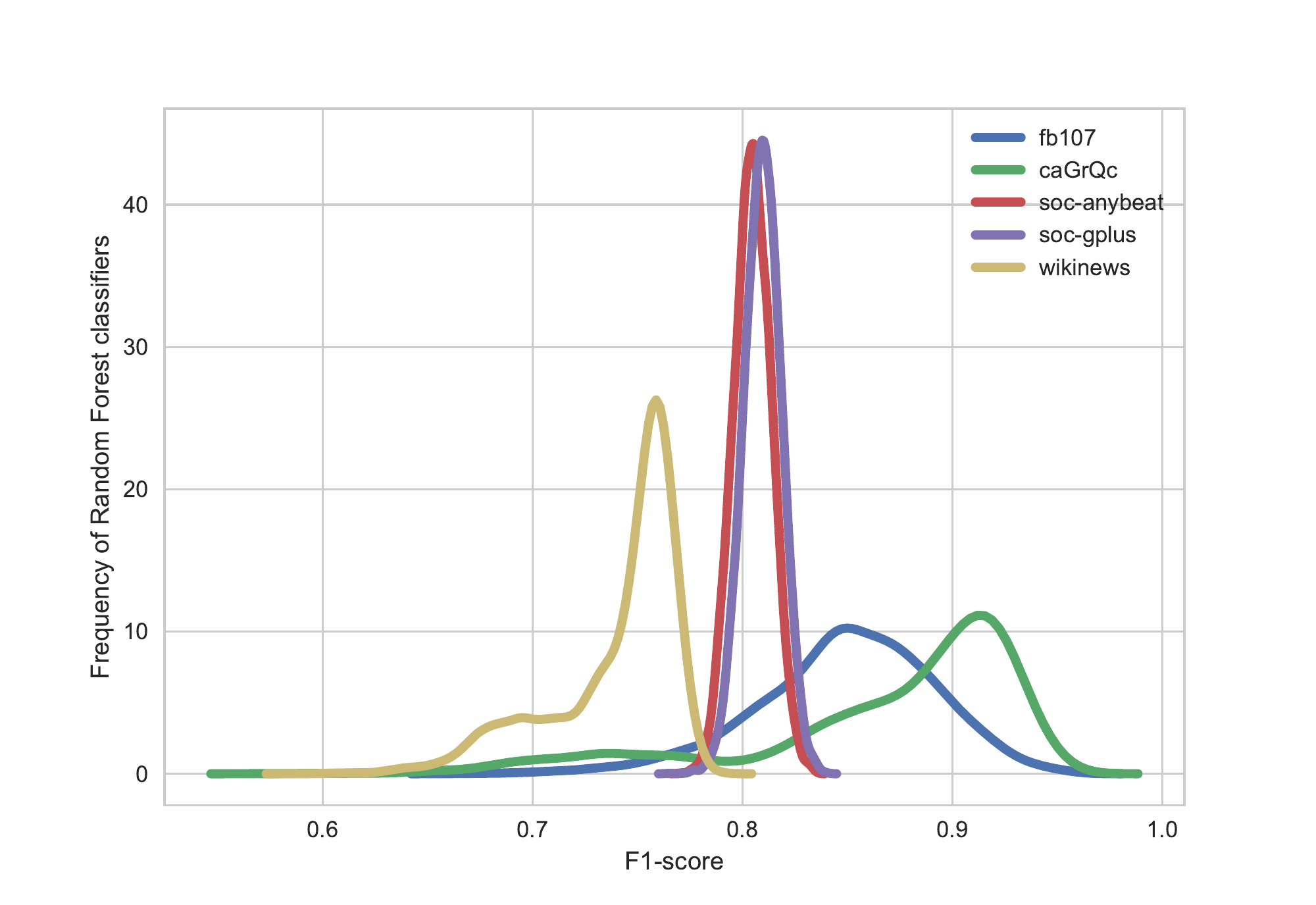}
		\label{fig:vulnerable_graphs_0-2_bfs1_25k}
	}
	\hspace{0mm}
	\subfloat[HD: GS]{
		\includegraphics[width=0.25\linewidth]{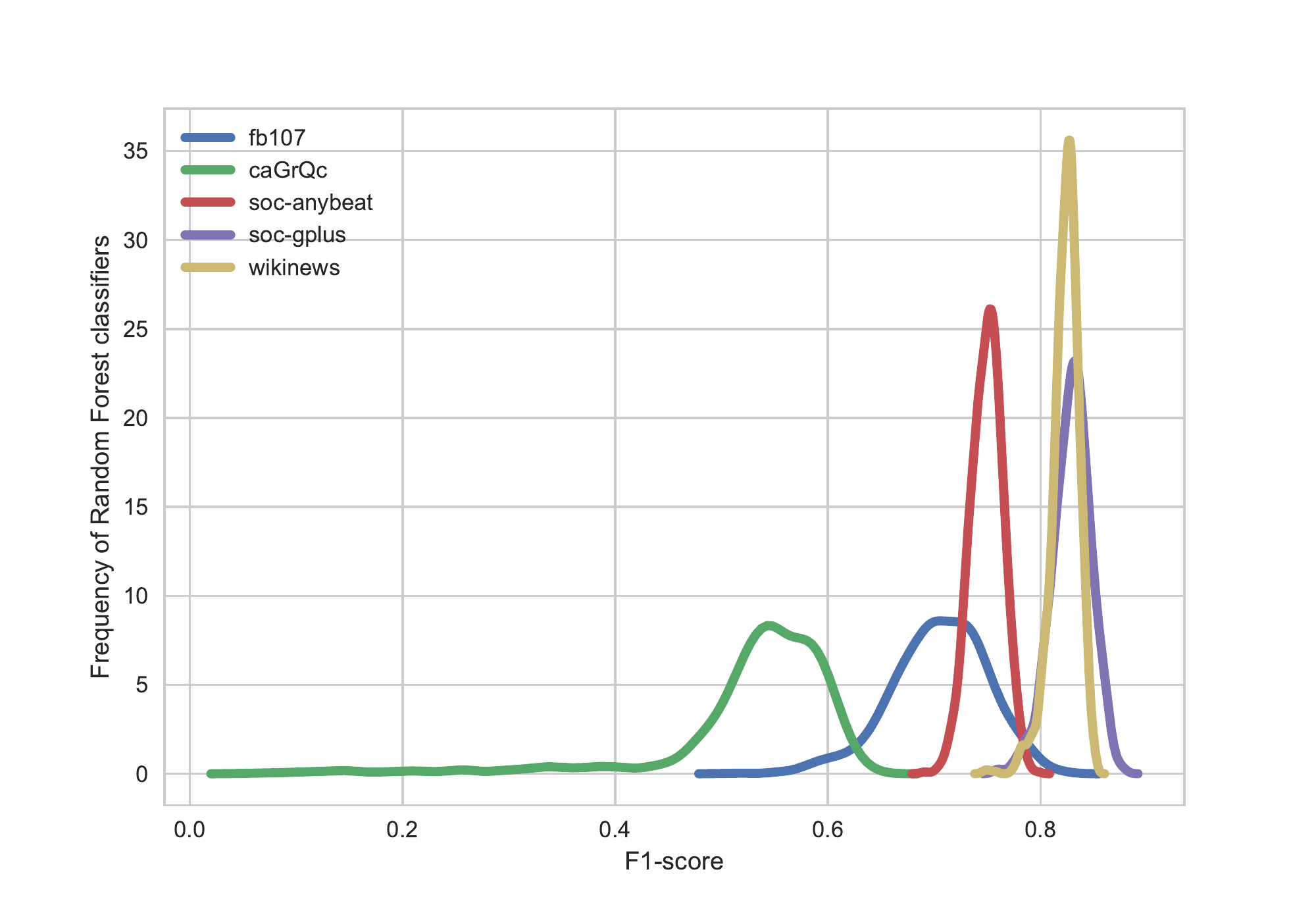}
		\label{fig:vulnerable_graphs_0-2_deg_gs}
	}
	\subfloat[HD: $1K$ space]{
		\includegraphics[width=0.25\linewidth]{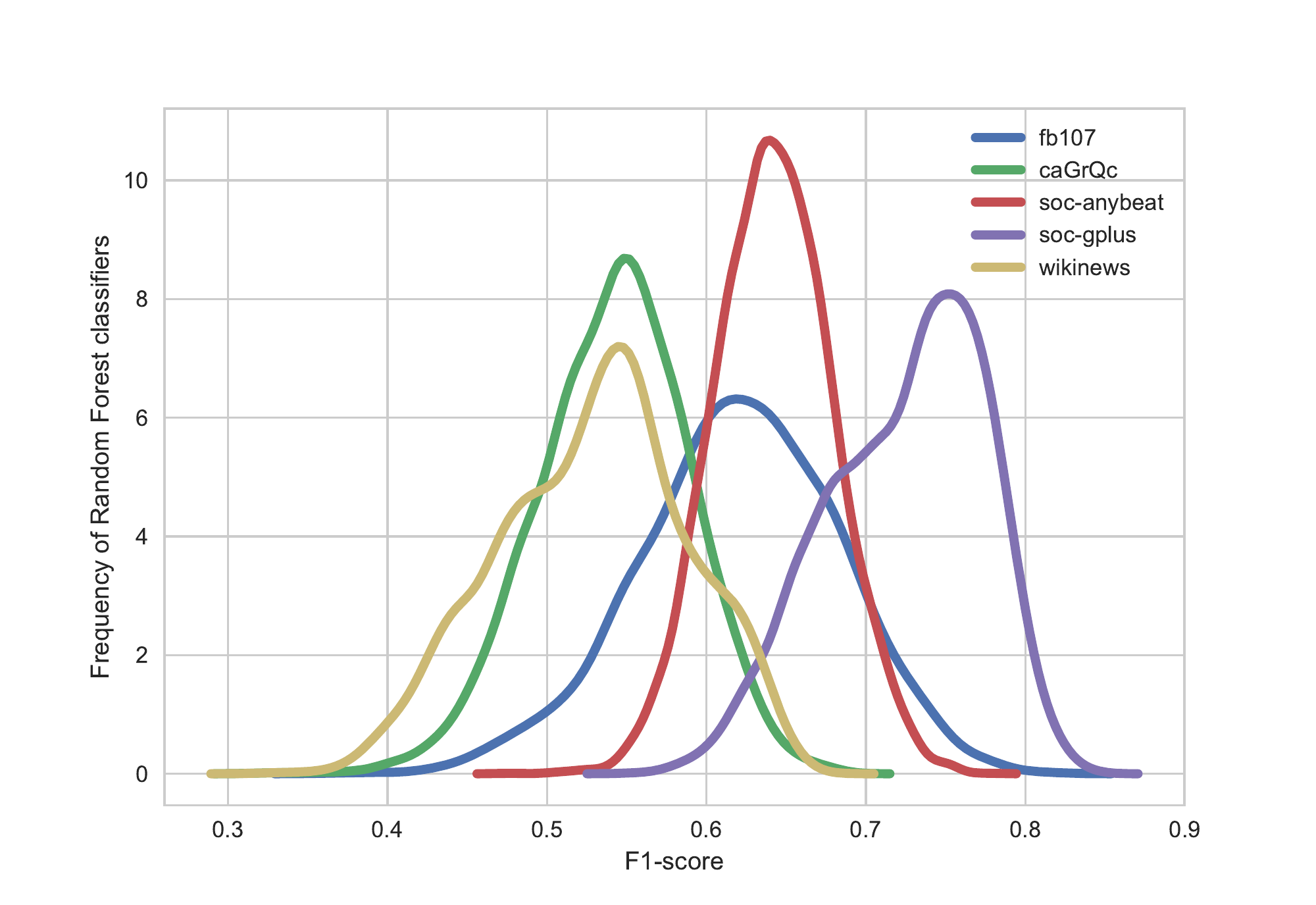}
		\label{fig:vulnerable_graphs_0-2_deg_1k}
	}
	\subfloat[HD: $2K$ space]{
		\includegraphics[width=0.25\linewidth]{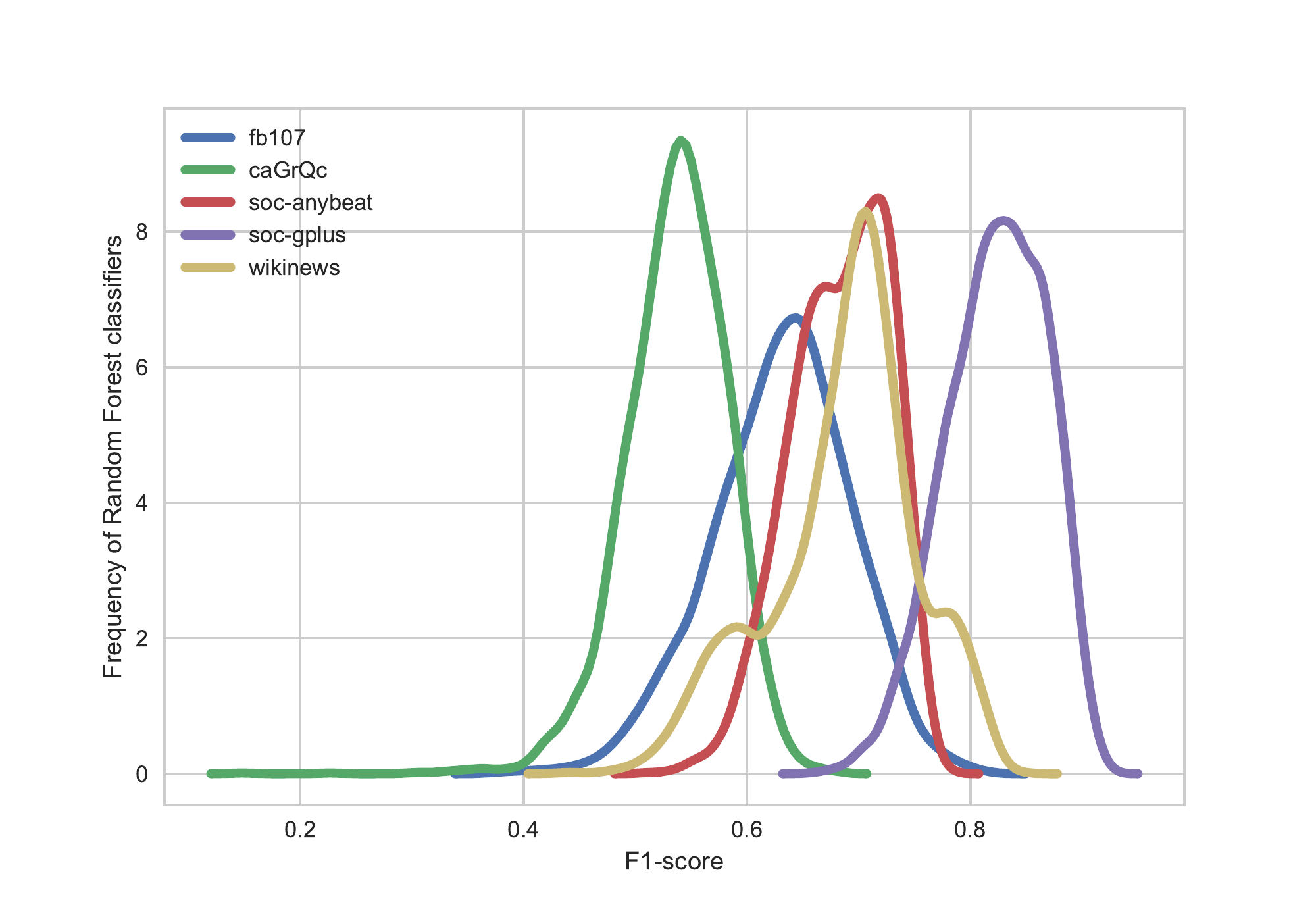}
		\label{fig:vulnerable_graphs_0-2_deg_2k}
	}
	\subfloat[HD: $2.5K$ space]{
		\includegraphics[width=0.25\linewidth]{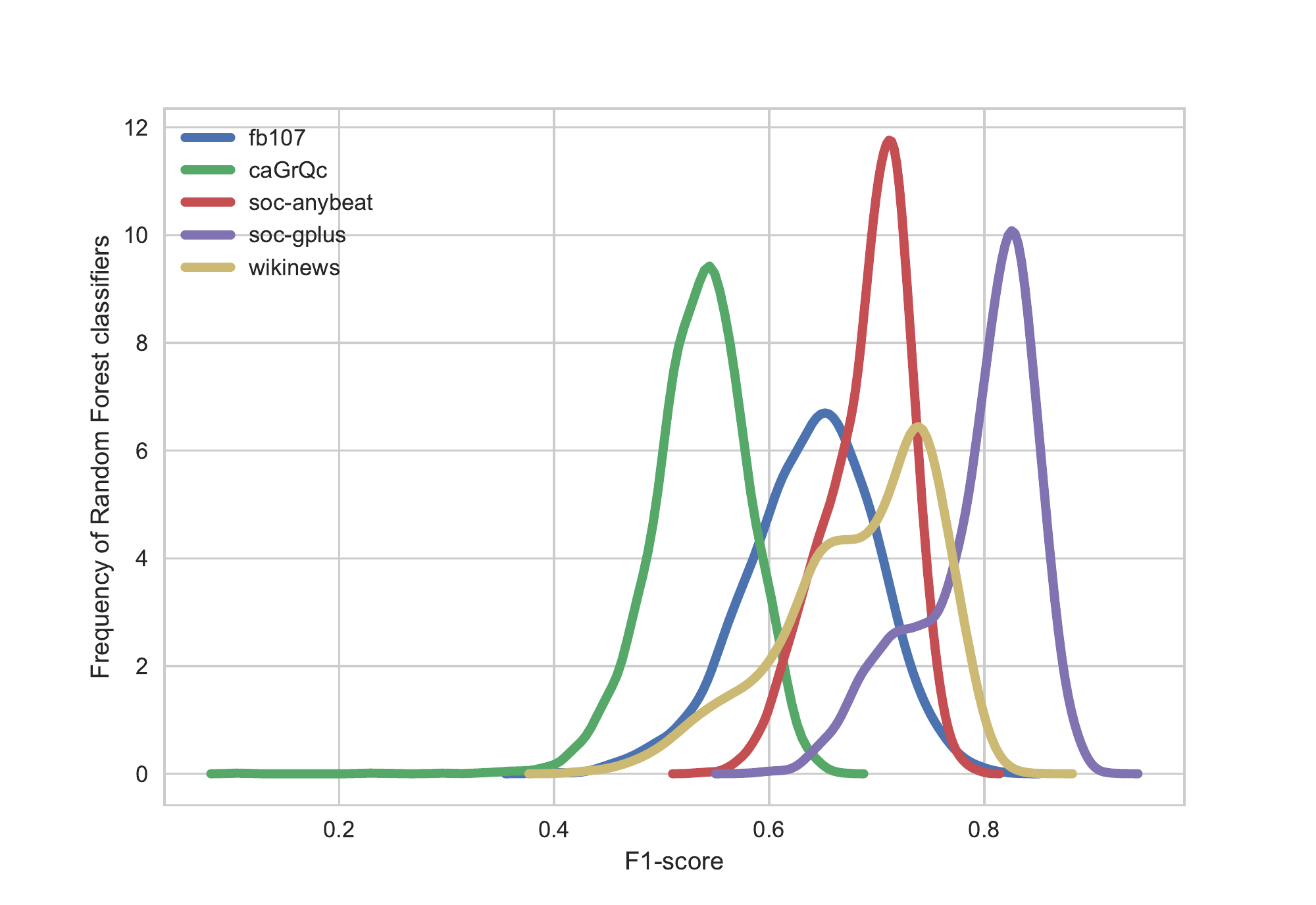}
		\label{fig:vulnerable_graphs_0-2_deg_25k}
	}
	\caption{Kernel Density Estimate (KDE) of the accuracy of predictions over identical pairs across several graphs. Results are presented in each $dK$-space (including GS) under the overlap choices of breadth-first-search tree starting from a degree based selection (BFS-HD) and High-Degree (HD).}
	\label{fig:vulnerable_graphs_0-2}
\end{figure*}

In this section we are asking: \emph{What makes some graph datasets more vulnerable to attacks than others?}
In order to isolate the graph datasets (as much as possible, given our methodology, see discussion in Summary section) from the threat model, we focus on only two choices of the attacker knowledge base, namely the BFS-HD and HD overlaps.
Because the overlap subgraphs in these two cases are deterministically generated, the $G_{san}$ and $G_{aux}$ graphs more accurately represent the structural properties of the original dataset. 
For example, the \texttt{fb107} original graph has density = 0.1 (as shown in Table~\ref{tbl:dataset}) and the corresponding generated $dK$ graphs approximate it by a mean error of 0.01 under the overlap choice of BFS-HD (as shown in Figure~\ref{fig:attack_strength}). 
Similarly, the degree assortativity in the original graph is approximated by the generated graphs in 2K and 2.5K spaces with a mean error of 0.03. 

Figures~\ref{fig:vulnerable_graphs_0-2_bfs1_gs}--\ref{fig:vulnerable_graphs_0-2_bfs1_25k} show the kernel density estimate (KDE) of the accuracy of predictions across graphs in each $dK$-space for the overlap choice of BFS-HD.
We observe that the vulnerability of the dataset is monotonically increasing with its transitivity in the $2K$ and $2.5K$ spaces: the higher the transitivity, the higher the vulnerability. 

For the graphs generated by constraining the overlapping subgraphs to high degree nodes, we preserve the assortativity of the original dataset in the $2K$ and $2.5K$ spaces. 
Figure~\ref{fig:vulnerable_graphs_0-2_deg_gs}--\ref{fig:vulnerable_graphs_0-2_deg_25k} shows that, consistently across all $dK$-spaces, \texttt{soc-gplus} is the most vulnerable dataset while the \texttt{caGrQc} is the least.
As Table~\ref{fig:attack_strength} shows, \texttt{soc-gplus} has the lowest degree assortativity ($r=-0.53$), while \texttt{caGrQc} has the largest ($r=0.6592$). 
Moreover, the relative order in vulnerability is preserved across all datasets with regard to degree assortativity.
We believe that degree assortativity is a telling sign for the vulnerability of the original dataset under the HD overlap construction scenario because of the ways nodes are represented as feature vectors.
Basically, for a high degree node, given that the degree assortativity is negative, most of its neighbors will be low degree. 
This likely means that all these low degree neighbors will be represented by the first bin in the node signature. 
The learning algorithm is thus likely to have seen many examples of values for the first bin, and thus it can easily recognize a node with a distinctive, unique signature (identified by the very first bin). 
We observe the same relative order in GS, supported by the same observation related to assortativity. 

\section{Summary and Discussions}
\label{sec:discussions}
We investigate empirically the power of $dK$-series for structural graph anonymization. 
The main contribution of this paper lies in the questions it answers empirically. 
Specifically, we evaluate the relative benefits of $dK$-random graphs for providing graph anonymity under a generalized attack supported by machine learning techniques. 
We quantify the strength of the attacker based on the structural graph properties of the knowledge it has access to in the training phase. 
Finally, we discover some of the structural graph properties that makes a dataset more vulnerable to de-anonymization attacks. 
To reach these conclusions, we implemented and augmented an experimental framework previously proposed in the literature~\cite{Sharad2016benchmark}.

Starting from five real datasets that represent social graphs from different contexts, we discover the following. 
First, we empirically measure the benefits of $dK$-based anonymization over the original datasets. 
For the very same attack scenario, the nodes of the original dataset (stripped of node identities) can be re-identified with the median accuracy of 83\%, compared to 74\% for a 1K-random graph, 79\% for a 2K-random graph, and 77\% for a 2.5K-random graph generated from the respective degree distributions of the original graph. 
Albeit limited to a small set of examples, this result is telling, as degree distribution is one of the main utility metrics in graph generation (and anonymization), thus of potential interest for related work, such as differential privacy~\cite{sala2011sharing,wang2013preserving}. 
As well understood, 1K graphs are (usually) less vulnerable than 2K and 2.5K graphs, under the accepted utility-anonymity tradeoff rule.

Second, we show that if the attacker starts with seed knowledge in the form of a subgraph with higher density and higher transitivity, re-identification is more potent.
On the other hand, a seed knowledge database that includes high degree nodes that are not strongly connected is less enabling.

Third, we discover that the tradeoff utility-anonymity is not always maintained: in particular cases, both utility and anonymity are lost, and this is due to the exploits of topology with less structural information by an attack. 
In other words, there is some defensive power in any graph depending on the amount of structural information exploited by an attack.
As an example, it's relatively hard to de-anonymize degree one nodes solely based on structural information~\cite{ji2014structural}.
When a large proportion of such nodes exists in any $dK$ graph, it's still difficult to de-anonymize the majority.

Fourth, we identified low (negative) degree assortativity as a canary for the vulnerability to the specific de-anonymization attacks we employed. 

This work has some caveats that we plan to address in future work. 
First, our empirical analysis is limited to five real datasets. 
We would like to extend this to gain more confidence on our observations. 
Second, our methodology allows for a blurred line between measuring the strength of the attacker and characterizing the original dataset. 
This is because the way we calibrate the strength of the attacker---by determining the topological structure that connects nodes that may be visible to the attacker as part of the training set---influences the topology of the sample graphs for testing. 
While we attempted to isolate the characterization of the two, we plan to improve our experimental framework to eliminate potential biases. 

We also would like to extend our work to other graph generation models (e.g., exponential random graph models~\cite{anderson1999p}), and compare the relative anonymization power with $dK$ models, especially related to more sophisticated \emph{attribute disclosure} attacks.

\bibliographystyle{ACM-Reference-Format}
\bibliography{main}

\end{document}